\documentclass[fleqn,showpacs,showkeys]{revtex4}
\usepackage{graphicx}
\usepackage{amssymb}
\usepackage{amsmath}
\usepackage{bm}
\usepackage[flushleft]{caption2}

\begin{document}
\setcounter{page}{1}
\title{Color Confinement and Spatial Dimensions in the Complex-sedenion Space}
\author{Zi-Hua Weng}
\email{xmuwzh@xmu.edu.cn.}
\affiliation{School of Physics and Mechanical \& Electrical Engineering, \\Xiamen University, Xiamen 361005, China}

\begin{abstract}
  The paper aims to apply the complex-sedenions to explore the wavefunctions and field equations of non-Abelian gauge fields, considering the spatial dimensions of a unit vector as the color degrees of freedom in the complex-quaternion wavefunctions, exploring the physical properties of the color confinement essentially. J. C. Maxwell was the first to employ the quaternions to study the physical properties of electromagnetic fields. His method inspires some subsequent scholars to introduce the quaternions, octonions, and sedenions to research the electromagnetic field, gravitational field, nuclear field, quantum mechanics, and gauge field. The application of complex-sedenions is capable of depicting not only the field equations of the classical mechanics on the macroscopic scale, but also the field equations of the quantum mechanics on the microscopic scale. The latter can be degenerated into the Dirac equation and Yang-Mills equation. In contrast to the complex-number wavefunction, the wavefunction in the complex-quaternion space possesses three new degrees of freedom, that is, three color degrees of freedom. One complex-quaternion wavefunction is equivalent to three conventional wavefunctions with the complex-numbers. It means that the three spatial dimensions of unit vector in the complex-quaternion wavefunction can be considered as the `three colors', naturally the color confinement will be effective. In other words, in the complex-quaternion space, the `three colors' are only the spatial dimensions, rather than any property of physical substance. The existing `three colors' can be merged into the wavefunction, described with the complex-quaternions.
\end{abstract}

\pacs{02.10.De; 03.65.-w; 12.10.-g; 12.15.-y; 24.85.+p; 25.75.Nq.}

\keywords{color degrees of freedom; color confinement; nuclear field; non-Abelian gauge field; quantum mechanics; octonion; sedenion.}

\maketitle

\section{\label{sec:level1}INTRODUCTION}

The essence of the color degrees of freedom in the strong-nuclear fields is attracting increasing attention of many researchers. Since a long time, this conundrum has been intriguing and puzzling numerous scholars. For many years, this puzzle urged some scholars to propose several sorts of hypotheses, attempting to reveal the essence of the color degrees of freedom. Until recently, the appearance of the quantum mechanics on the microscopic scale, described with the complex-sedenions, replies partially to this puzzle. The quantum mechanics, described with the complex-sedenions, is capable of deducing the wavefunctions, Dirac wave equations, Yang-Mills equations of the non-Abelian gauge field, and field equations of the electroweak field and so forth. Especially, their wavefunctions, described with the complex-quaternions, are able to elucidate directly the color degrees of freedom.

C.-N. Yang and R. L. Mills found the field equations for a non-Abelian gauge field in 1954. The field equations of non-Abelian gauge fields were supposed to be applicable to each of four fundamental fields. After the mass problem was finally conquered by means of the spontaneous symmetry breaking and Higgs mechanism, the Yang-Mill equations can be applied to construct successfully the unified theory of the electromagnetic field and weak-nuclear field. The triumph of the electroweak theory edified a part of scholars to extend the Yang-Mill equations into the strong-nuclear fields, establishing the quantum chromodynamics (QCD for short).

M. Gell-Mann and G. Zweig posited independently the model of quarks in 1964. In the same year, O. W. Greenberg introduced a sort of the degree of freedom for the subatomic particles, that is, the color degree of freedom. In the QCD, the scholars assume that a species of `charge' is capable of producing the strong-nuclear field, and possesses `three colors'. And it is called as the color charge, which is similar to the electric charge in the electromagnetic fields. The color charge is a fundamental and crucial assumption, and is the theoretical footstone of the QCD. On the basis of the assumption of color charge, it is able to conduct a variety of theoretical inferences and perform diverse experimental validations, for the QCD. Up to now, the color charge of quarks and gluons have not yet been observed directly in the experiments. As a result, a rule had been summarized to explain correlative physical phenomena. That is, only the `colorless' hadrons can be studied or observed in the physics. This rule was called as the quark confinement or color confinement.

However, the QCD is unable to elucidate the physical phenomena of color confinement by itself. Consequently it is necessary to introduce several hypotheses, including the lattice gauge theory, non-Abelian monopoles, microscopic resonance, hidden local symmetry, color charge, soliton model, string theory and so forth. Using the magnetic symmetry structure of non-Abelian gauge theories of the Yang-Mills type, Chandola \emph{et al.} \cite{color1} discussed the mathematical foundation of dual chromodynamics in fiber-bundle form. Nakamura \emph{et al.} \cite{color2} studied the long-range behavior of the heavy quark potential in Coulomb gauge, using a quenched SU(3) lattice gauge simulation with partial-length Polyakov line correlators. Chaichian \emph{et al.} \cite{color3} adopted that the colored states are unphysical, so the colored particles cannot be observed. And there are two phases in QCD distinguished by different choices of the gauge parameter. Eto \emph{et al.} \cite{color4} argued that the dual transformation of non-Abelian monopoles, resulting in the dual system being in the confinement phase. Wang \emph{et al.} \cite{color5} proposed a new kind microscopic resonance, the color confinement multi quark resonance. Suzuki \emph{et al.} \cite{color6} observed the Abelian mechanism of non-Abelian color confinement in a gauge-independent way, by the high precision lattice Monte Carlo simulations in the gluodynamics. The authors \cite{color7} studied the mechanism of non-Abelian color confinement in SU(2) lattice gauge theory, in terms of the Abelian fields and monopoles extracted from the non-Abelian link variables without adopting gauge fixing. Yamamoto \emph{et al.} \cite{color8} proposed a new lattice framework to extract the relevant gluonic energy scale of QCD phenomena, which is based on a `cut' on link variables in momentum space. Yu \emph{et al.} \cite{color9} studied the liquid properties of the strongly coupled quark-gluon plasma during the intermediate stage and after the end of crossover from the hadronic matter to the strongly coupled quark-gluon plasma, by means of the bond percolation model. Braun \emph{et al.} \cite{color10} identified a simple criterion for quark confinement, computing the order-parameter potential from the Landau-gauge correlation functions. Troshin \emph{et al.} \cite{color11} discussed how the confinement property of QCD results in the rational unitarization scheme, and how the unitarity saturation leads to the appearance of a hadron liquid phase at very high temperatures. Kitano \emph{et al.} \cite{color12} identified the gauge theory, with the hidden local symmetry, as the magnetic picture of QCD, enabling a linearly realized version of gauge theory to describe the color confinement and chiral symmetry breaking. Pandey \emph{et al.} \cite{color13} studied an effective theory of QCD, in which the color confinement is realized through the dynamical breaking of magnetic symmetry, leading to the magnetic condensation of QCD vacuum. Gates \emph{et al.} \cite{color14} presented the evidence in some examples that an Adinkra quantum number seems to play a role with regard to off-shell 4D, N = 2 supersymmetry similar to the role of color in QCD. Brodsky \emph{et al.} \cite{color15} showed that a mass gap and a fundamental color confinement scale will arise, when one extends the formalism of de Alfaro, Fubini and Furlan to the frame-independent light-front Hamiltonian theory. The author \cite{color16} studied the light-front wavefunctions and the functional form of the QCD running coupling in the nonperturbative domain, connecting the parameter in the QCD running coupling to the mass scale underlying confinement and hadron masses. Kharzeev \emph{et al.} \cite{color17} modified the gluon propagator to reconcile perturbation theory with the anomalous Ward identities for the topological current in the vacuum, making explicit the connection between confinement and topology of the QCD vacuum. The existence of so many hypotheses, to attempt to elucidate the color confinement, announces that our present understanding with respect to the strong-nuclear field is quite inadequate.

Some scholars proposed the Standard Model of elementary particles, attempting to further unify the QCD and Electroweak theory. This promising unification hypothesis is anticipated to be a huge success, although it does not include the gravitational field. Nevertheless, the Standard Model is either unable to unpuzzle the color confinement, color charge, and dark matter and so forth. Consequently, some scholars put the effort towards a few theoretical schemes in recent years, such as, `Beyond the Standard Model', `Superstring theory', and `Beyond the Relativity' and so forth.

Making a detailed comparison and analysis of preceding studies, a few primal problems of these theories are found as follows:

a) Four interactions. Either of the QCD and Electroweak theories is incapable of containing the gravitational interaction. Even in the theories regarding the `Beyond the Standard Model', there is not any tangible theoretical scheme, to include the gravitational interaction. In the four fundamental interactions, the gravitational interaction is the first to be discovered in the history. Unfortunately, the gravitational interaction still lies outside the mainstream of current unification theories. It means that the mainstream of current unification theories may be seized of some fatal defects essentially, so that they are unable to cover and describe the four fundamental interactions simultaneously. It is well known that an appropriate unification theory must be able to depict the four fundamental interactions simultaneously, especially the gravitational interaction.

b) Color confinement. The QCD and other existing theories are incapable of revealing the essence of color degrees of freedom. Either they cannot determine whether the color degrees of freedom belong to the physical substance or spatial dimension, and even others. Therefore they are unable to explain effectively the physical properties of color degrees of freedom. Under the circumstance that the essence of color degrees of freedom cannot be comprehended, the QCD assumed prematurely the color degrees of freedom are induced by one sort of physical substance, that is, the color charge. Undoubtedly this assumption is governed by expediency. As one fundamental assumption of QCD, the color charge may encounter certain setbacks or challenges, bringing negative effects on the subsequent development of theory. However, even though the color charge was deemed as one species of the physical substance, the QCD and other existing theories are still unable to account for the color confinement, laying themselves open to suspicion. The assumption of color charge in the QCD is appealing for more validation experiments. Up to now, the QCD may not be really perfect yet, especially its fundamental assumption.

c) Dark matter. The existence of the physical phenomena connected with the dark matters was firstly validated in the astronomy, and then was accepted generally by the whole academic circle. Nevertheless the Standard Model is blind to the existence of dark matters, and it is unable to elucidate the relevant physical phenomena either. Further the Standard Model and even the `Beyond the Standard Model' are incapable of predicting or inferring the confirmed dark matters. It means that the research scope, in the mainstream of current unification theories, is restricted and insufficient enough. For the unification theory, which is unable to explore the dark matter, there may be still something left to be improved. Apparently an appropriate unification theory must be capable of comprising and exploring the ordinary matter and dark matter simultaneously.

Presenting a striking contrast to the above is that it is able to account for a few problems, derived from the QCD and other existing theories, in the quantum mechanics described with the complex-sedenions, trying to improve the unification theory relevant to the four fundamental interactions to a certain extent. J. C. Maxwell was the first to employ the algebra of quaternions to explore the physical properties of electromagnetic fields. And it inspired the subsequent scholars to apply the quaternions, octonions, and sedenions to investigate the gravitational field, electromagnetic field, nuclear field, Dirac wave equation, Yang-Mills equation, electroweak field, color confinement, and dark matter field and so forth. When a part of coordinate values are complex-numbers, the quaternion, octonion, and sedenion are called as the complex-quaternion, complex-octonion, and complex-sedenion respectively.

In recent years, a part of scholars make use of the algebra of quaternions to research the electromagnetic field equations, quantum mechanics, gravitational fields, dark matters \cite{weng1}, and curved space and so on. Edmonds \cite{quaternion1} expressed the wave equation and gravitational theory with the quaternion in the curved spacetime. Doria \cite{quaternion2} researched the gravitational theory, by means of the quaternions. Winans \cite{quaternion3} described some physics quantities by means of the quaternion. Honig \cite{quaternion4} and Singh \cite{quaternion5} respectively applied the complex-quaternion to express the Maxwell's equations in the classical electromagnetic theory. Brumby \emph{et al.} \cite{quaternion6} studied the possibility that dark matter may arise as a consequence of the underlying quaternionic structure to the universe. And the authors \cite{quaternion7} explored some global consequences of a model quaternionic quantum field theory, enabling the quaternionic structure to induce the cosmic strings and nonbaryonic hot dark matter candidates. Majernik \cite{quaternion8} adopted the quaternions to deduce the modified Maxwell-like gravitational field equations. The author \cite{quaternion9} studied a model of the universe consisting of a mixture of the ordinary matter with cosmic quaternionic field, researching the interaction of the quaternionic field with the dark matter. Anastassiu \emph{et al.} \cite{quaternion10} explored the physics properties of the electromagnetic field with the quaternions. Grusky \emph{et al.} \cite{quaternion11} investigated the time-dependent electromagnetic field, by means of the quaternions. Morita \cite{quaternion12} studied the quaternion field theory, making use of the quaternions. Rawat \emph{et al.} \cite{quaternion13} explored the gravitational field, by means of the quaternion terminology. Davies adopted the quaternions to investigate the Dirac equation \cite{quaternion14}.

Some scholars introduce the algebra of octonions \cite{weng2} to explore the Dirac wave equation, curved space, electromagnetic field equations, gravitational fields, dark matters, and Yang-Mills equation and so forth. Penney \cite{octonion1} applied the octonions to extract the square root of the classical relativistic Hamiltonian, finding the resulting wave equation to be equivalent to a pair of coupled Dirac equations. Marques-Bonham \cite{octonion2} developed the geometrical properties of a flat tangent space-time local to the manifold of the Einstein-Schr\"{o}dinger non-symmetric theory on an octonionic curved space. Abdel-Khalek \emph{et al.} \cite{octonion3} introduced left-right barred operators to obtain a consistent formulation of octonionic quantum mechanics, developing an octonionic relativistic free wave equation. Koplinger \cite{octonion4} provided proof to a statement that the Dirac equation in physics can be found in conic sedenions. As the subalgebra of conic sedenions, the hyperbolic octonions are used to describe the Dirac equation. Gogberashvili \cite{octonion5} applied the octonions to investigate the electromagnetic field equations. Mironov \emph{et al.} \cite{octonion6} demonstrated a generalization of relativistic quantum mechanics using the octonionic wavefunction and octonionic spatial operators. The second-order equation for octonionic wavefunction can be reformulated in the form of a system of first-order equations for quantum fields. The authors \cite{octonion7} demonstrated a generalization of relativistic quantum mechanics using the octonions, generating an associative noncommutative spatial algebra. The octonionic second-order equation for the octonionic wavefunction may describe the particles with spin 1/2. The authors \cite{octonion8} made use of the octonion to describe the electromagnetic field equations and related features. Meanwhile Dundarer \cite{octonion9} defined a four-index antisymmetric and non-Abelian field, which satisfies a self-duality relation in eight-dimensional curved space. Demir \emph{et al.} proposed a new formulation for the massive gravi-electromagnetism with monopole terms \cite{octonion10}. The author formulated the Maxwell-Proca type field equations of linear gravity \cite{octonion11}, in terms of the hyperbolic octonions. The authors \cite{octonion12} made use of the octonion to discuss the gravitational field equations and relevant properties. Castro \cite{octonion13} proposed a non-associative octonionic ternary gauge field theory, based on a ternary bracket. Furui \cite{octonion14} expressed equivalently a Dirac fermion as a 4-component spinor, which is a combination of two quaternions. In terms of the quaternion in the Yang-Mills Lagrangian, the author \cite{octonion15} discussed the axial anomaly and the triality symmetry of octonion. The author \cite{octonion16} assumed the dark matter may be able to be interpreted as matter emitting photons in a different triality sector, rather than that of electromagnetic probes in the world. With the help of the algebraic properties of quaternions and octonions, Kalauni \emph{et al.} \cite{octonion17} obtained the fully symmetric Dirac-Maxwell's equations as one single equation. Negi \emph{et al.} \cite{octonion18} applied the octonion to express the consistent form of generalized Maxwell¡¯s equations in presence of electric and magnetic charges (dyons). In terms of the Zorn vector matrix realization, Chanyal \emph{et al.} \cite{octonion19} described the octonion formulation of Abelian/non-Abelian gauge theory. The authors \cite{octonion20} analyzed the role of octonions in various unified field theories associated with the dyon and dark matter, reconstructing the field equations of hot and cold dark matter by means of split octonions.

A few scholars adopt the sedenions to explore the Dirac wave equation, gravitational field equations, electromagnetic field equations, curved space, and dark matters and so forth. Koplinger \cite{sedenion1} applied the conic sedenions to express the Dirac equation in physics through the hyperbolic subalgebra, together with a counterpart on circular geometry that has been proposed for description of gravity. Making use of the conic sedenions, the author \cite{sedenion2} proposed the classical Minkowski and a new Euclidean space-time metric, in an alignment program for large body (non-quantum) physics. By transitioning between different geometries through genuine hypernumber rotation, the author \cite{sedenion3} demonstrated a method to extend the complex-number algebra using nonreal square roots of +1 to aid the mathematical description of physical laws. Mironov \emph{et al.} \cite{sedenion4} represented the sedenions to generate the associative noncommutative space-time algebra, and demonstrated a generalization of relativistic quantum mechanics using sedenionic wavefunctions and sedenionic space-time operators. Making use of the sedenions, the authors \cite{sedenion5} demonstrated some fundamental aspects of massive field's theory, including the gauge invariance, charge conservation, Poynting's theorem, potential of a stationary scalar point source, plane wave solution, and interactions between point sources. On the basis of the sedenionic space-time operators and sedenionic wavefunctions, the authors \cite{sedenion6} discussed the gauge invariance of generalized second-order and first-order wave equations for massive and massless fields. Demir \emph{et al.} \cite{sedenion7} presented the conic sedenionic formulation for the unification of generalized field equations of dyons (electromagnetic theory) and gravito-dyons (linear gravity). Chanyal \cite{sedenion8} formulated the sedenionic unified potential equations, field equations, and current equations of dyons and gravito-dyons, and developed the sedenionic unified theory of dyons and gravito-dyons, in terms of two eight-potentials leading to the structural symmetry. By means of the dual octonion algebra, the author \cite{sedenion9} formulated a new set of electrodynamic equations for massive dyons, and made an attempt to obtain the symmetrical form of generalized Proca-Dirac-Maxwell equations with respect to the dual octonion form.

The application of the complex-sedenions is able to express the relevant physical quantities as the wavefunctions in the paper, describing the quantum mechanics connected with the four fundamental interactions, including the non-Abelian gauge field, and electroweak field, and dark matter field and so forth. The quantum mechanics described with the complex-sedenions is capable of solving a part of conundrums, derived from the Standard Model and even `Beyond the Standard Model'.

a) Four fundamental fields. According to the basic postulates \cite{weng3}, one fundamental field extends its individual space. Each of fundamental fields can be described independently by one complex-quaternion space. The four complex-quaternion spaces for the four fundamental fields are perpendicular to each other. Consequently the complex-octonion space is able to study two fundamental fields simultaneously, such as, the electromagnetic and gravitational fields. Meanwhile the quantum mechanics in the complex-sedenion space is capable of exploring the four fundamental fields, including the electromagnetic field, gravitational field, and nuclear field.

b) Color confinement. In case the direction of a three-dimensional unit vector $\textbf{\emph{i}}_q$ , in the complex-quaternion wavefunction, is incapable of playing a major role, the unit vector $\textbf{\emph{i}}_q$ will be degenerated into the imaginary unit $i$, or one new three-dimensional unit vector $\textbf{\emph{i}}_q^\prime$ , which is independent of the unit vector $\textbf{\emph{i}}_q$ . The wavefunctions with the imaginary unit $i$ can be applied to study the quantum properties of some particles (such as, the electrons). Further the wavefunctions with the new three-dimensional unit vector, $\textbf{\emph{i}}_q^\prime$ , can be utilized to explore the quantum properties of some other particles (for instance, the quarks), which are in possession of `three colors'. The three-dimensional unit vector, $\textbf{\emph{i}}_q^\prime$ , embodies the physical properties of `three colors' relevant to the quarks. Consequently one complex-quaternion wavefunction is able to be degenerated into three complex-number wavefunctions, which are independent of each other. In other words, the color degrees of freedom can be merged into the wavefunctions, described with the complex-quaternions. And the color degrees of freedom are only the spatial dimensions, rather than any property of physical substance.

c) Dark matter fields. In the electromagnetic and gravitational fields described with the complex-octonions, there are two sorts of adjoint fields. One of them can be considered as the dark matter field, because the adjoint field possesses a few major properties of the fundamental fields. In the four fundamental interactions, described with the complex-sedenions, there are four fundamental fields and twelve adjoint fields. Three adjoint fields of them can be regarded as the dark matter fields. Moreover, the field sources of these three adjoint-fields can be combined with that of some other adjoint fields or fundamental fields to become certain comparatively complicated particles, producing several physical effects of dark matters.

In the paper, the complex-sedenions can be utilized to investigate simultaneously the four fundamental interactions, deducing the field equations of classical mechanics on the macroscopic scale (in Section 3), including the integrating function of field potential, field potential, field strength, field source, linear momentum, angular momentum, torque, and force and so forth. On the basis of the exponential form of complex-quaternions, either of complex-octonion and complex-sedenion can be written as the exponential form, expressing the relevant physical quantities as the wavefunctions. Making use of the composite operator and the wavefunction of physical quantity, it is able to derive similarly the field equations of quantum mechanics on the microscopic scale (in Section 5), from the preceding field equations of classical mechanics. Under few approximate conditions, the field equations of quantum mechanics, described with the complex-sedenions, can be degenerated into the Dirac wave equation or Yang-Mills equation and so forth. Further the unit vector of the complex-quaternion wavefunction is seized of three spatial dimensions, which can be interpreted as the color degrees of freedom. In other words, the color degrees of freedom are merely the spatial dimensions, rather than any property of physical substance, accounting for the rule of color confinement naturally.

\section{\label{sec:level1}Sedenion space}

By means of the algebraic property of sedenions, it is found that one complex-sedenion space consists of four complex-quaternion spaces. On the basis of the basic postulates (see Ref.[63]), the complex-sedenion space is able to describe the four sorts of fundamental fields simultaneously. However, according to the electroweak theory, the four fundamental fields in the paper include the gravitational field, electromagnetic field, and strong-nuclear field, except for the existing weak-nuclear field.

In the electroweak theory, the weak-nuclear field and electromagnetic field can be unified into the electroweak field. Either of the weak-nuclear field and electromagnetic field is merely one constituent of the electroweak field, so the weak-nuclear field would not be regarded as one fundamental field any more. Further, in the quantum mechanics described with the complex-sedenions, the weak-nuclear field can be considered as the adjoint field of electromagnetic field. As one significant component, the weak-nuclear field can be merged into the electromagnetic field described with the complex-sedenions (in Section 7). And the electromagnetic field, described with the complex-sedenions, can be degenerated naturally into the existing electroweak field.

However, it is inevitable that there must be simultaneously four sorts of fundamental fields in the complex-sedenion space, according to the multiplicative closure of sedenions. The four fundamental fields contain the gravitational field, electromagnetic field, strong-nuclear field, and one new species of unknown fundamental field. The last is called as W-nuclear field temporarily, marking with the initial of ¡®weak¡¯ traditionally. In other words, in the field theory described with the complex-sedenions, the four fundamental fields are the gravitational field, electromagnetic field, strong-nuclear field, and W-nuclear field. Apparently, the weak-nuclear field is replaced by the W-nuclear field, in the paper.

In the complex-quaternion space $\mathbb{H}_g$ for the gravitational fields, the coordinates are, $i R_{g0}$ and $R_{gk}$ , the basis vector is $\textbf{\emph{I}}_{gj}$ , the complex-quaternion radius vector is, $\mathbb{R}_g = i \textbf{\emph{I}}_{g0} R_{g0} + \Sigma \textbf{\emph{I}}_{gk} R_{gk}$ . Similarly, in the complex 2-quaternion (short for the second quaternion) space $\mathbb{H}_e$ for the electromagnetic fields, the coordinates are, $i R_{e0}$ and $R_{ek}$ , the basis vector is $\textbf{\emph{I}}_{ej}$ , the complex 2-quaternion radius vector is, $\mathbb{R}_e = i \textbf{\emph{I}}_{e0} R_{e0} + \Sigma \textbf{\emph{I}}_{ek} R_{ek}$ . In the complex 3-quaternion (short for the third quaternion) space $\mathbb{H}_w$ for the W-nuclear fields, the coordinates are, $i R_{w0}$ and $R_{wk}$ , the basis vector is $\textbf{\emph{I}}_{wj}$ , the complex 3-quaternion radius vector is, $\mathbb{R}_w = i \textbf{\emph{I}}_{w0} R_{w0} + \Sigma \textbf{\emph{I}}_{wk} R_{wk}$. In the complex 4-quaternion (short for the fourth quaternion) space $\mathbb{H}_s$ for the strong-nuclear fields, the coordinates are, $i R_{s0}$ and $R_{sk}$ , the basis vector is $\textbf{\emph{I}}_{sj}$ , the complex 4-quaternion radius vector is, $\mathbb{R}_s = i \textbf{\emph{I}}_{s0} R_{s0} + \Sigma \textbf{\emph{I}}_{sk} R_{sk}$ . Herein, it is convenient to apply the superscript or subscript, $g$, $e$, $w$, and $s$, to mark respectively the physical quantities in the gravitational field, electromagnetic field, W-nuclear field, and strong-nuclear field. $R_{gj}$ , $R_{ej}$ , $R_{wj}$ , and $R_{sj}$ are all real. $R_{g0} = v_0 t$. $v_0$ is the speed of light, and $t$ is the time. $\textbf{\emph{I}}_{g0} = 1$. $\textbf{\emph{I}}_{g0} \circ \textbf{\emph{I}}_{g0} = 1$. $\textbf{\emph{I}}_{gk} \circ \textbf{\emph{I}}_{gk} = - 1$. $\textbf{\emph{I}}_{ej} = \textbf{\emph{I}}_{gj} \circ \textbf{\emph{I}}_{e0}$ . $\textbf{\emph{I}}_{ej} \circ \textbf{\emph{I}}_{ej} = - 1$. $\textbf{\emph{I}}_{wj} = \textbf{\emph{I}}_{gj} \circ \textbf{\emph{I}}_{w0}$. $\textbf{\emph{I}}_{wj} \circ \textbf{\emph{I}}_{wj} = - 1$. $\textbf{\emph{I}}_{s0} = \textbf{\emph{I}}_{g0} \circ \textbf{\emph{I}}_{s0}$ . $\textbf{\emph{I}}_{sk} = - \textbf{\emph{I}}_{gk} \circ \textbf{\emph{I}}_{s0}$ . $\textbf{\emph{I}}_{sj} \circ \textbf{\emph{I}}_{sj} = - 1$. The symbol $\circ$ denotes the multiplication of sedenion. $j = 0, 1, 2, 3$. $k = 1, 2, 3$. $i$ is the imaginary unit.

In the above, the four independent complex-quaternion spaces, $\mathbb{H}_g$ , $\mathbb{H}_e$ , $\mathbb{H}_w$, and $\mathbb{H}_s$ , are perpendicular to each other. They can be combined together to become one complex-sedenion space $\mathbb{K}$ . In the complex-sedenion space $\mathbb{K}$ , the basis vectors are, $\textbf{\emph{I}}_{gj}$ , $\textbf{\emph{I}}_{ej}$ , $\textbf{\emph{I}}_{wj}$ , and $\textbf{\emph{I}}_{sj}$ . The complex-sedenion radius vector is, $\mathbb{R} = \mathbb{R}_g + k_{eg} \mathbb{R}_e + k_{wg} \mathbb{R}_w + k_{sg} \mathbb{R}_s$. Herein, $k_{eg}$ , $k_{wg}$ , and $k_{sg}$ are coefficients.

In the complex-quaternion space $\mathbb{H}_g$ for the gravitational fields, the complex-quaternion operator is, $\lozenge_g = i \textbf{\emph{I}}_{g0} \partial_{g0} + \Sigma \textbf{\emph{I}}_{gk} \partial_{gk}$ , with $\partial_{gj} = \partial / \partial R_{gj}$ . In the complex 2-quaternion space $\mathbb{H}_e$ for the electromagnetic fields, the complex 2-quaternion operator is, $\lozenge_e = i \textbf{\emph{I}}_{e0} \partial_{e0} + \Sigma \textbf{\emph{I}}_{ek} \partial_{ek}$ , with $\partial_{ej} = \partial / \partial R_{ej}$ . In the complex 3-quaternion space $\mathbb{H}_w$ for the W-nuclear fields, the complex 3-quaternion operator is,
$\lozenge_w = i \textbf{\emph{I}}_{w0} \partial_{w0} + \Sigma \textbf{\emph{I}}_{wk} \partial_{wk}$ , with $\partial_{wj} = \partial / \partial R_{wj}$ . In the complex 4-quaternion space $\mathbb{H}_s$ for the strong-nuclear fields, the complex 4-quaternion operator is, $\lozenge_s = i \textbf{\emph{I}}_{s0} \partial_{s0} + \Sigma \textbf{\emph{I}}_{sk} \partial_{sk}$, with $\partial_{sj} = \partial / \partial R_{sj}$ . As a result, the complex-sedenion operator is, $\lozenge = \lozenge_g + k_{eg}^{~~-1} \lozenge_e + k_{wg}^{~~-1} \lozenge_w + k_{sg}^{~~-1} \lozenge_s$ , in the complex-sedenion space $\mathbb{K}$ . Herein $\lozenge_g = i \textbf{\emph{I}}_{g0} \partial_{g0} + \nabla_g$ . $\lozenge_e = i \textbf{\emph{I}}_{e0} \partial_{e0} + \nabla_e$. $\lozenge_w = i \textbf{\emph{I}}_{w0} \partial_{w0} + \nabla_w$. $\lozenge_s = i \textbf{\emph{I}}_{s0} \partial_{s0} + \nabla_s$. $\nabla_g = \Sigma \textbf{\emph{I}}_{gk} \partial_{gk}$ . $\nabla_e = \Sigma \textbf{\emph{I}}_{ek} \partial_{ek}$ . $\nabla_w = \Sigma \textbf{\emph{I}}_{wk} \partial_{wk}$ . $\nabla_s = \Sigma \textbf{\emph{I}}_{sk} \partial_{sk}$ .

\section{\label{sec:level1}Classical field equations}

In contrast to the spaces in the Newtonian mechanics, the multidimensional spaces in the paper are differentiated. On the basis of the basic postulates, the spaces to research the four fundamental interactions have been extended from a single complex-quaternion space to four complex-quaternion spaces. Meanwhile the concepts of fields are differentiated also. According to the properties of the complex-sedenion space, the fields to study the four fundamental interactions have been extended from the four fundamental fields to the four fundamental fields and twelve adjoint fields. The adjoint fields possess merely a few major properties of the fundamental fields (see Ref.[18]).

Making use of the properties of the complex-sedenions, it is able to deduce the field equations relevant to the gravitational fields, electromagnetic fields, strong-nuclear fields, and W-nuclear fields on the macroscopic scale, including the field potential, field strength, field source, linear momentum, angular momentum, torque, and force and so forth. And some of these physical quantities can be separated and degenerated into the electromagnetic field equations, Newton's law of universal gravitation, mass continuity equation, current continuity equation, force equilibrium equation, and precessional equilibrium equation and so forth.

\subsection{\label{sec:level1}Field potential}

In the complex-sedenion space $\mathbb{K}$ , the complex-sedenion integrating function of field potential is, $\mathbb{X} = \mathbb{X}_g + k_{eg} \mathbb{X}_e + k_{wg} \mathbb{X}_w + k_{sg} \mathbb{X}_s$ . Herein $\mathbb{X}_g$ , $\mathbb{X}_e$ , $\mathbb{X}_w$ , and $\mathbb{X}_s$ are respectively the components of the integrating function of field potential $\mathbb{X}$ in the spaces, $\mathbb{H}_g$ , $\mathbb{H}_e$ , $\mathbb{H}_w$ , and $\mathbb{H}_s$ . $\mathbb{X}_g = i \textbf{\emph{I}}_{g0} X_{g0} + \Sigma \textbf{\emph{I}}_{gk} X_{gk}$ . $\mathbb{X}_e = i \textbf{\emph{I}}_{e0} X_{e0} + \Sigma \textbf{\emph{I}}_{ek} X_{ek}$. $\mathbb{X}_w = i \textbf{\emph{I}}_{w0} X_{w0} + \Sigma \textbf{\emph{I}}_{wk} X_{wk}$ . $\mathbb{X}_s = i \textbf{\emph{I}}_{s0} X_{s0} + \Sigma \textbf{\emph{I}}_{sk} X_{sk}$. $X_{gj}$ , $X_{ej}$ , $X_{wj}$ , and $X_{sj}$ are all real.

By means of the properties of the integrating function of field potential, the complex-sedenion field potential $\mathbb{A}$ can be defined as,
\begin{eqnarray}
\mathbb{A} = i \lozenge^\star \circ \mathbb{X}  ~ ,
\end{eqnarray}
where $\mathbb{A} = \mathbb{A}_g + k_{eg} \mathbb{A}_e + k_{wg} \mathbb{A}_w + k_{sg} \mathbb{A}_s$ . $\mathbb{A}_g$ , $\mathbb{A}_e$ , $\mathbb{A}_w$ , and $\mathbb{A}_s$ are respectively the components of the field potential $\mathbb{A}$ in the spaces, $\mathbb{H}_g$ , $\mathbb{H}_e$ , $\mathbb{H}_w$ , and $\mathbb{H}_s$ . $\lozenge^\star = \lozenge_g^\star + k_{eg}^{~~-1} \lozenge_e^\star + k_{wg}^{~~-1} \lozenge_w^\star + k_{sg}^{~~-1} \lozenge_s^\star $ . Especially, the operation of complex conjugate will be applied to some physical quantities, except for the coefficients, $k_{eg}$ , $k_{wg}$ , and $k_{sg}$ . The symbol $\star$ stands for the complex conjugate.

The complex-sedenion field potential $\mathbb{A}$ includes the field potentials of four fundamental fields and of twelve adjoint fields obviously. The ingredients of the field potentials in the four complex-quaternion spaces are as follows:

a) In the complex-quaternion space $\mathbb{H}_g$ , the component $\mathbb{A}_g$ contains the gravitational fundamental field potential $\mathbb{A}_g^g$ , electromagnetic adjoint field potential $\mathbb{A}_e^e$, W-nuclear adjoint field potential $\mathbb{A}_w^w$ , and strong-nuclear adjoint field potential $\mathbb{A}_s^s$. Herein $\mathbb{A}_g = \mathbb{A}_g^g + \mathbb{A}_e^e + \mathbb{A}_w^w + \mathbb{A}_s^s$ . $\mathbb{A}_g = i \textbf{\emph{I}}_{g0} A_{g0} + \Sigma \textbf{\emph{I}}_{gk} A_{gk}$. $\mathbb{A}_g^g = i \lozenge_g^\star \circ \mathbb{X}_g$. $\mathbb{A}_e^e = i \lozenge_e^\star \circ \mathbb{X}_e$ . $\mathbb{A}_w^w = i \lozenge_w^\star \circ \mathbb{X}_w$ . $\mathbb{A}_s^s = i \lozenge_s^\star \circ \mathbb{X}_s$ . $\mathbb{A}_g^g = i \textbf{\emph{I}}_{g0} A^g_{g0} + \Sigma \textbf{\emph{I}}_{gk} A^g_{gk}$ . $\mathbb{A}_e^e = i \textbf{\emph{I}}_{g0} A^e_{e0} + \Sigma \textbf{\emph{I}}_{gk} A^e_{ek}$ . $\mathbb{A}_w^w = i \textbf{\emph{I}}_{g0} A^w_{w0} + \Sigma \textbf{\emph{I}}_{gk} A^w_{wk}$ . $\mathbb{A}_s^s = i \textbf{\emph{I}}_{g0} A^s_{s0} + \Sigma \textbf{\emph{I}}_{gk} A^s_{sk}$ . $A_{gj}$ , $A^g_{gj}$ , $A^e_{ej}$ , $A^w_{wj}$ , and $A^s_{sj}$ are all real.

b) In the complex 2-quaternion space $\mathbb{H}_e$ , the component $\mathbb{A}_e$ consists of the electromagnetic fundamental field potential $\mathbb{A}_e^g$ , gravitational adjoint field potential $\mathbb{A}_g^e$ , W-nuclear adjoint field potential $\mathbb{A}_w^s$ , and strong-nuclear adjoint field potential $\mathbb{A}_s^w$ . Herein $\mathbb{A}_e = \mathbb{A}_e^g + k_{eg}^{~~-2} \mathbb{A}_g^e + ( k_{wg} k_{sg}^{~~-1} k_{eg}^{~~-1} ) \mathbb{A}_w^s + ( k_{sg} k_{wg}^{~~-1} k_{eg}^{~~-1} ) \mathbb{A}_s^w$ . $\mathbb{A}_e = i \textbf{\emph{I}}_{e0} A_{e0} + \Sigma \textbf{\emph{I}}_{ek} A_{ek}$ . $\mathbb{A}_e^g = i \lozenge_g^\star \circ \mathbb{X}_e$ . $\mathbb{A}_g^e = i \lozenge_e^\star \circ \mathbb{X}_g$ . $\mathbb{A}_w^s = i \lozenge_s^\star \circ \mathbb{X}_w$ . $\mathbb{A}_s^w = i \lozenge_w^\star \circ \mathbb{X}_s$ . $\mathbb{A}_e^g = i \textbf{\emph{I}}_{e0} A^g_{e0} + \Sigma \textbf{\emph{I}}_{ek} A^g_{ek}$ . $\mathbb{A}_g^e = i \textbf{\emph{I}}_{e0} A^e_{g0} + \Sigma \textbf{\emph{I}}_{ek} A^e_{gk}$. $\mathbb{A}_w^s = i \textbf{\emph{I}}_{e0} A^s_{w0} + \Sigma \textbf{\emph{I}}_{ek} A^s_{wk}$ . $\mathbb{A}_s^w = i \textbf{\emph{I}}_{e0} A^w_{s0} + \Sigma \textbf{\emph{I}}_{ek} A^w_{sk}$ . $A_{ej}$ , $A^g_{ej}$, $A^e_{gj}$ , $A^s_{wj}$ , and $A^w_{sj}$ are all real.

c) In the complex 3-quaternion space $\mathbb{H}_w$ , the component $\mathbb{A}_w$ includes the W-nuclear fundamental field potential $\mathbb{A}_w^g$ , gravitational adjoint field potential $\mathbb{A}_g^w$ , electromagnetic adjoint field potential $\mathbb{A}_e^s$ , and strong-nuclear adjoint field potential $\mathbb{A}_s^e$ . Herein $\mathbb{A}_w = \mathbb{A}_w^g + k_{wg}^{~~-2} \mathbb{A}_g^w + ( k_{eg} k_{wg}^{~~-1} k_{sg}^{~~-1} ) \mathbb{A}_e^s + ( k_{sg} k_{wg}^{~~-1} k_{eg}^{~~-1} ) \mathbb{A}_s^e$. $\mathbb{A}_w = i \textbf{\emph{I}}_{w0} A_{w0} + \Sigma \textbf{\emph{I}}_{wk} A_{wk}$ . $\mathbb{A}_w^g = i \lozenge_g^\star \circ \mathbb{X}_w$ . $\mathbb{A}_g^w = i \lozenge_w^\star \circ \mathbb{X}_g$ . $\mathbb{A}_e^s = i \lozenge_s^\star \circ \mathbb{X}_e$. $\mathbb{A}_s^e = i \lozenge_e^\star \circ \mathbb{X}_s$. $\mathbb{A}_w^g = i \textbf{\emph{I}}_{w0} A^g_{w0} + \Sigma \textbf{\emph{I}}_{wk} A^g_{wk}$ . $\mathbb{A}_g^w = i \textbf{\emph{I}}_{w0} A^w_{g0} + \Sigma \textbf{\emph{I}}_{wk} A^w_{gk}$. $\mathbb{A}_e^s = i \textbf{\emph{I}}_{w0} A^s_{e0} + \Sigma \textbf{\emph{I}}_{wk} A^s_{ek}$. $\mathbb{A}_s^e = i \textbf{\emph{I}}_{w0} A^e_{s0} + \Sigma \textbf{\emph{I}}_{wk} A^e_{sk}$ . $A_{wj}$ , $A^g_{wj}$ , $A^w_{gj}$, $A^s_{ej}$, and $A^e_{sj}$ are all real.

d) In the complex 4-quaternion space $\mathbb{H}_s$ , the component $\mathbb{A}_s$ covers the strong-nuclear fundamental field potential $\mathbb{A}_s^g$ , gravitational adjoint field potential $\mathbb{A}_g^s$ , electromagnetic adjoint field potential $\mathbb{A}_e^w$ , and W-nuclear adjoint field potential $\mathbb{A}_w^e$. Herein $\mathbb{A}_s = \mathbb{A}_s^g + k_{sg}^{~~-2} \mathbb{A}_g^s + ( k_{eg} k_{wg}^{~~-1} k_{sg}^{~~-1} ) \mathbb{A}_e^w + ( k_{wg} k_{sg}^{~~-1} k_{eg}^{~~-1} ) \mathbb{A}_w^e$ . $\mathbb{A}_s = i \textbf{\emph{I}}_{s0} A_{s0} + \Sigma \textbf{\emph{I}}_{sk} A_{sk}$ . $\mathbb{A}_s^g = i \lozenge_g^\star \circ \mathbb{X}_s$ . $\mathbb{A}_g^s = i \lozenge_s^\star \circ \mathbb{X}_g$. $\mathbb{A}_e^w = i \lozenge_w^\star \circ \mathbb{X}_e$ . $\mathbb{A}_w^e = i \lozenge_e^\star \circ \mathbb{X}_w$ . $\mathbb{A}_s^g = i \textbf{\emph{I}}_{s0} A^g_{s0} + \Sigma \textbf{\emph{I}}_{sk} A^g_{sk}$ . $\mathbb{A}_g^s = i \textbf{\emph{I}}_{s0} A^s_{g0} + \Sigma \textbf{\emph{I}}_{sk} A^s_{gk}$. $\mathbb{A}_e^w = i \textbf{\emph{I}}_{s0} A^w_{e0} + \Sigma \textbf{\emph{I}}_{sk} A^w_{ek}$ . $\mathbb{A}_w^e = i \textbf{\emph{I}}_{s0} A^e_{w0} + \Sigma \textbf{\emph{I}}_{sk} A^e_{wk}$ . $A_{sj}$ , $A^g_{sj}$ , $A^s_{gj}$ , $A^w_{ej}$ , and $A^e_{wj}$ are all real.

\subsection{\label{sec:level1}Field strength}

From the complex-sedenion field potential, it is able to define the complex-sedenion field strength $\mathbb{F}$ as,
\begin{eqnarray}
\mathbb{F} = \lozenge \circ \mathbb{A}  ~ ,
\end{eqnarray}
where $\mathbb{F} = \mathbb{F}_g + k_{eg} \mathbb{F}_e + k_{wg} \mathbb{F}_w + k_{sg} \mathbb{F}_s$ . $\mathbb{F}_g$ , $\mathbb{F}_e$ , $\mathbb{F}_w$ , and $\mathbb{F}_s$ are respectively the components of the field strength $\mathbb{F}$ in the spaces, $\mathbb{H}_g$ , $\mathbb{H}_e$ , $\mathbb{H}_w$ , and $\mathbb{H}_s$ .

The complex-sedenion field strength $\mathbb{F}$ comprises the field strengths of four fundamental fields and of twelve adjoint fields obviously. The ingredients of the field strengths in the four complex-quaternion spaces are as follows:

a) In the complex-quaternion space $\mathbb{H}_g$ , the component $\mathbb{F}_g$ comprises the gravitational fundamental field strength $\mathbb{F}_g^g$ , electromagnetic adjoint field strength $\mathbb{F}_e^e$ , W-nuclear adjoint field strength $\mathbb{F}_w^w$ , and strong-nuclear adjoint field strength $\mathbb{F}_s^s$ . Herein $\mathbb{F}_g = \mathbb{F}_g^g + \mathbb{F}_e^e + \mathbb{F}_w^w + \mathbb{F}_s^s$ . $\mathbb{F}_g = i \textbf{\emph{I}}_{g0} F_{g0} + \Sigma \textbf{\emph{I}}_{gk} F_{gk}$. $\mathbb{F}_g^g = \lozenge_g \circ \mathbb{A}_g$ . $\mathbb{F}_e^e = \lozenge_e \circ \mathbb{A}_e$. $\mathbb{F}_w^w = \lozenge_w \circ \mathbb{A}_w$ . $\mathbb{F}_s^s = \lozenge_s \circ \mathbb{A}_s$ . $\mathbb{F}_g^g = i \textbf{\emph{I}}_{g0} F^g_{g0} + \Sigma \textbf{\emph{I}}_{gk} F^g_{gk}$ . $\mathbb{F}_e^e = i \textbf{\emph{I}}_{g0} F^e_{e0} + \Sigma \textbf{\emph{I}}_{gk} F^e_{ek}$. $\mathbb{F}_w^w = i \textbf{\emph{I}}_{g0} F^w_{w0} + \Sigma \textbf{\emph{I}}_{gk} F^w_{wk}$ . $\mathbb{F}_s^s = i \textbf{\emph{I}}_{g0} F^s_{s0} + \Sigma \textbf{\emph{I}}_{gk} F^s_{sk}$. $F_{g0}$ , $F^g_{g0}$ , $F^e_{e0}$ , $F^w_{w0}$ , and $F^s_{s0}$ are all real. $F_{gk}$ , $F^g_{gk}$ , $F^e_{ek}$ , $F^w_{wk}$ , and $F^s_{sk}$ are complex-numbers.

b) In the complex 2-quaternion space $\mathbb{H}_e$ , the component $\mathbb{F}_e$ contains the electromagnetic fundamental field strength $\mathbb{F}_e^g$ , gravitational adjoint field strength $\mathbb{F}_g^e$ , W-nuclear adjoint field strength $\mathbb{F}_w^s$ , and strong-nuclear adjoint field strength $\mathbb{F}_s^w$ . Herein $\mathbb{F}_e = \mathbb{F}_e^g + k_{eg}^{~~-2} \mathbb{F}_g^e + ( k_{wg} k_{sg}^{~~-1} k_{eg}^{~~-1} ) \mathbb{F}_w^s + ( k_{sg} k_{wg}^{~~-1} k_{eg}^{~~-1} ) \mathbb{F}_s^w$. $\mathbb{F}_e = i \textbf{\emph{I}}_{e0} F_{e0} + \Sigma \textbf{\emph{I}}_{ek} F_{ek}$. $\mathbb{F}_e^g = \lozenge_g \circ \mathbb{A}_e$. $\mathbb{F}_g^e = \lozenge_e \circ \mathbb{A}_g$ . $\mathbb{F}_w^s = \lozenge_s \circ \mathbb{A}_w$. $\mathbb{F}_s^w = \lozenge_w \circ \mathbb{A}_s$ . $\mathbb{F}_e^g = i \textbf{\emph{I}}_{e0} F^g_{e0} + \Sigma \textbf{\emph{I}}_{ek} F^g_{ek}$. $\mathbb{F}_g^e = i \textbf{\emph{I}}_{e0} F^e_{g0} + \Sigma \textbf{\emph{I}}_{ek} F^e_{gk}$ . $\mathbb{F}_w^s = i \textbf{\emph{I}}_{e0} F^s_{w0} + \Sigma \textbf{\emph{I}}_{ek} F^s_{wk}$ . $\mathbb{F}_s^w = i \textbf{\emph{I}}_{e0} F^w_{s0} + \Sigma \textbf{\emph{I}}_{ek} F^w_{sk}$. $F_{e0}$ , $F^g_{e0}$ , $F^e_{g0}$ , $F^s_{w0}$ , and $F^w_{s0}$ are all real. $F_{ek}$ , $F^g_{ek}$ , $F^e_{gk}$ , $F^s_{wk}$ , and $F^w_{sk}$ are complex-numbers.

c) In the complex 3-quaternion space $\mathbb{H}_w$ , the component $\mathbb{F}_w$ covers the W-nuclear fundamental field strength $\mathbb{F}_w^g$, gravitational adjoint field strength $\mathbb{F}_g^w$ , electromagnetic adjoint field strength $\mathbb{F}_e^s$ , and strong-nuclear adjoint field strength $\mathbb{F}_s^e$ . Herein $\mathbb{F}_w = \mathbb{F}_w^g + k_{wg}^{~~-2} \mathbb{F}_g^w + ( k_{eg} k_{wg}^{~~-1} k_{sg}^{~~-1} ) \mathbb{F}_e^s + ( k_{sg} k_{wg}^{~~-1} k_{eg}^{~~-1} ) \mathbb{F}_s^e$ . $\mathbb{F}_w = i \textbf{\emph{I}}_{w0} F_{w0} + \Sigma \textbf{\emph{I}}_{wk} F_{wk}$ . $\mathbb{F}_w^g = \lozenge_g \circ \mathbb{A}_w$. $\mathbb{F}_g^w = \lozenge_w \circ \mathbb{A}_g$ . $\mathbb{F}_e^s = \lozenge_s \circ \mathbb{A}_e$. $\mathbb{F}_s^e = \lozenge_e \circ \mathbb{A}_s$ . $\mathbb{F}_w^g = i \textbf{\emph{I}}_{w0} F^g_{w0} + \Sigma \textbf{\emph{I}}_{wk} F^g_{wk}$ . $\mathbb{F}_g^w = i \textbf{\emph{I}}_{w0} F^w_{g0} + \Sigma \textbf{\emph{I}}_{wk} F^w_{gk}$. $\mathbb{F}_e^s = i \textbf{\emph{I}}_{w0} F^s_{e0} + \Sigma \textbf{\emph{I}}_{wk} F^s_{ek}$ . $\mathbb{F}_s^e = i \textbf{\emph{I}}_{w0} F^e_{s0} + \Sigma \textbf{\emph{I}}_{wk} F^e_{sk}$ . $F_{w0}$ , $F^g_{w0}$ , $F^w_{g0}$ , $F^s_{e0}$ , and $F^e_{s0}$ are all real. $F_{wk}$ , $F^g_{wk}$ , $F^w_{gk}$ , $F^s_{ek}$ , and $F^e_{sk}$ are complex-numbers.

d) In the complex 4-quaternion space $\mathbb{H}_s$ , the component $\mathbb{F}_s$ includes the strong-nuclear fundamental field strength $\mathbb{F}_s^g$ , gravitational adjoint field strength $\mathbb{F}_g^s$ , electromagnetic adjoint field strength $\mathbb{F}_e^w$ , and W-nuclear adjoint field strength $\mathbb{F}_w^e$ . Herein $\mathbb{F}_s = \mathbb{F}_s^g + k_{sg}^{~~-2} \mathbb{F}_g^s + ( k_{eg} k_{wg}^{~~-1} k_{sg}^{~~-1} ) \mathbb{F}_e^w + ( k_{wg} k_{sg}^{~~-1} k_{eg}^{~~-1} ) \mathbb{F}_w^e$. $\mathbb{F}_s = i \textbf{\emph{I}}_{s0} F_{s0} + \Sigma \textbf{\emph{I}}_{sk} F_{sk}$. $\mathbb{F}_s^g = \lozenge_g \circ \mathbb{A}_s$. $\mathbb{F}_g^s = \lozenge_s \circ \mathbb{A}_g$ . $\mathbb{F}_e^w = \lozenge_w \circ \mathbb{A}_e$. $\mathbb{F}_w^e = \lozenge_e \circ \mathbb{A}_w$ . $\mathbb{F}_s^g = i \textbf{\emph{I}}_{s0} F^g_{s0} + \Sigma \textbf{\emph{I}}_{sk} F^g_{sk}$. $\mathbb{F}_g^s = i \textbf{\emph{I}}_{s0} F^s_{g0} + \Sigma \textbf{\emph{I}}_{sk} F^s_{gk}$ . $\mathbb{F}_e^w = i \textbf{\emph{I}}_{s0} F^w_{e0} + \Sigma \textbf{\emph{I}}_{sk} F^w_{ek}$ . $\mathbb{F}_w^e = i \textbf{\emph{I}}_{s0} F^e_{w0} + \Sigma \textbf{\emph{I}}_{sk} F^e_{wk}$. $F_{s0}$ , $F^g_{s0}$ , $F^s_{g0}$ , $F^w_{e0}$ , and $F^e_{w0}$ are all real. $F_{sk}$ , $F^g_{sk}$ , $F^s_{gk}$ , $F^w_{ek}$ , and $F^e_{wk}$ are complex-numbers.

\subsection{\label{sec:level1}Field source}

Making use of the physical properties of the field strength, the complex-sedenion field source $\mathbb{S}$ can be defined as,
\begin{eqnarray}
\mu \mathbb{S} = - ( i \mathbb{F} / v_0 + \lozenge )^\ast \circ \mathbb{F}  ~ ,
\end{eqnarray}
where $\mu \mathbb{S} = \mu_g \mathbb{S}_g + k_{eg} \mu_e \mathbb{S}_e + k_{wg} \mu_w \mathbb{S}_w + k_{sg} \mu_s \mathbb{S}_s - i \mathbb{F}^\ast \circ \mathbb{F} / v_0$ . $\mathbb{S}_g$ , $\mathbb{S}_e$ , $\mathbb{S}_w$, and $\mathbb{S}_s$ are respectively the components of the field source $\mathbb{S}$ in the spaces, $\mathbb{H}_g$, $\mathbb{H}_e$ , $\mathbb{H}_w$, and $\mathbb{H}_s$ . $\mu$ , $\mu_g$ , $\mu_e$ , $\mu_w$ , and $\mu_s$ are coefficients. The symbol $\ast$ denotes the sedenion conjugate.

The complex-sedenion field source $\mathbb{S}$ consists of the field sources of four fundamental fields and of twelve adjoint fields obviously. The ingredients of the field sources in the four complex-quaternion spaces are as follows:

a) In the complex-quaternion space $\mathbb{H}_g$ , the component $\mathbb{S}_g$ covers the gravitational fundamental field source $\mathbb{S}_g^g$ , electromagnetic adjoint field source $\mathbb{S}_e^e$ , W-nuclear adjoint field source $\mathbb{S}_w^w$ , and strong-nuclear adjoint field source $\mathbb{S}_s^s$ . Herein $\mu_g \mathbb{S}_g = \mu_g^g \mathbb{S}_g^g + \mu_e^e \mathbb{S}_e^e + \mu_w^w \mathbb{S}_w^w + \mu_s^s \mathbb{S}_s^s$ . $\mathbb{S}_g = i \textbf{\emph{I}}_{g0} S_{g0} + \Sigma \textbf{\emph{I}}_{gk} S_{gk}$ . $\mu_g^g \mathbb{S}_g^g = - \lozenge_g^\ast \circ \mathbb{F}_g$ . $\mu_e^e \mathbb{S}_e^e = - \lozenge_e^\ast \circ \mathbb{F}_e$ . $\mu_w^w \mathbb{S}_w^w = - \lozenge_w^\ast \circ \mathbb{F}_w$ . $\mu_s^s \mathbb{S}_s^s = - \lozenge_s^\ast \circ \mathbb{F}_s$ . $\mathbb{S}_g^g = i \textbf{\emph{I}}_{g0} S^g_{g0} + \Sigma \textbf{\emph{I}}_{gk} S^g_{gk}$ . $\mathbb{S}_e^e = i \textbf{\emph{I}}_{g0} S^e_{e0} + \Sigma \textbf{\emph{I}}_{gk} S^e_{ek}$ . $\mathbb{S}_w^w = i \textbf{\emph{I}}_{g0} S^w_{w0} + \Sigma \textbf{\emph{I}}_{gk} S^w_{wk}$. $\mathbb{S}_s^s = i \textbf{\emph{I}}_{g0} S^s_{s0} + \Sigma \textbf{\emph{I}}_{gk} S^s_{sk}$ . $\mu_g^g$ , $\mu_e^e$ , $\mu_w^w$ , and $\mu_s^s$ are coefficients. $\mu_g^g$ is the conventional gravitational constant, and $\mu_g^g < 0$, in the paper. $S_{gj}$ , $S^g_{gj}$ , $S^e_{ej}$, $S^w_{wj}$ , and $S^s_{sj}$ are all real.

b) In the complex 2-quaternion space $\mathbb{H}_e$ , the component $\mathbb{S}_e$ contains the electromagnetic fundamental field source $\mathbb{S}_e^g$ , gravitational adjoint field source $\mathbb{S}_g^e$ , W-nuclear adjoint field source $\mathbb{S}_w^s$ , and strong-nuclear adjoint field source $\mathbb{S}_s^w$. Herein $\mu_e \mathbb{S}_e = \mu_e^g \mathbb{S}_e^g + k_{eg}^{~~-2} \mu_g^e \mathbb{S}_g^e + ( k_{wg} k_{sg}^{~~-1} k_{eg}^{~~-1} ) \mu_w^s \mathbb{S}_w^s + ( k_{sg} k_{wg}^{~~-1} k_{eg}^{~~-1} ) \mu_s^w \mathbb{S}_s^w$ . $\mathbb{S}_e = i \textbf{\emph{I}}_{e0} S_{e0} + \Sigma \textbf{\emph{I}}_{ek} S_{ek}$. $\mu_e^g \mathbb{S}_e^g = - \lozenge_g^\ast \circ \mathbb{F}_e$. $\mu_g^e \mathbb{S}_g^e = - \lozenge_e^\ast \circ \mathbb{F}_g$ . $\mu_w^s \mathbb{S}_w^s = - \lozenge_s^\ast \circ \mathbb{F}_w$. $\mu_s^w \mathbb{S}_s^w = - \lozenge_w^\ast \circ \mathbb{F}_s$ . $\mathbb{S}_e^g = i \textbf{\emph{I}}_{e0} S^g_{e0} + \Sigma \textbf{\emph{I}}_{ek} S^g_{ek}$. $\mathbb{S}_g^e = i \textbf{\emph{I}}_{e0} S^e_{g0} + \Sigma \textbf{\emph{I}}_{ek} S^e_{gk}$ . $\mathbb{S}_w^s = i \textbf{\emph{I}}_{e0} S^s_{w0} + \Sigma \textbf{\emph{I}}_{ek} S^s_{wk}$ . $\mathbb{S}_s^w = i \textbf{\emph{I}}_{e0} S^w_{s0} + \Sigma \textbf{\emph{I}}_{ek} S^w_{sk}$ . $\mu_e^g$ , $\mu_g^e$ , $\mu_w^s$ , and $\mu_s^w$ are coefficients. $\mu_e^g$ is the conventional electromagnetic constant, and $\mu_e^g > 0$, in the paper. $S_{ej}$ , $S^g_{ej}$ , $S^e_{gj}$ , $S^s_{wj}$ , and $S^w_{sj}$ are all real.

c) In the complex 3-quaternion space $\mathbb{H}_w$ , the component $\mathbb{S}_w$ comprises the W-nuclear fundamental field source $\mathbb{S}_w^g$, gravitational adjoint field source $\mathbb{S}_g^w$ , electromagnetic adjoint field source $\mathbb{S}_e^s$ , and strong-nuclear adjoint field source $\mathbb{S}_s^e$. Herein $\mu_w \mathbb{S}_w = \mu_w^g \mathbb{S}_w^g + k_{wg}^{~~-2} \mu_g^w \mathbb{S}_g^w + ( k_{eg} k_{wg}^{~~-1} k_{sg}^{~~-1} ) \mu_e^s \mathbb{S}_e^s + ( k_{sg} k_{wg}^{~~-1} k_{eg}^{~~-1} ) \mu_s^e \mathbb{S}_s^e$. $\mathbb{S}_w = i \textbf{\emph{I}}_{w0} S_{w0} + \Sigma \textbf{\emph{I}}_{wk} S_{wk}$. $\mu_w^g \mathbb{S}_w^g = - \lozenge_g^\ast \circ \mathbb{F}_w$ . $\mu_g^w \mathbb{S}_g^w = - \lozenge_w^\ast \circ \mathbb{F}_g$ . $\mu_e^s \mathbb{S}_e^s = - \lozenge_s^\ast \circ \mathbb{F}_e$. $\mu_s^e \mathbb{S}_s^e = - \lozenge_e^\ast \circ \mathbb{F}_s$ . $\mathbb{S}_w^g = i \textbf{\emph{I}}_{w0} S^g_{w0} + \Sigma \textbf{\emph{I}}_{wk} S^g_{wk}$ . $\mathbb{S}_g^w = i \textbf{\emph{I}}_{w0} S^w_{g0} + \Sigma \textbf{\emph{I}}_{wk} S^w_{gk}$ . $\mathbb{S}_e^s = i \textbf{\emph{I}}_{w0} S^s_{e0} + \Sigma \textbf{\emph{I}}_{wk} S^s_{ek}$ . $\mathbb{S}_s^e = i \textbf{\emph{I}}_{w0} S^e_{s0} + \Sigma \textbf{\emph{I}}_{wk} S^e_{sk}$ . $\mu_w^g$ , $\mu_g^w$ , $\mu_e^s$ , and $\mu_s^e$ are coefficients. $S_{wj}$ , $S^g_{wj}$ , $S^w_{gj}$ , $S^s_{ej}$ , and $S^e_{sj}$ are all real.

d) In the complex 4-quaternion space $\mathbb{H}_s$ , the component $\mathbb{S}_s$ includes the strong-nuclear fundamental field source $\mathbb{S}_s^g$, gravitational adjoint field source $\mathbb{S}_g^s$, electromagnetic adjoint field source $\mathbb{S}_e^w$ , and W-nuclear adjoint field source $\mathbb{S}_w^e$ . Herein $\mu_s \mathbb{S}_s = \mu_s^g \mathbb{S}_s^g + k_{sg}^{~~-2} \mu_g^s \mathbb{S}_g^s + ( k_{eg} k_{wg}^{~~-1} k_{sg}^{~~-1} ) \mu_e^w \mathbb{S}_e^w + ( k_{wg} k_{sg}^{~~-1} k_{eg}^{~~-1} ) \mu_w^e \mathbb{S}_w^e$ . $\mathbb{S}_s = i \textbf{\emph{I}}_{s0} S_{s0} + \Sigma \textbf{\emph{I}}_{sk} S_{sk}$. $\mu_s^g \mathbb{S}_s^g = - \lozenge_g^\ast \circ \mathbb{F}_s$. $\mu_g^s \mathbb{S}_g^s = - \lozenge_s^\ast \circ \mathbb{F}_g$ . $\mu_e^w \mathbb{S}_e^w = - \lozenge_w^\ast \circ \mathbb{F}_e$. $\mu_w^e \mathbb{S}_w^e = - \lozenge_e^\ast \circ \mathbb{F}_w$ . $\mathbb{S}_s^g = i \textbf{\emph{I}}_{s0} S^g_{s0} + \Sigma \textbf{\emph{I}}_{sk} S^g_{sk}$ . $\mathbb{S}_g^s = i \textbf{\emph{I}}_{s0} S^s_{g0} + \Sigma \textbf{\emph{I}}_{sk} S^s_{gk}$ . $\mathbb{S}_e^w = i \textbf{\emph{I}}_{s0} S^w_{e0} + \Sigma \textbf{\emph{I}}_{sk} S^w_{ek}$ . $\mathbb{S}_w^e = i \textbf{\emph{I}}_{s0} S^e_{w0} + \Sigma \textbf{\emph{I}}_{sk} S^e_{wk}$ . $\mu_s^g$ , $\mu_g^s$ , $\mu_w^e$ , and $\mu_e^w$ are coefficients. $S_{sj}$ , $S^g_{sj}$ , $S^s_{gj}$ , $S^w_{ej}$ , and $S^e_{wj}$ are all real.

In the complex-quaternion space $\mathbb{H}_g$ , the field equation, $\mu_g^g \mathbb{S}_g^g = - \lozenge_g^\ast \circ \mathbb{F}_g$ , is capable of determining the `charge' (or mass, $m_g^g$ ) of the gravitational fields. Meanwhile the field equation can be expanded into the classical gravitational field equations. And the latter can be degenerated into the Newton's law of universal gravitation. In the complex 2-quaternion space $\mathbb{H}_e$ , the field equation, $\mu_e^g \mathbb{S}_e^g = - \lozenge_g^\ast \circ \mathbb{F}_e$ , is able to determine the `charge' (or electric charge, $m_e^g$ ) of the electromagnetic fields. And the field equation can be expanded into the classical electromagnetic field equations. Similarly, in the complex 3-quaternion space $\mathbb{H}_w$ , from the field equation, $\mu_w^g \mathbb{S}_w^g = - \lozenge_g^\ast \circ \mathbb{F}_w$ , it is able to determine the `charge' (or W charge, $m_w^g$ ) of the W-nuclear fields. In the complex 4-quaternion space $\mathbb{H}_s$ , from the field equation, $\mu_s^g \mathbb{S}_s^g = - \lozenge_g^\ast \circ \mathbb{F}_s$ , it is capable of determining the `charge' (or strong charge, $m_s^g$ ) of the strong-nuclear fields. Further this method can be extended from the fundamental fields into the adjoint fields, defining twelve sorts of `charges' of adjoint fields (Table 1).

In case the strong-nuclear field and W-nuclear field can be neglected, the definitions of field potential, field strength, and field source, described with the complex-sedenions, can be reduced respectively into that described with the complex-octonions. Further the definitions of field sources, described with the complex-octonions, can be simplified into the Newton's law of universal gravitation and Maxwell equations, described with the complex-quaternions. Meanwhile, a species of electromagnetic adjoint field can be chosen as the dark matter field, in the complex-octonion space.

\subsection{\label{sec:level1}Angular momentum}

In the complex-sedenion space $\mathbb{K}$ , from the complex-sedenion field source, it is able to define the complex-sedenion linear momentum $\mathbb{P}$ as,
\begin{eqnarray}
\mathbb{P} = \mu \mathbb{S} / \mu_g^g  ~ ,
\end{eqnarray}
where $\mathbb{P} = \mathbb{P}_g + k_{eg} \mathbb{P}_e + k_{wg} \mathbb{P}_w + k_{sg} \mathbb{P}_s$ . $\mathbb{P}_g$ , $\mathbb{P}_e$ , $\mathbb{P}_w$ , and $\mathbb{P}_s$ are respectively the components of the linear momentum $\mathbb{P}$ in the spaces, $\mathbb{H}_g$ , $\mathbb{H}_e$ , $\mathbb{H}_w$ , and $\mathbb{H}_s$. $\mathbb{P}_g = \{ \mu_g \mathbb{S}_g - i \mathbb{F}^\ast \circ \mathbb{F} / v_0 \} / \mu_g^g$ . $\mathbb{P}_e = \mu_e \mathbb{S}_e / \mu_g^g$ . $\mathbb{P}_w = \mu_w \mathbb{S}_w / \mu_g^g$ . $\mathbb{P}_s = \mu_s \mathbb{S}_s / \mu_g^g$. $\mathbb{P}_g = i \textbf{\emph{I}}_{g0} P_{g0} + \Sigma \textbf{\emph{I}}_{gk} P_{gk}$ . $\mathbb{P}_e = i \textbf{\emph{I}}_{e0} P_{e0} + \Sigma \textbf{\emph{I}}_{ek} P_{ek}$ . $\mathbb{P}_w = i \textbf{\emph{I}}_{w0} P_{w0} + \Sigma \textbf{\emph{I}}_{wk} P_{wk}$. $\mathbb{P}_s = i \textbf{\emph{I}}_{s0} P_{s0} + \Sigma \textbf{\emph{I}}_{sk} P_{sk}$ . $P_{gj}$ , $P_{ej}$, $P_{wj}$ , and $P_{sj}$ are all real.

From the complex-sedenion radius vector $\mathbb{R}$ , linear momentum $\mathbb{P}$ , and integrating function of field potential $\mathbb{X}$ , it is capable of defining the complex-sedenion angular momentum $\mathbb{L}$ as ,
\begin{eqnarray}
\mathbb{L} = \mathbb{U}^\star \circ \mathbb{P}    ~ ,
\end{eqnarray}
where $\mathbb{L} = \mathbb{L}_g + k_{eg} \mathbb{L}_e + k_{wg} \mathbb{L}_w + k_{sg} \mathbb{L}_s$ . $\mathbb{L}_g$ , $\mathbb{L}_e$ , $\mathbb{L}_w$ , and $\mathbb{L}_s$ are respectively the components of the angular momentum $\mathbb{L}$ in the spaces, $\mathbb{H}_g$ , $\mathbb{H}_e$ , $\mathbb{H}_w$ , and $\mathbb{H}_s$ . $\mathbb{R}^\star = \mathbb{R}^\star_g + k_{eg} \mathbb{R}^\star_e + k_{wg} \mathbb{R}^\star_w + k_{sg} \mathbb{R}^\star_s$ . $\mathbb{X}^\star = \mathbb{X}^\star_g + k_{eg} \mathbb{X}^\star_e + k_{wg} \mathbb{X}^\star_w + k_{sg} \mathbb{X}^\star_s$. $\mathbb{U} = \mathbb{R} + k_{rx} \mathbb{X}$. $\mathbb{U}_g = \mathbb{R}_g + k_{rx} \mathbb{X}_g$ . $\mathbb{U}_e = \mathbb{R}_e + k_{rx} \mathbb{X}_e$ . $\mathbb{U}_w = \mathbb{R}_w + k_{rx} \mathbb{X}_w$ . $\mathbb{U}_s = \mathbb{R}_s + k_{rx} \mathbb{X}_s$ . $k_{rx}$ is a coefficient, to meet the requirement of the dimensional homogeneity (see Ref.[63]). And that the ingredients of the angular momentum in the four complex-quaternion spaces are as follows,
\begin{eqnarray}
&&  \mathbb{L}_g = \mathbb{U}_g^\star \circ \mathbb{P}_g + k_{eg}^{~~2} \mathbb{U}_e^\star \circ \mathbb{P}_e
%\nonumber
%\\
%&&~~~~~~~~~~~~~~~
+ k_{wg}^{~~2} \mathbb{U}_w^\star \circ \mathbb{P}_w + k_{sg}^{~~2} \mathbb{U}_s^\star \circ \mathbb{P}_s  ~,
\\
&&  \mathbb{L}_e = \mathbb{U}_g^\star \circ \mathbb{P}_e + \mathbb{U}_e^\star \circ \mathbb{P}_g
%\nonumber
%\\
%&&~~~~~~~~~~~~~~~
+ ( k_{wg} k_{sg} k_{eg}^{~~-1} ) \{ \mathbb{U}_w^\star \circ \mathbb{P}_s + \mathbb{U}_s^\star \circ \mathbb{P}_w \}  ~,
\\
&&  \mathbb{L}_w = \mathbb{U}_g^\star \circ \mathbb{P}_w + \mathbb{U}_w^\star \circ \mathbb{P}_g
%\nonumber
%\\
%&&~~~~~~~~~~~~~~~
+ ( k_{eg} k_{sg} k_{wg}^{~~-1} ) \{ \mathbb{U}_e^\star \circ \mathbb{P}_s + \mathbb{U}_s^\star \circ \mathbb{P}_e \}  ~,
\\
&&  \mathbb{L}_s = \mathbb{U}_g^\star \circ \mathbb{P}_s + \mathbb{U}_s^\star \circ \mathbb{P}_g
%\nonumber
%\\
%&&~~~~~~~~~~~~~~~
+ ( k_{eg} k_{wg} k_{sg}^{~~-1} ) \{ \mathbb{U}_e^\star \circ \mathbb{P}_w + \mathbb{U}_w^\star \circ \mathbb{P}_e \}  ~.
\end{eqnarray}

\subsection{\label{sec:level1}Torque}

From the complex-sedenion angular momentum, it is able to define the complex-sedenion torque $\mathbb{W}$ as,
\begin{eqnarray}
\mathbb{W} = - v_0 ( i \mathbb{F} / v_0 + \lozenge ) \circ \mathbb{L}   ~,
\end{eqnarray}
where $\mathbb{W} = \mathbb{W}_g + k_{eg} \mathbb{W}_e + k_{wg} \mathbb{W}_w + k_{sg} \mathbb{W}_s$ . $\mathbb{W}_g$ , $\mathbb{W}_e$ , $\mathbb{W}_w$ , and $\mathbb{W}_s$ are respectively the components of the torque $\mathbb{W}$ in the spaces, $\mathbb{H}_g$ , $\mathbb{H}_e$ , $\mathbb{H}_w$, and $\mathbb{H}_s$ .
$\mathbb{D}_F = i \mathbb{F} / v_0 + \lozenge$. $\mathbb{D}_F = \mathbb{D}_{Fg} + k_{eg} \mathbb{D}_{Fe} + k_{wg} \mathbb{D}_{Fw} + k_{sg} \mathbb{D}_{Fs}$ . $\mathbb{D}_{Fg} = i \mathbb{F}_g / v_0 + \lozenge_g$. $\mathbb{D}_{Fe} = i \mathbb{F}_e / v_0 + k_{eg}^{~~-2} \lozenge_e$ . $\mathbb{D}_{Fw} = i \mathbb{F}_w / v_0 + k_{wg}^{~~-2} \lozenge_w$ . $\mathbb{D}_{Fs} = i \mathbb{F}_s / v_0 + k_{sg}^{~~-2} \lozenge_s$ . And that the ingredients of the torque in the four complex-quaternion spaces are as follows,
\begin{eqnarray}
&&  \mathbb{W}_g = - v_0 \mathbb{D}_{Fg} \circ \mathbb{L}_g - v_0 k_{eg}^{~~2} \mathbb{D}_{Fe} \circ \mathbb{L}_e
- v_0 k_{wg}^{~~2} \mathbb{D}_{Fw} \circ \mathbb{L}_w - v_0 k_{sg}^{~~2} \mathbb{D}_{Fs} \circ \mathbb{L}_s  ~,
\\
&&  \mathbb{W}_e = - v_0 \mathbb{D}_{Fg} \circ \mathbb{L}_e - v_0 \mathbb{D}_{Fe} \circ \mathbb{L}_g
- v_0 ( k_{wg} k_{sg} k_{eg}^{~~-1} ) ( \mathbb{D}_{Fw} \circ \mathbb{L}_s + \mathbb{D}_{Fs} \circ \mathbb{L}_w )  ~,
\\
&&  \mathbb{W}_w = - v_0 \mathbb{D}_{Fg} \circ \mathbb{L}_w - v_0 \mathbb{D}_{Fw} \circ \mathbb{L}_g
- v_0 ( k_{eg} k_{sg} k_{wg}^{~~-1} ) ( \mathbb{D}_{Fe} \circ \mathbb{L}_s + \mathbb{D}_{Fs} \circ \mathbb{L}_e )  ~,
\\
&&  \mathbb{W}_s = - v_0 \mathbb{D}_{Fg} \circ \mathbb{L}_s - v_0 \mathbb{D}_{Fs} \circ \mathbb{L}_g
- v_0 ( k_{eg} k_{wg} k_{sg}^{~~-1} ) ( \mathbb{D}_{Fe} \circ \mathbb{L}_w + \mathbb{D}_{Fw} \circ \mathbb{L}_e )  ~.
\end{eqnarray}

\begin{table}[h]
\caption{The `charges' of fundamental fields and of adjoint fields in the four complex-quaternion spaces, relevant to the complex-sedenion space.}
%\begin{ruledtabular}
\centering
\begin{tabular}{@{}llll@{}}
\hline\hline
space~~~~        &   field~equation                                                         &   charge                          &    field                  \\
\hline
$\mathbb{H}_g$   &   $\mu_g^g \mathbb{S}_g^g = - \lozenge_g^\ast \circ \mathbb{F}_g$        &   $m_g^g$ (mass)                  &    fundamental~field      \\
$\mathbb{H}_g$   &   $\mu_e^e \mathbb{S}_e^e = - \lozenge_e^\ast \circ \mathbb{F}_e$        &   $m_e^e$                         &    adjoint~field          \\
$\mathbb{H}_g$   &   $\mu_w^w \mathbb{S}_w^w = - \lozenge_w^\ast \circ \mathbb{F}_w$~~~~    &   $m_w^w$                         &    adjoint~field          \\
$\mathbb{H}_g$   &   $\mu_s^s \mathbb{S}_s^s = - \lozenge_s^\ast \circ \mathbb{F}_s$        &   $m_s^s$                         &    adjoint~field          \\
$\mathbb{H}_e$   &   $\mu_e^g \mathbb{S}_e^g = - \lozenge_g^\ast \circ \mathbb{F}_e$        &   $m_e^g$ (electric charge)~~~~   &    fundamental~field      \\
$\mathbb{H}_e$   &   $\mu_g^e \mathbb{S}_g^e = - \lozenge_e^\ast \circ \mathbb{F}_g$        &   $m_g^e$                         &    adjoint~field          \\
$\mathbb{H}_e$   &   $\mu_w^s \mathbb{S}_w^s = - \lozenge_s^\ast \circ \mathbb{F}_w$        &   $m_w^s$                         &    adjoint~field          \\
$\mathbb{H}_e$   &   $\mu_s^w \mathbb{S}_s^w = - \lozenge_w^\ast \circ \mathbb{F}_s$        &   $m_s^w$                         &    adjoint~field          \\
$\mathbb{H}_w$   &   $\mu_w^g \mathbb{S}_w^g = - \lozenge_g^\ast \circ \mathbb{F}_w$        &   $m_w^g$ (W charge)              &    fundamental~field      \\
$\mathbb{H}_w$   &   $\mu_g^w \mathbb{S}_g^w = - \lozenge_w^\ast \circ \mathbb{F}_g$        &   $m_g^w$                         &    adjoint~field          \\
$\mathbb{H}_w$   &   $\mu_e^s \mathbb{S}_e^s = - \lozenge_s^\ast \circ \mathbb{F}_e$        &   $m_e^s$                         &    adjoint~field          \\
$\mathbb{H}_w$   &   $\mu_s^e \mathbb{S}_s^e = - \lozenge_e^\ast \circ \mathbb{F}_s$        &   $m_s^e$                         &    adjoint~field          \\
$\mathbb{H}_s$   &   $\mu_s^g \mathbb{S}_s^g = - \lozenge_g^\ast \circ \mathbb{F}_s$        &   $m_s^g$ (strong charge)         &    fundamental~field      \\
$\mathbb{H}_s$   &   $\mu_g^s \mathbb{S}_g^s = - \lozenge_s^\ast \circ \mathbb{F}_g$        &   $m_g^s$                         &    adjoint~field          \\
$\mathbb{H}_s$   &   $\mu_e^w \mathbb{S}_e^w = - \lozenge_w^\ast \circ \mathbb{F}_e$        &   $m_e^w$                         &    adjoint~field          \\
$\mathbb{H}_s$   &   $\mu_w^e \mathbb{S}_w^e = - \lozenge_e^\ast \circ \mathbb{F}_w$        &   $m_w^e$                         &    adjoint~field          \\
\hline\hline
\end{tabular}
%\end{ruledtabular}
\end{table}

\subsection{\label{sec:level1}Force}

From the complex-sedenion torque, it is capable of defining the complex-sedenion force $\mathbb{N}$ as,
\begin{eqnarray}
\mathbb{N} = - ( i \mathbb{F} / v_0 + \lozenge ) \circ \mathbb{W}   ~,
\end{eqnarray}
where $\mathbb{N} = \mathbb{N}_g + k_{eg} \mathbb{N}_e + k_{wg} \mathbb{N}_w + k_{sg} \mathbb{N}_s$ . $\mathbb{N}_g$ , $\mathbb{N}_e$ , $\mathbb{N}_w$ , and $\mathbb{N}_s$ are respectively the components of the force $\mathbb{N}$ in the spaces, $\mathbb{H}_g$ , $\mathbb{H}_e$ , $\mathbb{H}_w$ , and $\mathbb{H}_s$ .
And that the ingredients of the torque in the four complex-quaternion spaces are as follows,
\begin{eqnarray}
&&  \mathbb{N}_g = - \mathbb{D}_{Fg} \circ \mathbb{W}_g - k_{eg}^{~~2} \mathbb{D}_{Fe} \circ \mathbb{W}_e
- k_{wg}^{~~2} \mathbb{D}_{Fw} \circ \mathbb{W}_w - k_{sg}^{~~2} \mathbb{D}_{Fs} \circ \mathbb{W}_s  ~,
\\
&&  \mathbb{N}_e = - \mathbb{D}_{Fg} \circ \mathbb{W}_e - \mathbb{D}_{Fe} \circ \mathbb{W}_g
- ( k_{wg} k_{sg} k_{eg}^{~~-1} ) ( \mathbb{D}_{Fw} \circ \mathbb{W}_s + \mathbb{D}_{Fs} \circ \mathbb{W}_w )  ~,
\\
&&  \mathbb{N}_w = - \mathbb{D}_{Fg} \circ \mathbb{W}_w - \mathbb{D}_{Fw} \circ \mathbb{W}_g
- ( k_{eg} k_{sg} k_{wg}^{~~-1} ) ( \mathbb{D}_{Fe} \circ \mathbb{W}_s + \mathbb{D}_{Fs} \circ \mathbb{W}_e )  ~,
\\
&&  \mathbb{N}_s = - \mathbb{D}_{Fg} \circ \mathbb{W}_s - \mathbb{D}_{Fs} \circ \mathbb{W}_g
- ( k_{eg} k_{wg} k_{sg}^{~~-1} ) ( \mathbb{D}_{Fe} \circ \mathbb{W}_w + \mathbb{D}_{Fw} \circ \mathbb{W}_e )  ~.
\end{eqnarray}

The above shows that the application of complex-sedenion is able to infer some field equations, relevant to fundamental fields and adjoint fields, in the classical mechanics on the macroscopic scale (Table 2). In the complex-sedenion space, the complex-sedenion angular momentum comprises the orbital angular momentum, magnetic moment, and electric moment and so forth. The complex-sedenion torque consists of the conventional torque and energy and so on. In most cases, the complex-sedenion force will be equal to zero. From the equation, it is able to deduce the force equilibrium equation, precessional equilibrium equation, power equation, mass continuity equation, and current continuity equation and so forth. The force equilibrium equation includes the inertial force, gravity, electromagnetic force, and energy gradient and so forth. As a force term, the energy gradient can be applied to explore the astrophysical jets (see Ref.[33]) , condensed dark matters, and new principle of the accelerator and so forth. Meanwhile the precessional equilibrium equation can be utilized to account for certain precessional motions, deducing the angular velocity of Larmor precession and so forth.

In case the strong-nuclear field and W-nuclear field can be neglected, the definitions of angular momentum, torque, and force, described with the complex-sedenions, can be degenerated into that described with the complex-octonions respectively.

\begin{table}[h]
\caption{Some physical quantities and definitions relevant to the four fundamental interactions, in the classical mechanics described with the complex-sedenions.}
%\begin{ruledtabular}
\centering
\begin{tabular}{@{}ll@{}}
\hline\hline
sedenion~physics~quantity ~~~~~~~~    &  definition                                                                                                 \\
\hline
radius~vector                         &  $\mathbb{R} = \mathbb{R}_g + k_{eg} \mathbb{R}_e + k_{wg} \mathbb{R}_w + k_{sg} \mathbb{R}_s$              \\
sedenion~operator                     &  $\lozenge = \lozenge_g + k_{eg}^{~~-1} \lozenge_e + k_{wg}^{~~-1} \lozenge_w + k_{sg}^{~~-1} \lozenge_s$   \\
integrating~function                  &  $\mathbb{X} = \mathbb{X}_g + k_{eg} \mathbb{X}_e + k_{wg} \mathbb{X}_w + k_{sg} \mathbb{X}_s$              \\
%%%
field~potential                       &  $\mathbb{A} = i \lozenge^\star \circ \mathbb{X}  $                                                         \\
field~strength                        &  $\mathbb{F} = \lozenge \circ \mathbb{A}  $                                                                 \\
field~source                          &  $\mu \mathbb{S} = - ( i \mathbb{F} / v_0 + \lozenge )^* \circ \mathbb{F} $                                 \\
%%%
linear~momentum                       &  $\mathbb{P} = \mu \mathbb{S} / \mu_g^g $                                                                   \\
angular~momentum                      &  $\mathbb{L} = \mathbb{U}^\star \circ \mathbb{P} $                                                          \\
sedenion~torque                       &  $\mathbb{W} = - v_0 ( i \mathbb{F} / v_0 + \lozenge ) \circ \mathbb{L} $                                   \\
sedenion~force                        &  $\mathbb{N} = - ( i \mathbb{F} / v_0 + \lozenge ) \circ \mathbb{W} $                                       \\
\hline\hline
\end{tabular}
%\end{ruledtabular}
\end{table}

\section{\label{sec:level1}Wave function}

In the quantum mechanics, what plays an important role is the wavefunction, connected with the physical quantities in the classical mechanics, rather than any pure physical quantity in the classical mechanics. Therefore, in the classical mechanics described with the complex-sedenions, it is necessary to multiply the physical quantity, in the classical mechanics, with the dimensionless complex-sedenion auxiliary quantity, transforming it to become the wavefunction. For instance, the wavefunction of the complex-sedenion angular momentum $\mathbb{L}$ is, $\Psi_{ZL} = \mathbb{Z}_L \circ \mathbb{L} / \hbar$ . Herein $\mathbb{Z}_L$ is a dimensionless auxiliary quantity, and $(2 \pi \hbar)$ is the Planck constant. Apparently, the wavefunctions or quantum physical quantities in the quantum mechanics are some functions of the physical quantities in the classical mechanics (Table 3). In the Table 3, the complex-sedenion quantum physical quantities include the quantum integrating function of field potential, quantum-field potential, quantum-field strength, quantum-field source, quantum linear momentum, quantum angular momentum, quantum torque, and quantum force and so on. In other words, in terms of the concept of function, the quantum-fields of the quantum mechanics on the microscopic scale are just the functions of the classical fields of the classical mechanics on the macroscopic scale. There are some similarities as well as differences between the quantum physical quantity and classical physical quantity.

In the subspace $\mathbb{H}_g$ of the complex-octonion space $\mathbb{O}$ , the complex-quaternion wavefunction is defined as, $\Psi_{Lg} = \mathbb{L}_g / \hbar$ , in terms of the complex-quaternion angular momentum $\mathbb{L}_g$ . The major ingredient of $\Psi_{Lg}$ is, $\Psi_{Lg}^\prime = L_q exp ( \textbf{\emph{i}}_q \alpha_q )$. By means of certain appropriate transformations (Appendix A), it is able to achieve two new wavefunctions, $\Psi = - \textbf{\emph{i}}_q \circ \Psi_{Lg}^\prime$ and $\Psi^{\prime\prime} = - \textbf{z} \circ \Psi_{Lg}^\prime$ . Herein $L_q$ and $\alpha_q$ are all real. $\textbf{\emph{i}}_q$ is a three-dimensional unit vector, and $\textbf{z}$ is a vector.

In the complex-octonion space $\mathbb{O}$ , which consists of two complex-quaternion spaces, $\mathbb{H}_g$ and $\mathbb{H}_e$ , any complex-quaternion wavefunction can be degenerated into two correlative wavefunctions. a) Imaginary unit. When the direction of unit vector $\textbf{\emph{i}}_q$ is incapable of playing a major role in the complex-quaternion wavefunction, the unit vector $\textbf{\emph{i}}_q$ will be degenerated into the imaginary unit $i$. If the unit vector $\textbf{\emph{i}}_q$ can be replaced by the imaginary unit $i$ , the above wavefunctions, $\Psi$ and $\Psi^{\prime\prime}$ , will be degraded into the wavefunction, described with the complex-number, in the conventional quantum mechanics. b) Unit vector. When the vector property of the unit vector $\textbf{\emph{i}}_q$ can be neglected partially in the complex-quaternion wavefunction, the unit vector $\textbf{\emph{i}}_q$ will be degenerated into a new three-dimensional unit vector $\textbf{\emph{i}}_q^\prime$ , which is independent of the unit vector $\textbf{\emph{i}}_q$ . In case the unit vector $\textbf{\emph{i}}_q$ can be replaced by the unit vector $\textbf{\emph{i}}_q^\prime$, the three-dimensional unit vector $\textbf{\emph{i}}_q^\prime$ will possess three new degrees of freedom, in contrast to the imaginary unit $i$ . That is, one wavefunction with the unit vector $\textbf{\emph{i}}_q^\prime$ is equivalent to three conventional wavefunctions with the complex-numbers. In a general way, a complex-quaternion wavefunction may be degraded into three complex-number wavefunctions, which are independent of each other. If we mistake the three-dimensional unit vector, $\textbf{\emph{i}}_q^\prime$ , for the imaginary unit, either of two wavefunctions, $\Psi$ and $\Psi^{\prime\prime}$ , must be considered equivalently as the complex-number wavefunctions, which are in possession of three new degrees of freedom. Herein $i^2 = -1$, $\textbf{\emph{i}}_q^{~2} = -1$, $(\textbf{\emph{i}}_q^\prime)^2 = -1$.

Similarly, in the complex-quaternion space $\mathbb{H}_g$ , in terms of some other physical quantities, also it is able to define their corresponding complex-quaternion wavefunctions. What is more interesting is that, it is capable of defining further some complex-quaternion wavefunctions, connected with the physical quantities, in the rest of complex-quaternion spaces, $\mathbb{H}_e$ , $\mathbb{H}_w$ , and $\mathbb{H}_s$ . Moreover, under certain different circumstances, one should choose different approximate results (imaginary unit $i$ , or unit vector $\textbf{\emph{i}}_q^\prime$ and so forth) for the complex-quaternion wavefunctions, in order to facilitate these complex-quaternion wavefunctions to be degenerated into a few wavefunctions, described with complex-numbers, in the conventional quantum mechanics.

\begin{table}[h]
\caption{Some classical physical quantities, auxiliary quantities, wavefunctions, and quantum physical quantities, in the complex-sedenion space.}
%\begin{ruledtabular}
\centering
\begin{tabular}{@{}lclc@{}}
\hline\hline
classical~physical~quantity~~~~    & auxiliary~quantity~~~~  & wave~function                                    & quantum~physical~quantity                                \\
\hline
composite~radius~vector, $\mathbb{U}$ & $\mathbb{Z}_U$   & $\Psi_{ZU} = \mathbb{Z}_U \circ \mathbb{U} / \hbar$  & $\mathbb{U}_{(\Psi)} = \mathbb{Z}_U \circ \mathbb{U}$
\\
integrating~function, $\mathbb{X}$ & $\mathbb{Z}_X$      & $\Psi_{ZX} = \mathbb{Z}_X \circ \mathbb{X} / \hbar$  & $\mathbb{X}_{(\Psi)} = \mathbb{Z}_X \circ \mathbb{X}$ \\
field~potential,  $\mathbb{A}$     & $\mathbb{Z}_A$      & $\Psi_{ZA} = \mathbb{Z}_A \circ \mathbb{A} / \hbar$  & $\mathbb{A}_{(\Psi)} = \mathbb{Z}_A \circ \mathbb{A}$ \\
field~strength,   $\mathbb{F}$     & $\mathbb{Z}_F$      & $\Psi_{ZF} = \mathbb{Z}_F \circ \mathbb{F} / \hbar$  & $\mathbb{F}_{(\Psi)} = \mathbb{Z}_F \circ \mathbb{F}$ \\
field~source,     $\mathbb{S}$     & $\mathbb{Z}_S$      & $\Psi_{ZS} = \mathbb{Z}_S \circ \mathbb{S} / \hbar$  & $\mathbb{S}_{(\Psi)} = \mathbb{Z}_S \circ \mathbb{S}$ \\
linear~momentum,  $\mathbb{P}$     & $\mathbb{Z}_P$      & $\Psi_{ZP} = \mathbb{Z}_P \circ \mathbb{P} / \hbar$  & $\mathbb{P}_{(\Psi)} = \mathbb{Z}_P \circ \mathbb{P}$ \\
angular~momentum, $\mathbb{L}$     & $\mathbb{Z}_L$      & $\Psi_{ZL} = \mathbb{Z}_L \circ \mathbb{L} / \hbar$  & $\mathbb{L}_{(\Psi)} = \mathbb{Z}_L \circ \mathbb{L}$ \\
torque,           $\mathbb{W}$     & $\mathbb{Z}_W$      & $\Psi_{ZW} = \mathbb{Z}_W \circ \mathbb{W} / \hbar$~~~~  & $\mathbb{W}_{(\Psi)} = \mathbb{Z}_W \circ \mathbb{W}$ \\
force,            $\mathbb{N}$     & $\mathbb{Z}_N$      & $\Psi_{ZN} = \mathbb{Z}_N \circ \mathbb{N} / \hbar$  & $\mathbb{N}_{(\Psi)} = \mathbb{Z}_N \circ \mathbb{N}$ \\
\hline\hline
\end{tabular}
%\end{ruledtabular}
\end{table}

\section{\label{sec:level1}Quantum field equations}

In the complex-sedenion space $\mathbb{K}$ , making use of the exponential forms and wavefunctions of  the complex-sedenions, it is able to deduce the quantum-field equations, on the microscopic scale, for the gravitational fields, electromagnetic fields, W-nuclear fields, and strong-nuclear fields, including the quantum-field potential, quantum-field strength, quantum-field source, quantum linear momentum, angular momentum, quantum torque, and quantum force and so forth.

The quantum-fields of the quantum mechanics on the microscopic scale are just the functions or transformations of the classical fields of the classical mechanics on the macroscopic scale, in the complex-sedenion space $\mathbb{K}$ . In a similar way to the method of the mathematical inference, for the field equations of classical mechanics on the macroscopic scale, it is capable of deducing the field equations of the quantum mechanics on the microscopic scale. Starting from the integrating-function of field potential $\mathbb{X}$ , of the field equations in the classical mechanics, one can infer gradually the field equations in the quantum mechanics (Table 4). Obviously, a majority of physical quantities and field equations in the Table 4 are different from that in the Table 2. For instance, the quantum-field source $\mathbb{S}_{(\Psi)}$ in the Table 4 is independent of the classical field source $\mathbb{S}$ in the Table 2.

\subsection{\label{sec:level1}Quantum field potential}

In the complex-sedenion space $\mathbb{K}$ , the complex-sedenion quantum integrating-function of field potential is, $\mathbb{X}_{(\Psi)} = \mathbb{X}_{(\Psi)g} + k_{eg} \mathbb{X}_{(\Psi)e} + k_{wg} \mathbb{X}_{(\Psi)w} + k_{sg} \mathbb{X}_{(\Psi)s}$. $\mathbb{X}_{(\Psi)g}$ , $\mathbb{X}_{(\Psi)e}$ , $\mathbb{X}_{(\Psi)w}$ , and $\mathbb{X}_{(\Psi)s}$ are respectively the components of the quantum integrating-function of field potential $\mathbb{X}_{(\Psi)}$ in the spaces, $\mathbb{H}_g$ , $\mathbb{H}_e$ , $\mathbb{H}_w$ , and $\mathbb{H}_s$. Herein $\mathbb{X}_{(\Psi)} = \mathbb{Z}_X \circ \mathbb{X}$, and $\mathbb{Z}_X$ is one auxiliary quantity. $\mathbb{X}_{(\Psi)g} = i \textbf{\emph{I}}_{g0} X_{(\Psi)g0} + \Sigma \textbf{\emph{I}}_{gk} X_{(\Psi)gk}$ . $\mathbb{X}_{(\Psi)e} = i \textbf{\emph{I}}_{e0} X_{(\Psi)e0} + \Sigma \textbf{\emph{I}}_{ek} X_{(\Psi)ek}$ . $\mathbb{X}_{(\Psi)w} = i \textbf{\emph{I}}_{w0} X_{(\Psi)w0} + \Sigma \textbf{\emph{I}}_{wk} X_{(\Psi)wk}$ . $\mathbb{X}_{(\Psi)s} = i \textbf{\emph{I}}_{s0} X_{(\Psi)s0} + \Sigma \textbf{\emph{I}}_{sk} X_{(\Psi)sk}$ . $X_{(\Psi)gj}$ , $X_{(\Psi)ej}$ , $X_{(\Psi)wj}$ , and $X_{(\Psi)sj}$ are all real.

In the quantum mechanics, we may encounter a new composite operator, $\mathbb{D}_Z = i \mathbb{Z}_W \circ \mathbb{W}^\star / ( \hbar v_0 ) + \lozenge $ . Herein
$\mathbb{D}_Z = \mathbb{D}_{Zg} + k_{eg} \mathbb{D}_{Ze} + k_{wg} \mathbb{D}_{Zw} + k_{sg} \mathbb{D}_{Zs}$ . $\mathbb{D}_{(ZW)} = \mathbb{Z}_W \circ \mathbb{W}^\star$ . $\mathbb{D}_{Zg} = i \mathbb{D}_{(ZW)g} / ( \hbar v_0 ) + \lozenge_g$ . $\mathbb{D}_{Ze} = i \mathbb{D}_{(ZW)e} / ( \hbar v_0 ) + k_{eg}^{~~-2} \lozenge_e$. $\mathbb{D}_{Zw} = i \mathbb{D}_{(ZW)w} / ( \hbar v_0 ) + k_{wg}^{~~-2} \lozenge_w$ . $\mathbb{D}_{Zs} = i \mathbb{D}_{(ZW)s} / ( \hbar v_0 ) + k_{sg}^{~~-2} \lozenge_s$ . Meanwhile, $\mathbb{W}^\star = \mathbb{W}^\star_g + k_{eg} \mathbb{W}^\star_e + k_{wg} \mathbb{W}^\star_w + k_{sg} \mathbb{W}^\star_s$. $\mathbb{Z}_W = \mathbb{Z}_{Wg} + k_{eg} \mathbb{Z}_{We} + k_{wg} \mathbb{Z}_{Ww} + k_{sg} \mathbb{Z}_{Ws}$. $\mathbb{Z}^\star_W = \mathbb{Z}^\star_{Wg} + k_{eg} \mathbb{Z}^\star_{We} + k_{wg} \mathbb{Z}^\star_{Ww} + k_{sg} \mathbb{Z}^\star_{Ws}$ . $\mathbb{D}_{(ZW)} = \mathbb{D}_{(ZW)g} + k_{eg} \mathbb{D}_{(ZW)e} + k_{wg} \mathbb{D}_{(ZW)w} + k_{sg} \mathbb{D}_{(ZW)s}$ . Herein $\mathbb{Z}_{Wg}$, $\mathbb{Z}_{We}$, $\mathbb{Z}_{Ww}$, and $\mathbb{Z}_{Ws}$ are respectively the components of the auxiliary quantity, $\mathbb{Z}_W$ , in the complex-quaternion spaces, $\mathbb{H}_g$ , $\mathbb{H}_e$ , $\mathbb{H}_w$ and $\mathbb{H}_s$ . And that the ingredients of this new composite operator in the four complex-quaternion spaces are as follows,
\begin{eqnarray}
&& \mathbb{D}_{(ZW)g} = \mathbb{Z}_{Wg} \circ \mathbb{W}_g^\star + k_{eg}^{~~2} \mathbb{Z}_{We} \circ \mathbb{W}_e^\star
+ k_{wg}^{~~2} \mathbb{Z}_{Ww} \circ \mathbb{W}_w^\star + k_{sg}^{~~2} \mathbb{Z}_{Ws} \circ \mathbb{W}_s^\star ~ ,  \\
&& \mathbb{D}_{(ZW)e} = \mathbb{Z}_{Wg} \circ \mathbb{W}_e^\star + \mathbb{Z}_{We} \circ \mathbb{W}_g^\star
+ (k_{wg} k_{sg} k_{eg}^{~~-1}) \mathbb{Z}_{Ww} \circ \mathbb{W}_s^\star + (k_{sg} k_{wg} k_{eg}^{~~-1}) \mathbb{Z}_{Ws} \circ \mathbb{W}_w^\star ~ ,  \\
&& \mathbb{D}_{(ZW)w} = \mathbb{Z}_{Wg} \circ \mathbb{W}_w^\star + \mathbb{Z}_{Ww} \circ \mathbb{W}_g^\star
+ (k_{eg} k_{sg} k_{wg}^{~~-1}) \mathbb{Z}_{We} \circ \mathbb{W}_s^\star + (k_{sg} k_{eg} k_{wg}^{~~-1}) \mathbb{Z}_{Ws} \circ \mathbb{W}_e^\star ~ ,  \\
&& \mathbb{D}_{(ZW)s} = \mathbb{Z}_{Wg} \circ \mathbb{W}_s^\star + \mathbb{Z}_{Ws} \circ \mathbb{W}_g^\star
+ (k_{eg} k_{wg} k_{sg}^{~~-1}) \mathbb{Z}_{We} \circ \mathbb{W}_w^\star + (k_{wg} k_{eg} k_{sg}^{~~-1}) \mathbb{Z}_{Ww} \circ \mathbb{W}_e^\star ~ .
\end{eqnarray}

In the complex-sedenion space $\mathbb{K}$ , by means of the physical quantities of the complex-sedenion quantum integrating-function of field potential $\mathbb{X}_{(\Psi)}$ , the complex-sedenion quantum-field potential $\mathbb{A}_{(\Psi)}$ is defined as,
\begin{eqnarray}
\mathbb{A}_{(\Psi)} = i \mathbb{D}_Z^\star \circ \mathbb{X}_{(\Psi)}  ~ ,
\end{eqnarray}
where $\mathbb{A}_{(\Psi)} = \mathbb{Z}_A \circ \mathbb{A}$ , and $\mathbb{Z}_A$ is one auxiliary quantity. $\mathbb{A}_{(\Psi)} = \mathbb{A}_{(\Psi)g} + k_{eg} \mathbb{A}_{(\Psi)e} + k_{wg} \mathbb{A}_{(\Psi)w} + k_{sg} \mathbb{A}_{(\Psi)s}$ . $\mathbb{A}_{(\Psi)g}$ , $\mathbb{A}_{(\Psi)e}$, $\mathbb{A}_{(\Psi)w}$, and $\mathbb{A}_{(\Psi)s}$ are respectively the components of the quantum-field potential $\mathbb{A}_{(\Psi)}$ in the  complex-quaternion spaces, $\mathbb{H}_g$ , $\mathbb{H}_e$ , $\mathbb{H}_w$ , and $\mathbb{H}_s$ . $\mathbb{D}_Z^\star = \mathbb{D}_{Zg}^\star + k_{eg} \mathbb{D}_{Ze}^\star + k_{wg} \mathbb{D}_{Zw}^\star + k_{sg} \mathbb{D}_{Zs}^\star$ .

According to the multiplication rule of the sedenions, the complex-sedenion quantum-field potential $\mathbb{A}_{(\Psi)}$ includes the quantum-field potentials of the four fundamental quantum-fields (that is, gravitational fundamental quantum-field, electromagnetic fundamental quantum-field, W-nuclear fundamental quantum-field, and strong-nuclear fundamental quantum-field), and twelve adjoint quantum-fields. And that the ingredients of the quantum-field potential in the four complex-quaternion spaces are as follows:

a) In the complex-quaternion space $\mathbb{H}_g$ , the component $\mathbb{A}_{(\Psi)g}$ contains the gravitational fundamental quantum-field potential $\mathbb{A}_{(\Psi)g}^g$ , electromagnetic adjoint quantum-field potential $\mathbb{A}_{(\Psi)e}^e$ , W-nuclear adjoint quantum-field potential $\mathbb{A}_{(\Psi)w}^w$ , and strong-nuclear adjoint quantum-field potential $\mathbb{A}_{(\Psi)s}^s$ . Herein $\mathbb{A}_{(\Psi)g} = \mathbb{A}_{(\Psi)g}^g + \mathbb{A}_{(\Psi)e}^e + \mathbb{A}_{(\Psi)w}^w + \mathbb{A}_{(\Psi)s}^s$. $\mathbb{A}_{(\Psi)g} = i \textbf{\emph{I}}_{g0} A_{(\Psi)g0} + \Sigma \textbf{\emph{I}}_{gk} A_{(\Psi)gk}$. $\mathbb{A}_{(\Psi)g}^g = i \mathbb{D}_{Zg}^\star \circ \mathbb{X}_{(\Psi)g}$. $\mathbb{A}_{(\Psi)e}^e = i \mathbb{D}_{Ze}^\star \circ \mathbb{X}_{(\Psi)e}$ . $\mathbb{A}_{(\Psi)w}^w = i \mathbb{D}_{Zw}^\star \circ \mathbb{X}_{(\Psi)w}$. $\mathbb{A}_{(\Psi)s}^s = i \mathbb{D}_{Zs}^\star \circ \mathbb{X}_{(\Psi)s}$ . $\mathbb{A}_{(\Psi)g}^g = i \textbf{\emph{I}}_{g0} A^g_{(\Psi)g0} + \Sigma \textbf{\emph{I}}_{gk} A^g_{(\Psi)gk}$ . $\mathbb{A}_{(\Psi)e}^e = i \textbf{\emph{I}}_{g0} A^e_{(\Psi)e0} + \Sigma \textbf{\emph{I}}_{gk} A^e_{(\Psi)ek}$ . $\mathbb{A}_{(\Psi)w}^w = i \textbf{\emph{I}}_{g0} A^w_{(\Psi)w0} + \Sigma \textbf{\emph{I}}_{gk} A^w_{(\Psi)wk}$ . $\mathbb{A}_{(\Psi)s}^s = i \textbf{\emph{I}}_{g0} A^s_{(\Psi)s0} + \Sigma \textbf{\emph{I}}_{gk} A^s_{(\Psi)sk}$ . $A_{(\Psi)gj}$ , $A^g_{(\Psi)gj}$ , $A^e_{(\Psi)ej}$ , $A^w_{(\Psi)wj}$, and $A^s_{(\Psi)sj}$ are all real.

b) In the complex 2-quaternion space $\mathbb{H}_e$ , the component $\mathbb{A}_{(\Psi)e}$ consists of the electromagnetic fundamental quantum-field potential $\mathbb{A}_{(\Psi)e}^g$ , gravitational adjoint quantum-field potential $\mathbb{A}_{(\Psi)g}^e$ , W-nuclear adjoint quantum-field potential $\mathbb{A}_{(\Psi)w}^s$ , and strong-nuclear adjoint quantum-field potential $\mathbb{A}_{(\Psi)s}^w$ . Herein $\mathbb{A}_{(\Psi)e} = \mathbb{A}_{(\Psi)e}^g + k_{eg}^{~~-2} \mathbb{A}_{(\Psi)g}^e + ( k_{wg} k_{sg}^{~~-1} k_{eg}^{~~-1} ) \mathbb{A}_{(\Psi)w}^s + ( k_{sg} k_{wg}^{~~-1} k_{eg}^{~~-1} ) \mathbb{A}_{(\Psi)s}^w$ . $\mathbb{A}_{(\Psi)e} = i \textbf{\emph{I}}_{e0} A_{(\Psi)e0} + \Sigma \textbf{\emph{I}}_{ek} A_{(\Psi)ek}$ . $\mathbb{A}_{(\Psi)e}^g = i \mathbb{D}_{Zg}^\star \circ \mathbb{X}_{(\Psi)e}$ . $\mathbb{A}_{(\Psi)g}^e = i \mathbb{D}_{Ze}^\star \circ \mathbb{X}_{(\Psi)g}$ . $\mathbb{A}_{(\Psi)w}^s = i \mathbb{D}_{Zs}^\star \circ \mathbb{X}_{(\Psi)w}$ . $\mathbb{A}_{(\Psi)s}^w = i \mathbb{D}_{Zw}^\star \circ \mathbb{X}_{(\Psi)s}$ . $\mathbb{A}_{(\Psi)e}^g = i \textbf{\emph{I}}_{e0} A^g_{(\Psi)e0} + \Sigma \textbf{\emph{I}}_{ek} A^g_{(\Psi)ek}$. $\mathbb{A}_{(\Psi)g}^e = i \textbf{\emph{I}}_{e0} A^e_{(\Psi)g0} + \Sigma \textbf{\emph{I}}_{ek} A^e_{(\Psi)gk}$ . $\mathbb{A}_{(\Psi)w}^s = i \textbf{\emph{I}}_{e0} A^s_{(\Psi)w0} + \Sigma \textbf{\emph{I}}_{ek} A^s_{(\Psi)wk}$ . $\mathbb{A}_{(\Psi)s}^w = i \textbf{\emph{I}}_{e0} A^w_{(\Psi)s0} + \Sigma \textbf{\emph{I}}_{ek} A^w_{(\Psi)sk}$ . $A_{(\Psi)ej}$ , $A^g_{(\Psi)ej}$ , $A^e_{(\Psi)gj}$ , $A^s_{(\Psi)wj}$ , and $A^w_{(\Psi)sj}$ are all real.

c) In the complex 3-quaternion space $\mathbb{H}_w$ , the component $\mathbb{A}_{(\Psi)w}$ includes the W-nuclear fundamental quantum-field potential $\mathbb{A}_{(\Psi)w}^g$ , gravitational adjoint quantum-field potential $\mathbb{A}_{(\Psi)g}^w$ , electromagnetic adjoint quantum-field potential $\mathbb{A}_{(\Psi)e}^s$ , and strong-nuclear adjoint quantum-field potential $\mathbb{A}_{(\Psi)s}^e$ . Herein $\mathbb{A}_{(\Psi)w} = \mathbb{A}_{(\Psi)w}^g + k_{wg}^{~~-2} \mathbb{A}_{(\Psi)g}^w + ( k_{eg} k_{wg}^{~~-1} k_{sg}^{~~-1} ) \mathbb{A}_{(\Psi)e}^s + ( k_{sg} k_{wg}^{~~-1} k_{eg}^{~~-1} ) \mathbb{A}_{(\Psi)s}^e$. $\mathbb{A}_{(\Psi)w} = i \textbf{\emph{I}}_{w0} A_{(\Psi)w0} + \Sigma \textbf{\emph{I}}_{wk} A_{(\Psi)wk}$ . $\mathbb{A}_{(\Psi)w}^g = i \mathbb{D}_{Zg}^\star \circ \mathbb{X}_{(\Psi)w}$ . $\mathbb{A}_{(\Psi)g}^w = i \mathbb{D}_{Zw}^\star \circ \mathbb{X}_{(\Psi)g}$ . $\mathbb{A}_{(\Psi)e}^s = i \mathbb{D}_{Zs}^\star \circ \mathbb{X}_{(\Psi)e}$ . $\mathbb{A}_{(\Psi)s}^e = i \mathbb{D}_{Ze}^\star \circ \mathbb{X}_{(\Psi)s}$ . $\mathbb{A}_{(\Psi)w}^g = i \textbf{\emph{I}}_{w0} A^g_{(\Psi)w0} + \Sigma \textbf{\emph{I}}_{wk} A^g_{(\Psi)wk}$. $\mathbb{A}_{(\Psi)g}^w = i \textbf{\emph{I}}_{w0} A^w_{(\Psi)g0} + \Sigma \textbf{\emph{I}}_{wk} A^w_{(\Psi)gk}$. $\mathbb{A}_{(\Psi)e}^s = i \textbf{\emph{I}}_{w0} A^s_{(\Psi)e0} + \Sigma \textbf{\emph{I}}_{wk} A^s_{(\Psi)ek}$ . $\mathbb{A}_{(\Psi)s}^e = i \textbf{\emph{I}}_{w0} A^e_{(\Psi)s0} + \Sigma \textbf{\emph{I}}_{wk} A^e_{(\Psi)sk}$ . $A_{(\Psi)wj}$ , $A^g_{(\Psi)wj}$, $A^w_{(\Psi)gj}$ , $A^s_{(\Psi)ej}$ , and $A^e_{(\Psi)sj}$ are all real.

d) In the complex 4-quaternion space $\mathbb{H}_s$ , the component $\mathbb{A}_{(\Psi)s}$ covers the strong-nuclear fundamental quantum-field potential $\mathbb{A}_{(\Psi)s}^g$ , gravitational adjoint quantum-field potential $\mathbb{A}_{(\Psi)g}^s$ , electromagnetic adjoint quantum-field potential $\mathbb{A}_{(\Psi)e}^w$ , and W-nuclear adjoint quantum-field potential $\mathbb{A}_{(\Psi)w}^e$ . Herein $\mathbb{A}_{(\Psi)s} = \mathbb{A}_{(\Psi)s}^g + k_{sg}^{~~-2} \mathbb{A}_{(\Psi)g}^s + ( k_{eg} k_{wg}^{~~-1} k_{sg}^{~~-1} ) \mathbb{A}_{(\Psi)e}^w + ( k_{wg} k_{sg}^{~~-1} k_{eg}^{~~-1} ) \mathbb{A}_{(\Psi)w}^e$ . $\mathbb{A}_{(\Psi)s} = i \textbf{\emph{I}}_{s0} A_{(\Psi)s0} + \Sigma \textbf{\emph{I}}_{sk} A_{(\Psi)sk}$ . $\mathbb{A}_{(\Psi)s}^g = i \mathbb{D}_{Zg}^\star \circ \mathbb{X}_{(\Psi)s}$ . $\mathbb{A}_{(\Psi)g}^s = i \mathbb{D}_{Zs}^\star \circ \mathbb{X}_{(\Psi)g}$ . $\mathbb{A}_{(\Psi)e}^w = i \mathbb{D}_{Zw}^\star \circ \mathbb{X}_{(\Psi)e}$. $\mathbb{A}_{(\Psi)w}^e = i \mathbb{D}_{Ze}^\star \circ \mathbb{X}_{(\Psi)w}$. $\mathbb{A}_{(\Psi)s}^g = i \textbf{\emph{I}}_{s0} A^g_{(\Psi)s0} + \Sigma \textbf{\emph{I}}_{sk} A^g_{(\Psi)sk}$ . $\mathbb{A}_{(\Psi)g}^s = i \textbf{\emph{I}}_{s0} A^s_{(\Psi)g0} + \Sigma \textbf{\emph{I}}_{sk} A^s_{(\Psi)gk}$. $\mathbb{A}_{(\Psi)e}^w = i \textbf{\emph{I}}_{s0} A^w_{(\Psi)e0} + \Sigma \textbf{\emph{I}}_{sk} A^w_{(\Psi)ek}$ . $\mathbb{A}_{(\Psi)w}^e = i \textbf{\emph{I}}_{s0} A^e_{(\Psi)w0} + \Sigma \textbf{\emph{I}}_{sk} A^e_{(\Psi)wk}$. $A_{(\Psi)sj}$ , $A^g_{(\Psi)sj}$ , $A^s_{(\Psi)gj}$ , $A^w_{(\Psi)ej}$ , and $A^e_{(\Psi)wj}$ are all real.

\subsection{\label{sec:level1}Quantum field strength}

From the complex-sedenion quantum-field potential, it is able to define the complex-sedenion quantum-field strength $\mathbb{F}_{(\Psi)}$ as,
\begin{eqnarray}
\mathbb{F}_{(\Psi)} = \mathbb{D}_Z \circ \mathbb{A}_{(\Psi)}  ~ ,
\end{eqnarray}
where $\mathbb{F}_{(\Psi)} = \mathbb{Z}_F \circ \mathbb{F}$ , and $\mathbb{Z}_F$ is one auxiliary quantity. $\mathbb{F}_{(\Psi)} = \mathbb{F}_{(\Psi)g} + k_{eg} \mathbb{F}_{(\Psi)e} + k_{wg} \mathbb{F}_{(\Psi)w} + k_{sg} \mathbb{F}_{(\Psi)s}$ . $\mathbb{F}_{(\Psi)g}$ , $\mathbb{F}_{(\Psi)e}$, $\mathbb{F}_{(\Psi)w}$ , and $\mathbb{F}_{(\Psi)s}$ are respectively the components of the quantum-field strength $\mathbb{F}_{(\Psi)}$ in the spaces, $\mathbb{H}_g$ , $\mathbb{H}_e$ , $\mathbb{H}_w$ , and $\mathbb{H}_s$ .

The complex-sedenion quantum-field strength $\mathbb{F}_{(\Psi)}$ comprises the quantum-field strengths of four fundamental quantum-fields and of twelve adjoint quantum-fields obviously. The ingredients of the quantum-field strengths in the four complex-quaternion spaces are as follows:

a) In the complex-quaternion space $\mathbb{H}_g$ , the component $\mathbb{F}_{(\Psi)}$ comprises the gravitational fundamental quantum-field strength $\mathbb{F}_{(\Psi)g}^g$ , electromagnetic adjoint quantum-field strength $\mathbb{F}_{(\Psi)e}^e$ , W-nuclear adjoint quantum-field strength $\mathbb{F}_{(\Psi)w}^w$ , and strong-nuclear adjoint quantum-field strength $\mathbb{F}_{(\Psi)s}^s$ . Herein $\mathbb{F}_{(\Psi)g} = \mathbb{F}_{(\Psi)g}^g + \mathbb{F}_{(\Psi)e}^e + \mathbb{F}_{(\Psi)w}^w + \mathbb{F}_{(\Psi)s}^s$ . $\mathbb{F}_{(\Psi)g} = i \textbf{\emph{I}}_{g0} F_{(\Psi)g0} + \Sigma \textbf{\emph{I}}_{gk} F_{(\Psi)gk}$ . $\mathbb{F}_{(\Psi)g}^g = \mathbb{D}_{Zg} \circ \mathbb{A}_{(\Psi)g}$ . $\mathbb{F}_{(\Psi)e}^e = \mathbb{D}_{Ze} \circ \mathbb{A}_{(\Psi)e}$ . $\mathbb{F}_{(\Psi)w}^w = \mathbb{D}_{Zw} \circ \mathbb{A}_{(\Psi)w}$ . $\mathbb{F}_{(\Psi)s}^s = \mathbb{D}_{Zs} \circ \mathbb{A}_{(\Psi)s}$ . $\mathbb{F}_{(\Psi)g}^g = i \textbf{\emph{I}}_{g0} F^g_{(\Psi)g0} + \Sigma \textbf{\emph{I}}_{gk} F^g_{(\Psi)gk}$ . $\mathbb{F}_{(\Psi)e}^e = i \textbf{\emph{I}}_{g0} F^e_{(\Psi)e0} + \Sigma \textbf{\emph{I}}_{gk} F^e_{(\Psi)ek}$ . $\mathbb{F}_{(\Psi)w}^w = i \textbf{\emph{I}}_{g0} F^w_{(\Psi)w0} + \Sigma \textbf{\emph{I}}_{gk} F^w_{(\Psi)wk}$. $\mathbb{F}_{(\Psi)s}^s = i \textbf{\emph{I}}_{g0} F^s_{(\Psi)s0} + \Sigma \textbf{\emph{I}}_{gk} F^s_{(\Psi)sk}$ . $F_{(\Psi)g0}$, $F^g_{(\Psi)g0}$ , $F^e_{(\Psi)e0}$ , $F^w_{(\Psi)w0}$ , and $F^s_{(\Psi)s0}$ are all real. $F_{(\Psi)gk}$, $F^g_{(\Psi)gk}$ , $F^e_{(\Psi)ek}$ , $F^w_{(\Psi)wk}$, and $F^s_{(\Psi)sk}$ are complex-numbers.

b) In the complex 2-quaternion space $\mathbb{H}_e$ , the component $\mathbb{F}_{(\Psi)e}$ contains the electromagnetic fundamental quantum-field strength $\mathbb{F}_{(\Psi)e}^g$ , gravitational adjoint quantum-field strength $\mathbb{F}_{(\Psi)g}^e$ , W-nuclear adjoint quantum-field strength $\mathbb{F}_{(\Psi)w}^s$ , and strong-nuclear adjoint quantum-field strength $\mathbb{F}_{(\Psi)s}^w$ . Herein $\mathbb{F}_{(\Psi)e} = \mathbb{F}_{(\Psi)e}^g + k_{eg}^{~~-2} \mathbb{F}_{(\Psi)g}^e + ( k_{wg} k_{sg}^{~~-1} k_{eg}^{~~-1} ) \mathbb{F}_{(\Psi)w}^s + ( k_{sg} k_{wg}^{~~-1} k_{eg}^{~~-1} ) \mathbb{F}_{(\Psi)s}^w$ . $\mathbb{F}_{(\Psi)e} = i \textbf{\emph{I}}_{e0} F_{(\Psi)e0} + \Sigma \textbf{\emph{I}}_{ek} F_{(\Psi)ek}$. $\mathbb{F}_{(\Psi)e}^g = \mathbb{D}_{Zg} \circ \mathbb{A}_{(\Psi)e}$ . $\mathbb{F}_{(\Psi)g}^e = \mathbb{D}_{Ze} \circ \mathbb{A}_{(\Psi)g}$. $\mathbb{F}_{(\Psi)w}^s = \mathbb{D}_{Zs} \circ \mathbb{A}_{(\Psi)w}$ . $\mathbb{F}_{(\Psi)s}^w = \mathbb{D}_{Zw} \circ \mathbb{A}_{(\Psi)s}$ . $\mathbb{F}_{(\Psi)e}^g = i \textbf{\emph{I}}_{e0} F^g_{(\Psi)e0} + \Sigma \textbf{\emph{I}}_{ek} F^g_{(\Psi)ek}$. $\mathbb{F}_{(\Psi)g}^e = i \textbf{\emph{I}}_{e0} F^e_{(\Psi)g0} + \Sigma \textbf{\emph{I}}_{ek} F^e_{(\Psi)gk}$. $\mathbb{F}_{(\Psi)w}^s = i \textbf{\emph{I}}_{e0} F^s_{(\Psi)w0} + \Sigma \textbf{\emph{I}}_{ek} F^s_{(\Psi)wk}$ . $\mathbb{F}_{(\Psi)s}^w = i \textbf{\emph{I}}_{e0} F^w_{(\Psi)s0} + \Sigma \textbf{\emph{I}}_{ek} F^w_{(\Psi)sk}$ . $F_{(\Psi)e0}$ , $F^g_{(\Psi)e0}$ , $F^e_{(\Psi)g0}$ , $F^s_{(\Psi)w0}$ , and $F^w_{(\Psi)s0}$ are all real. $F_{(\Psi)ek}$ , $F^g_{(\Psi)ek}$ , $F^e_{(\Psi)gk}$ , $F^s_{(\Psi)wk}$ , and $F^w_{(\Psi)sk}$ are complex-numbers.

c) In the complex 3-quaternion space $\mathbb{H}_w$ , the component $\mathbb{F}_{(\Psi)w}$ covers the W-nuclear fundamental quantum-field strength $\mathbb{F}_{(\Psi)w}^g$ , gravitational adjoint quantum-field strength $\mathbb{F}_{(\Psi)g}^w$ , electromagnetic adjoint quantum-field strength $\mathbb{F}_{(\Psi)e}^s$, and strong-nuclear adjoint quantum-field strength $\mathbb{F}_{(\Psi)s}^e$ . Herein $\mathbb{F}_{(\Psi)w} = \mathbb{F}_{(\Psi)w}^g + k_{wg}^{~~-2} \mathbb{F}_{(\Psi)g}^w + ( k_{eg} k_{wg}^{~~-1} k_{sg}^{~~-1} ) \mathbb{F}_{(\Psi)e}^s + ( k_{sg} k_{wg}^{~~-1} k_{eg}^{~~-1} ) \mathbb{F}_{(\Psi)s}^e$ . $\mathbb{F}_{(\Psi)w} = i \textbf{\emph{I}}_{w0} F_{(\Psi)w0} + \Sigma \textbf{\emph{I}}_{wk} F_{(\Psi)wk}$ . $\mathbb{F}_{(\Psi)w}^g = \mathbb{D}_{Zg} \circ \mathbb{A}_{(\Psi)w}$ . $\mathbb{F}_{(\Psi)g}^w = \mathbb{D}_{Zw} \circ \mathbb{A}_{(\Psi)g}$. $\mathbb{F}_{(\Psi)e}^s = \mathbb{D}_{Zs} \circ \mathbb{A}_{(\Psi)e}$ . $\mathbb{F}_{(\Psi)s}^e = \mathbb{D}_{Ze} \circ \mathbb{A}_{(\Psi)s}$ . $\mathbb{F}_{(\Psi)w}^g = i \textbf{\emph{I}}_{w0} F^g_{(\Psi)w0} + \Sigma \textbf{\emph{I}}_{wk} F^g_{(\Psi)wk}$. $\mathbb{F}_{(\Psi)g}^w = i \textbf{\emph{I}}_{w0} F^w_{(\Psi)g0} + \Sigma \textbf{\emph{I}}_{wk} F^w_{(\Psi)gk}$. $\mathbb{F}_{(\Psi)e}^s = i \textbf{\emph{I}}_{w0} F^s_{(\Psi)e0} + \Sigma \textbf{\emph{I}}_{wk} F^s_{(\Psi)ek}$ . $\mathbb{F}_{(\Psi)s}^e = i \textbf{\emph{I}}_{w0} F^e_{(\Psi)s0} + \Sigma \textbf{\emph{I}}_{wk} F^e_{(\Psi)sk}$. $F_{(\Psi)w0}$ , $F^g_{(\Psi)w0}$ , $F^w_{(\Psi)g0}$ , $F^s_{(\Psi)e0}$, and $F^e_{(\Psi)s0}$ are all real. $F_{(\Psi)wk}$ , $F^g_{(\Psi)wk}$ , $F^w_{(\Psi)gk}$, $F^s_{(\Psi)ek}$ , and $F^e_{(\Psi)sk}$ are complex-numbers.

d) In the complex 4-quaternion space $\mathbb{H}_s$ , the component $\mathbb{F}_{(\Psi)s}$ includes the strong-nuclear fundamental quantum-field strength $\mathbb{F}_{(\Psi)s}^g$ , gravitational adjoint quantum-field strength $\mathbb{F}_{(\Psi)g}^s$ , electromagnetic adjoint quantum-field strength $\mathbb{F}_{(\Psi)e}^w$, and W-nuclear adjoint quantum-field strength $\mathbb{F}_{(\Psi)w}^e$ . Herein $\mathbb{F}_{(\Psi)s} = \mathbb{F}_{(\Psi)s}^g + k_{sg}^{~~-2} \mathbb{F}_{(\Psi)g}^s + ( k_{eg} k_{wg}^{~~-1} k_{sg}^{~~-1} ) \mathbb{F}_{(\Psi)e}^w + ( k_{wg} k_{sg}^{~~-1} k_{eg}^{~~-1} ) \mathbb{F}_{(\Psi)w}^e$ . $\mathbb{F}_{(\Psi)s} = i \textbf{\emph{I}}_{s0} F_{(\Psi)s0} + \Sigma \textbf{\emph{I}}_{sk} F_{(\Psi)sk}$. $\mathbb{F}_{(\Psi)s}^g = \mathbb{D}_{Zg} \circ \mathbb{A}_{(\Psi)s}$ . $\mathbb{F}_{(\Psi)g}^s = \mathbb{D}_{Zs} \circ \mathbb{A}_{(\Psi)g}$. $\mathbb{F}_{(\Psi)e}^w = \mathbb{D}_{Zw} \circ \mathbb{A}_{(\Psi)e}$ . $\mathbb{F}_{(\Psi)w}^e = \mathbb{D}_{Ze} \circ \mathbb{A}_{(\Psi)w}$ . $\mathbb{F}_{(\Psi)s}^g = i \textbf{\emph{I}}_{s0} F^g_{(\Psi)s0} + \Sigma \textbf{\emph{I}}_{sk} F^g_{(\Psi)sk}$ . $\mathbb{F}_{(\Psi)g}^s = i \textbf{\emph{I}}_{s0} F^s_{(\Psi)g0} + \Sigma \textbf{\emph{I}}_{sk} F^s_{(\Psi)gk}$. $\mathbb{F}_{(\Psi)e}^w = i \textbf{\emph{I}}_{s0} F^w_{(\Psi)e0} + \Sigma \textbf{\emph{I}}_{sk} F^w_{(\Psi)ek}$ . $\mathbb{F}_{(\Psi)w}^e = i \textbf{\emph{I}}_{s0} F^e_{(\Psi)w0} + \Sigma \textbf{\emph{I}}_{sk} F^e_{(\Psi)wk}$ . $F_{(\Psi)s0}$ , $F^g_{(\Psi)s0}$ , $F^s_{(\Psi)g0}$ , $F^w_{(\Psi)e0}$ , and $F^e_{(\Psi)w0}$ are all real. $F_{(\Psi)sk}$ , $F^g_{(\Psi)sk}$ , $F^s_{(\Psi)gk}$ , $F^w_{(\Psi)ek}$ , and $F^e_{(\Psi)wk}$ are complex-numbers.

\subsection{\label{sec:level1}Quantum field source}

Making use of the physical properties of the quantum-field strength, the complex-sedenion quantum-field source $\mathbb{S}_{(\Psi)}$ can be defined as,
\begin{eqnarray}
\mu \mathbb{S}_{(\Psi)} = - \mathbb{D}_Z^\ast \circ \mathbb{F}_{(\Psi)}  ~ ,
\end{eqnarray}
where $\mathbb{S}_{(\Psi)} = \mathbb{Z}_S \circ \mathbb{S}$ , and $\mathbb{Z}_S$ is one auxiliary quantity. $\mu \mathbb{S}_{(\Psi)} = \mu_g \mathbb{S}_{(\Psi)g} + k_{eg} \mu_e \mathbb{S}_{(\Psi)e} + k_{wg} \mu_w \mathbb{S}_{(\Psi)w} + k_{sg} \mu_s \mathbb{S}_{(\Psi)s} $ . $\mathbb{S}_{(\Psi)g}$ , $\mathbb{S}_{(\Psi)e}$ , $\mathbb{S}_{(\Psi)w}$ , and $\mathbb{S}_{(\Psi)s}$ are respectively the components of the quantum-field source $\mathbb{S}_{(\Psi)}$ in the spaces, $\mathbb{H}_g$ , $\mathbb{H}_e$ , $\mathbb{H}_w$ , and $\mathbb{H}_s$ .

The complex-sedenion quantum-field source $\mathbb{S}_{(\Psi)s}$ consists of the field sources of four fundamental quantum-fields and of twelve adjoint quantum-fields obviously. The ingredients of the quantum-field sources in the four complex-quaternion spaces are as follows:

a) In the complex-quaternion space $\mathbb{H}_g$ , the component $\mathbb{S}_{(\Psi)g}$ covers the gravitational fundamental quantum-field source $\mathbb{S}_{(\Psi)g}^g$ , electromagnetic adjoint quantum-field source $\mathbb{S}_{(\Psi)e}^e$ , W-nuclear adjoint quantum-field source $\mathbb{S}_{(\Psi)w}^w$ , and strong-nuclear adjoint quantum-field source $\mathbb{S}_{(\Psi)s}^s$ . Herein $\mu_g \mathbb{S}_{(\Psi)g} = \mu_g^g \mathbb{S}_{(\Psi)g}^g + \mu_e^e \mathbb{S}_{(\Psi)e}^e + \mu_w^w \mathbb{S}_{(\Psi)w}^w + \mu_s^s \mathbb{S}_{(\Psi)s}^s$ . $\mathbb{S}_{(\Psi)g} = i \textbf{\emph{I}}_{g0} S_{(\Psi)g0} + \Sigma \textbf{\emph{I}}_{gk} S_{(\Psi)gk}$ . $\mu_g^g \mathbb{S}_{(\Psi)g}^g = - \mathbb{D}_{Zg}^\ast \circ \mathbb{F}_{(\Psi)g}$. $\mu_e^e \mathbb{S}_{(\Psi)e}^e = - \mathbb{D}_{Ze}^\ast \circ \mathbb{F}_{(\Psi)e}$ . $\mu_w^w \mathbb{S}_{(\Psi)w}^w = - \mathbb{D}_{Zw}^\ast \circ \mathbb{F}_{(\Psi)w}$. $\mu_s^s \mathbb{S}_{(\Psi)s}^s = - \mathbb{D}_{Zs}^\ast \circ \mathbb{F}_{(\Psi)s}$ . $\mathbb{S}_{(\Psi)g}^g = i \textbf{\emph{I}}_{g0} S^g_{(\Psi)g0} + \Sigma \textbf{\emph{I}}_{gk} S^g_{(\Psi)gk}$ . $\mathbb{S}_{(\Psi)e}^e = i \textbf{\emph{I}}_{g0} S^e_{(\Psi)e0} + \Sigma \textbf{\emph{I}}_{gk} S^e_{(\Psi)ek}$ . $\mathbb{S}_{(\Psi)w}^w = i \textbf{\emph{I}}_{g0} S^w_{(\Psi)w0} + \Sigma \textbf{\emph{I}}_{gk} S^w_{(\Psi)wk}$ . $\mathbb{S}_{(\Psi)s}^s = i \textbf{\emph{I}}_{g0} S^s_{(\Psi)s0} + \Sigma \textbf{\emph{I}}_{gk} S^s_{(\Psi)sk}$. $S_{(\Psi)gj}$ , $S^g_{(\Psi)gj}$ , $S^e_{(\Psi)ej}$, $S^w_{(\Psi)wj}$ , and $S^s_{(\Psi)sj}$ are all real.

b) In the complex 2-quaternion space $\mathbb{H}_e$ , the component $\mathbb{S}_{(\Psi)e}$ contains the electromagnetic fundamental quantum-field source $\mathbb{S}_{(\Psi)e}^g$ , gravitational adjoint quantum-field source $\mathbb{S}_{(\Psi)g}^e$ , W-nuclear adjoint quantum-field source $\mathbb{S}_{(\Psi)w}^s$, and strong-nuclear adjoint quantum-field source $\mathbb{S}_{(\Psi)s}^w$ . Herein $\mu_e \mathbb{S}_{(\Psi)e} = \mu_e^g \mathbb{S}_{(\Psi)e}^g + k_{eg}^{~~-2} \mu_g^e \mathbb{S}_{(\Psi)g}^e + ( k_{wg} k_{sg}^{~~-1} k_{eg}^{~~-1} ) \mu_w^s \mathbb{S}_{(\Psi)w}^s + ( k_{sg} k_{wg}^{~~-1} k_{eg}^{~~-1} ) \mu_s^w \mathbb{S}_{(\Psi)s}^w$ . $\mathbb{S}_{(\Psi)e} = i \textbf{\emph{I}}_{e0} S_{(\Psi)e0} + \Sigma \textbf{\emph{I}}_{ek} S_{(\Psi)ek}$ . $\mu_e^g \mathbb{S}_{(\Psi)e}^g = - \mathbb{D}_{Zg}^\ast \circ \mathbb{F}_{(\Psi)e}$ . $\mu_g^e \mathbb{S}_{(\Psi)g}^e = - \mathbb{D}_{Ze}^\ast \circ \mathbb{F}_{(\Psi)g}$. $\mu_w^s \mathbb{S}_{(\Psi)w}^s = - \mathbb{D}_{Zs}^\ast \circ \mathbb{F}_{(\Psi)w}$ . $\mu_s^w \mathbb{S}_{(\Psi)s}^w = - \mathbb{D}_{Zw}^\ast \circ \mathbb{F}_{(\Psi)s}$ . $\mathbb{S}_{(\Psi)e}^g = i \textbf{\emph{I}}_{e0} S^g_{(\Psi)e0} + \Sigma \textbf{\emph{I}}_{ek} S^g_{(\Psi)ek}$. $\mathbb{S}_{(\Psi)g}^e = i \textbf{\emph{I}}_{e0} S^e_{(\Psi)g0} + \Sigma \textbf{\emph{I}}_{ek} S^e_{(\Psi)gk}$ . $\mathbb{S}_{(\Psi)w}^s = i \textbf{\emph{I}}_{e0} S^s_{(\Psi)w0} + \Sigma \textbf{\emph{I}}_{ek} S^s_{(\Psi)wk}$. $\mathbb{S}_{(\Psi)s}^w = i \textbf{\emph{I}}_{e0} S^w_{(\Psi)s0} + \Sigma \textbf{\emph{I}}_{ek} S^w_{(\Psi)sk}$ . $S_{(\Psi)ej}$ , $S^g_{(\Psi)ej}$ , $S^e_{(\Psi)gj}$ , $S^s_{(\Psi)wj}$ , and $S^w_{(\Psi)sj}$ are all real.

c) In the complex 3-quaternion space $\mathbb{H}_w$ , the component $\mathbb{S}_{(\Psi)w}$ comprises the W-nuclear fundamental quantum-field source $\mathbb{S}_{(\Psi)w}^g$ , gravitational adjoint quantum-field source $\mathbb{S}_{(\Psi)g}^w$ , electromagnetic adjoint quantum-field source $\mathbb{S}_{(\Psi)e}^s$ , and strong-nuclear adjoint quantum-field source $\mathbb{S}_{(\Psi)s}^e$ . Herein $\mu_w \mathbb{S}_{(\Psi)w} = \mu_w^g \mathbb{S}_{(\Psi)w}^g + k_{wg}^{~~-2} \mu_g^w \mathbb{S}_{(\Psi)g}^w + ( k_{eg} k_{wg}^{~~-1} k_{sg}^{~~-1} ) \mu_e^s \mathbb{S}_{(\Psi)e}^s + ( k_{sg} k_{wg}^{~~-1} k_{eg}^{~~-1} ) \mu_s^e \mathbb{S}_{(\Psi)s}^e$. $\mathbb{S}_{(\Psi)w} = i \textbf{\emph{I}}_{w0} S_{(\Psi)w0} + \Sigma \textbf{\emph{I}}_{wk} S_{(\Psi)wk}$. $\mu_w^g \mathbb{S}_{(\Psi)w}^g = - \mathbb{D}_{Zg}^\ast \circ \mathbb{F}_{(\Psi)w}$. $\mu_g^w \mathbb{S}_{(\Psi)g}^w = - \mathbb{D}_{Zw}^\ast \circ \mathbb{F}_{(\Psi)g}$ . $\mu_e^s \mathbb{S}_{(\Psi)e}^s = - \mathbb{D}_{Zs}^\ast \circ \mathbb{F}_{(\Psi)e}$ . $\mu_s^e \mathbb{S}_{(\Psi)s}^e = - \mathbb{D}_{Ze}^\ast \circ \mathbb{F}_{(\Psi)s}$ . $\mathbb{S}_{(\Psi)w}^g = i \textbf{\emph{I}}_{w0} S^g_{(\Psi)w0} + \Sigma \textbf{\emph{I}}_{wk} S^g_{(\Psi)wk}$. $\mathbb{S}_{(\Psi)g}^w = i \textbf{\emph{I}}_{w0} S^w_{(\Psi)g0} + \Sigma \textbf{\emph{I}}_{wk} S^w_{(\Psi)gk}$ . $\mathbb{S}_{(\Psi)e}^s = i \textbf{\emph{I}}_{w0} S^s_{(\Psi)e0} + \Sigma \textbf{\emph{I}}_{wk} S^s_{(\Psi)ek}$ . $\mathbb{S}_{(\Psi)s}^e = i \textbf{\emph{I}}_{w0} S^e_{(\Psi)s0} + \Sigma \textbf{\emph{I}}_{wk} S^e_{(\Psi)sk}$ . $S_{(\Psi)wj}$ , $S^g_{(\Psi)wj}$, $S^w_{(\Psi)gj}$ , $S^s_{(\Psi)ej}$ , and $S^e_{(\Psi)sj}$ are all real.

d) In the complex 4-quaternion space $\mathbb{H}_s$ , the component $\mathbb{S}_{(\Psi)s}$ includes the strong-nuclear fundamental quantum-field source $\mathbb{S}_{(\Psi)s}^g$ , gravitational adjoint quantum-field source $\mathbb{S}_{(\Psi)g}^s$ , electromagnetic adjoint quantum-field source $\mathbb{S}_{(\Psi)e}^w$, and W-nuclear adjoint quantum-field source $\mathbb{S}_{(\Psi)w}^e$ . Herein $\mu_s \mathbb{S}_{(\Psi)s} = \mu_s^g \mathbb{S}_{(\Psi)s}^g + k_{sg}^{~~-2} \mu_g^s \mathbb{S}_{(\Psi)g}^s + ( k_{eg} k_{wg}^{~~-1} k_{sg}^{~~-1} ) \mu_e^w \mathbb{S}_{(\Psi)e}^w + ( k_{wg} k_{sg}^{~~-1} k_{eg}^{~~-1} ) \mu_w^e \mathbb{S}_{(\Psi)w}^e$ . $\mathbb{S}_{(\Psi)s} = i \textbf{\emph{I}}_{s0} S_{(\Psi)s0} + \Sigma \textbf{\emph{I}}_{sk} S_{(\Psi)sk}$ . $\mu_s^g \mathbb{S}_{(\Psi)s}^g = - \mathbb{D}_{Zg}^\ast \circ \mathbb{F}_{(\Psi)s}$ . $\mu_g^s \mathbb{S}_{(\Psi)g}^s = - \mathbb{D}_{Zs}^\ast \circ \mathbb{F}_{(\Psi)g}$. $\mu_e^w \mathbb{S}_{(\Psi)e}^w = - \mathbb{D}_{Zw}^\ast \circ \mathbb{F}_{(\Psi)e}$ . $\mu_w^e \mathbb{S}_{(\Psi)w}^e = - \mathbb{D}_{Ze}^\ast \circ \mathbb{F}_{(\Psi)w}$ . $\mathbb{S}_{(\Psi)s}^g = i \textbf{\emph{I}}_{s0} S^g_{(\Psi)s0} + \Sigma \textbf{\emph{I}}_{sk} S^g_{(\Psi)sk}$. $\mathbb{S}_{(\Psi)g}^s = i \textbf{\emph{I}}_{s0} S^s_{(\Psi)g0} + \Sigma \textbf{\emph{I}}_{sk} S^s_{(\Psi)gk}$ . $\mathbb{S}_{(\Psi)e}^w = i \textbf{\emph{I}}_{s0} S^w_{(\Psi)e0} + \Sigma \textbf{\emph{I}}_{sk} S^w_{(\Psi)ek}$. $\mathbb{S}_{(\Psi)w}^e = i \textbf{\emph{I}}_{s0} S^e_{(\Psi)w0} + \Sigma \textbf{\emph{I}}_{sk} S^e_{(\Psi)wk}$ . $S_{(\Psi)sj}$ , $S^g_{(\Psi)sj}$ , $S^s_{(\Psi)gj}$ , $S^w_{(\Psi)ej}$ , and $S^e_{(\Psi)wj}$ are all real.

In the complex-sedenion space $\mathbb{K}$ , the quantum-fields in the quantum mechanics are the functions of the classical fields in the classical mechanics. Especially, the quantum-field sources in the quantum mechanics can be considered as the functions of the classical field sources in the classical mechanics. This situation is similar to that of the stationary wave or solitary wave, which can be found in the rivers sometimes.

Under certain circumstances, the term, $\mathbb{Z}_W \circ \mathbb{W}^\star / ( \hbar v_0 )$ , in the operator $\mathbb{D}_Z$ may be degraded into the term, $g_A \mathbb{Z}_W^\prime \circ \mathbb{A}_{(\Psi)}$ . Herein $g_A$ is a coefficient, and $\mathbb{Z}_W^\prime$ is an auxiliary quantity. By means of the transformations of complex-quaternion spaces, the definitions of the complex-sedenion quantum-field potential, quantum-field strength, and quantum-field source are able to be degenerated respectively into that of the field potential, field strength, and field source in the Yang-Mills equations for the non-Abelian gauge field (in Section 7).

\begin{table}[h]
\caption{Some field equations of the quantum mechanics in the complex-sedenion space.}
%\begin{ruledtabular}
\centering
\begin{tabular}{@{}ll@{}}
\hline\hline
quantum~physics~quantity~~~~~~~~~~&  definition                                                                                                                       \\
\hline
quantum~field~potential      &  $\mathbb{A}_{(\Psi)} = i \{ i \mathbb{Z}_W \circ \mathbb{W}^\star / ( \hbar v_0 ) + \lozenge \}^\star \circ \mathbb{X}_{(\Psi)}  $    \\
quantum~field~strength       &  $\mathbb{F}_{(\Psi)} = \{ i \mathbb{Z}_W \circ \mathbb{W}^\star / ( \hbar v_0 ) + \lozenge \} \circ \mathbb{A}_{(\Psi)}  $            \\
quantum~field~source         &  $\mu \mathbb{S}_{(\Psi)} = - \{ i \mathbb{Z}_W \circ \mathbb{W}^\star / ( \hbar v_0 ) + \lozenge \}^\ast \circ \mathbb{F}_{(\Psi)} $  \\
%%%
quantum~linear~momentum      &  $\mathbb{P}_{(\Psi)} = \mu \mathbb{S}_{(\Psi)} / \mu_g^g  $                                                                           \\
quantum~angular~momentum     &  $\mathbb{L}_{(\Psi)} = ( \mathbb{U}_{(\Psi)} )^\star \circ \mathbb{P}_{(\Psi)}$                                                       \\
quantum~torque               &  $\mathbb{W}_{(\Psi)} = - v_0 \{ i \mathbb{Z}_W \circ \mathbb{W}^\star / ( \hbar v_0 ) + \lozenge \} \circ \mathbb{L}_{(\Psi)}  $      \\
quantum~force                &  $\mathbb{N}_{(\Psi)} = - \{ i \mathbb{Z}_W \circ \mathbb{W}^\star / ( \hbar v_0 ) + \lozenge \} \circ \mathbb{W}_{(\Psi)} $           \\
\hline\hline
\end{tabular}
%\end{ruledtabular}
\end{table}

\subsection{\label{sec:level1}Quantum angular momentum}

In the complex-sedenion space $\mathbb{K}$ , from the complex-sedenion quantum-field source, it is able to define the complex-sedenion quantum linear momentum $\mathbb{P}_{(\Psi)}$ as,
\begin{eqnarray}
\mathbb{P}_{(\Psi)} = \mu \mathbb{S}_{(\Psi)} / \mu_g^g  ~ ,
\end{eqnarray}
where $\mathbb{P}_{(\Psi)} = \mathbb{Z}_P \circ \mathbb{P}$ , and $\mathbb{Z}_P$ is one auxiliary quantity. $\mathbb{P}_{(\Psi)} = \mathbb{P}_{(\Psi)g} + k_{eg} \mathbb{P}_{(\Psi)e} + k_{wg} \mathbb{P}_{(\Psi)w} + k_{sg} \mathbb{P}_{(\Psi)s}$ . $\mathbb{P}_{(\Psi)g}$, $\mathbb{P}_{(\Psi)e}$ , $\mathbb{P}_{(\Psi)w}$ , and $\mathbb{P}_{(\Psi)s}$ are respectively the components of the quantum linear momentum $\mathbb{P}_{(\Psi)}$ in the spaces, $\mathbb{H}_g$, $\mathbb{H}_e$ , $\mathbb{H}_w$ , and $\mathbb{H}_s$ . $\mathbb{P}_{(\Psi)g} = \mu_g \mathbb{S}_{(\Psi)g} / \mu_g^g$ . $\mathbb{P}_{(\Psi)e} = \mu_e \mathbb{S}_{(\Psi)e} / \mu_g^g$. $\mathbb{P}_{(\Psi)w} = \mu_w \mathbb{S}_{(\Psi)w} / \mu_g^g$ . $\mathbb{P}_{(\Psi)s} = \mu_s \mathbb{S}_{(\Psi)s} / \mu_g^g$. $\mathbb{P}_{(\Psi)g} = i \textbf{\emph{I}}_{g0} P_{(\Psi)g0} + \Sigma \textbf{\emph{I}}_{gk} P_{(\Psi)gk}$. $\mathbb{P}_{(\Psi)e} = i \textbf{\emph{I}}_{e0} P_{(\Psi)e0} + \Sigma \textbf{\emph{I}}_{ek} P_{(\Psi)ek}$ . $\mathbb{P}_{(\Psi)w} = i \textbf{\emph{I}}_{w0} P_{(\Psi)w0} + \Sigma \textbf{\emph{I}}_{wk} P_{(\Psi)wk}$ . $\mathbb{P}_{(\Psi)s} = i \textbf{\emph{I}}_{s0} P_{(\Psi)s0} + \Sigma \textbf{\emph{I}}_{sk} P_{(\Psi)sk}$ . $P_{(\Psi)gj}$ , $P_{(\Psi)ej}$ , $P_{(\Psi)wj}$ , and $P_{(\Psi)sj}$ are all real.

From the complex-sedenion composite radius vector $\mathbb{U}$ and quantum linear momentum $\mathbb{P}_{(\Psi)}$ , it is capable of defining the complex-sedenion quantum angular momentum $\mathbb{L}_{(\Psi)}$ as ,
\begin{eqnarray}
\mathbb{L}_{(\Psi)} = ( \mathbb{U}_{(\Psi)} )^\star \circ \mathbb{P}_{(\Psi)}    ~ ,
\end{eqnarray}
where $\mathbb{L}_{(\Psi)} = \mathbb{Z}_L \circ \mathbb{L}$ , and $\mathbb{U}_{(\Psi)} = \mathbb{Z}_U \circ \mathbb{U}$ . $\mathbb{Z}_L$ and $\mathbb{Z}_U$ both are auxiliary quantities. $\mathbb{L}_{(\Psi)} = \mathbb{L}_{(\Psi)g} + k_{eg} \mathbb{L}_{(\Psi)e} + k_{wg} \mathbb{L}_{(\Psi)w} + k_{sg} \mathbb{L}_{(\Psi)s}$ . $\mathbb{L}_{(\Psi)g}$ , $\mathbb{L}_{(\Psi)e}$ , $\mathbb{L}_{(\Psi)w}$ , and $\mathbb{L}_{(\Psi)s}$ are the components of the quantum angular momentum $\mathbb{L}_{(\Psi)}$ in the spaces, $\mathbb{H}_g$ , $\mathbb{H}_e$ , $\mathbb{H}_w$ , and $\mathbb{H}_s$ , respectively. And that the ingredients of the quantum angular momentum in the four complex-quaternion spaces are as follows,
\begin{eqnarray}
&& \mathbb{L}_{(\Psi)g} = ( \mathbb{U}_{(\Psi)g} )^\star \circ \mathbb{P}_{(\Psi)g} + k_{eg}^{~~2} ( \mathbb{U}_{(\Psi)e} )^\star \circ \mathbb{P}_{(\Psi)e}
+ k_{wg}^{~~2} ( \mathbb{U}_{(\Psi)w} )^\star \circ \mathbb{P}_{(\Psi)w} + k_{sg}^{~~2} ( \mathbb{U}_{(\Psi)s} )^\star \circ \mathbb{P}_{(\Psi)s}  ~,
\\
&& \mathbb{L}_{(\Psi)e} = ( \mathbb{U}_{(\Psi)g} )^\star \circ \mathbb{P}_{(\Psi)e} + ( \mathbb{U}_{(\Psi)e} )^\star \circ \mathbb{P}_{(\Psi)g}
+ ( k_{wg} k_{sg} k_{eg}^{~~-1} ) \{ ( \mathbb{U}_{(\Psi)w} )^\star \circ \mathbb{P}_{(\Psi)s} + ( \mathbb{U}_{(\Psi)s} )^\star \circ \mathbb{P}_{(\Psi)w} \}  ~,
\\
&& \mathbb{L}_{(\Psi)w} = ( \mathbb{U}_{(\Psi)g} )^\star \circ \mathbb{P}_{(\Psi)w} + ( \mathbb{U}_{(\Psi)w} )^\star \circ \mathbb{P}_{(\Psi)g}
+ ( k_{eg} k_{sg} k_{wg}^{~~-1} ) \{ ( \mathbb{U}_{(\Psi)e} )^\star \circ \mathbb{P}_{(\Psi)s} + ( \mathbb{U}_{(\Psi)s} )^\star \circ \mathbb{P}_{(\Psi)e} \}  ~,
\\
&& \mathbb{L}_{(\Psi)s} = ( \mathbb{U}_{(\Psi)g} )^\star \circ \mathbb{P}_{(\Psi)s} + ( \mathbb{U}_{(\Psi)s} )^\star \circ \mathbb{P}_{(\Psi)g}
+ ( k_{eg} k_{wg} k_{sg}^{~~-1} ) \{ ( \mathbb{U}_{(\Psi)e} )^\star \circ \mathbb{P}_{(\Psi)w} + ( \mathbb{U}_{(\Psi)w} )^\star \circ \mathbb{P}_{(\Psi)e} \}  ~.
\end{eqnarray}

\subsection{\label{sec:level1}Quantum torque}

From the complex-sedenion quantum angular momentum, it is able to define the complex-sedenion quantum torque $\mathbb{W}_{(\Psi)}$ as,
\begin{eqnarray}
\mathbb{W}_{(\Psi)} = - v_0 \mathbb{D}_Z \circ \mathbb{L}_{(\Psi)}   ~,
\end{eqnarray}
where $\mathbb{W}_{(\Psi)} = \mathbb{Z}_W \circ \mathbb{W}$ , and $\mathbb{Z}_W$ is one auxiliary quantity. $\mathbb{W}_{(\Psi)} = \mathbb{W}_{(\Psi)g} + k_{eg} \mathbb{W}_{(\Psi)e} + k_{wg} \mathbb{W}_{(\Psi)w} + k_{sg} \mathbb{W}_{(\Psi)s}$ . $\mathbb{W}_{(\Psi)g}$ , $\mathbb{W}_{(\Psi)e}$ , $\mathbb{W}_{(\Psi)w}$, and $\mathbb{W}_{(\Psi)s}$ are respectively the components of the quantum torque $\mathbb{W}_{(\Psi)}$ in the spaces, $\mathbb{H}_g$, $\mathbb{H}_e$ , $\mathbb{H}_w$ , and $\mathbb{H}_s$ . And that the ingredients of the quantum torque in the four complex-quaternion spaces are,
\begin{eqnarray}
&&  \mathbb{W}_{(\Psi)g} = - v_0 \mathbb{D}_{Zg} \circ \mathbb{L}_{(\Psi)g} - v_0 k_{eg}^{~~2} \mathbb{D}_{Ze} \circ \mathbb{L}_{(\Psi)e}
- v_0 k_{wg}^{~~2} \mathbb{D}_{Zw} \circ \mathbb{L}_{(\Psi)w} - v_0 k_{sg}^{~~2} \mathbb{D}_{Zs} \circ \mathbb{L}_{(\Psi)s}  ~,
\\
&&  \mathbb{W}_{(\Psi)e} = - v_0 \mathbb{D}_{Zg} \circ \mathbb{L}_{(\Psi)e} - v_0 \mathbb{D}_{Ze} \circ \mathbb{L}_{(\Psi)g}
- v_0 ( k_{wg} k_{sg} k_{eg}^{~~-1} ) ( \mathbb{D}_{Zw} \circ \mathbb{L}_{(\Psi)s} + \mathbb{D}_{Zs} \circ \mathbb{L}_{(\Psi)w} )  ~,
\\
&&  \mathbb{W}_{(\Psi)w} = - v_0 \mathbb{D}_{Zg} \circ \mathbb{L}_{(\Psi)w} - v_0 \mathbb{D}_{Zw} \circ \mathbb{L}_{(\Psi)g}
- v_0 ( k_{eg} k_{sg} k_{wg}^{~~-1} ) ( \mathbb{D}_{Ze} \circ \mathbb{L}_{(\Psi)s} + \mathbb{D}_{Zs} \circ \mathbb{L}_{(\Psi)e} )  ~,
\\
&&  \mathbb{W}_{(\Psi)s} = - v_0 \mathbb{D}_{Zg} \circ \mathbb{L}_{(\Psi)s} - v_0 \mathbb{D}_{Zs} \circ \mathbb{L}_{(\Psi)g}
- v_0 ( k_{eg} k_{wg} k_{sg}^{~~-1} ) ( \mathbb{D}_{Ze} \circ \mathbb{L}_{(\Psi)w} + \mathbb{D}_{Zw} \circ \mathbb{L}_{(\Psi)e} )  ~.
\end{eqnarray}

In the complex-sedenion space $\mathbb{K}$ , under certain circumstances, the term, $\mathbb{Z}_W \circ \mathbb{W}^\star$ , in the operator $\mathbb{D}_Z$ may be degenerated into the term, $\mathbb{W}^\star$, and then the quantum torque equation, $\mathbb{W}_{(\Psi)} = 0$ , is able to be degraded into the Dirac wave equation. Especially, in the complex-octonion space $\mathbb{O}$ , the quantum torque equation, $\mathbb{W}_{(\Psi)} = 0$ , can be reduced into the Dirac wave equation described with the complex-numbers (Appendix B), and even Sch\"{o}dinger wave equation (Appendix C).

\subsection{\label{sec:level1}Quantum force}

From the complex-sedenion quantum torque, it is capable of defining the complex-sedenion quantum force $\mathbb{N}_{(\Psi)}$ as,
\begin{eqnarray}
\mathbb{N}_{(\Psi)} = - \mathbb{D}_Z \circ \mathbb{W}_{(\Psi)}   ~,
\end{eqnarray}
where $\mathbb{N}_{(\Psi)} = \mathbb{Z}_N \circ \mathbb{N}$ , and $\mathbb{Z}_N$ is one auxiliary quantity. $\mathbb{N}_{(\Psi)} = \mathbb{N}_{(\Psi)g} + k_{eg} \mathbb{N}_{(\Psi)e} + k_{wg} \mathbb{N}_{(\Psi)w} + k_{sg} \mathbb{N}_{(\Psi)s}$ . $\mathbb{N}_{(\Psi)g}$, $\mathbb{N}_{(\Psi)e}$ , $\mathbb{N}_{(\Psi)w}$, and $\mathbb{N}_{(\Psi)s}$ are respectively the components of the quantum force $\mathbb{N}_{(\Psi)}$ in the spaces, $\mathbb{H}_g$, $\mathbb{H}_e$ , $\mathbb{H}_w$ , and $\mathbb{H}_s$ . And that the ingredients of the quantum force in the four complex-quaternion spaces are as follows,
\begin{eqnarray}
&&  \mathbb{N}_{(\Psi)g} = - \mathbb{D}_{Zg} \circ \mathbb{W}_{(\Psi)g} - k_{eg}^{~~2} \mathbb{D}_{Ze} \circ \mathbb{W}_{(\Psi)e}
- k_{wg}^{~~2} \mathbb{D}_{Zw} \circ \mathbb{W}_{(\Psi)w} - k_{sg}^{~~2} \mathbb{D}_{Zs} \circ \mathbb{W}_{(\Psi)s}  ~,
\\
&&  \mathbb{N}_{(\Psi)e} = - \mathbb{D}_{Zg} \circ \mathbb{W}_{(\Psi)e} - \mathbb{D}_{Ze} \circ \mathbb{W}_{(\Psi)g}
- ( k_{wg} k_{sg} k_{eg}^{~~-1} ) ( \mathbb{D}_{Zw} \circ \mathbb{W}_{(\Psi)s} + \mathbb{D}_{Zs} \circ \mathbb{W}_{(\Psi)w} )  ~,
\\
&&  \mathbb{N}_{(\Psi)w} = - \mathbb{D}_{Zg} \circ \mathbb{W}_{(\Psi)w} - \mathbb{D}_{Zw} \circ \mathbb{W}_{(\Psi)g}
- ( k_{eg} k_{sg} k_{wg}^{~~-1} ) ( \mathbb{D}_{Ze} \circ \mathbb{W}_{(\Psi)s} + \mathbb{D}_{Zs} \circ \mathbb{W}_{(\Psi)e} )  ~,
\\
&&  \mathbb{N}_{(\Psi)s} = - \mathbb{D}_{Zg} \circ \mathbb{W}_{(\Psi)s} - \mathbb{D}_{Zs} \circ \mathbb{W}_{(\Psi)g}
- ( k_{eg} k_{wg} k_{sg}^{~~-1} ) ( \mathbb{D}_{Ze} \circ \mathbb{W}_{(\Psi)w} + \mathbb{D}_{Zw} \circ \mathbb{W}_{(\Psi)e} )  ~.
\end{eqnarray}

In the complex-sedenion space $\mathbb{K}$ , it is found that there are some similarities as well as differences between the quantum physical quantity in the Table 4 and classical physical quantity in the Table 2. For instance, in terms of the definition of quantum-field strengths in the quantum mechanics, it should satisfy the following three equations simultaneously,
\begin{eqnarray}
&& \mathbb{F}_{(\Psi)} = \{ i \mathbb{Z}_W \circ \mathbb{W}^\star / ( \hbar v_0 ) + \lozenge \} \circ \mathbb{A}_{(\Psi)}  ~ , \\
&& \mathbb{A}_{(\Psi)} = \mathbb{Z}_A \circ \mathbb{A}   ~,  \\
&& \mathbb{F}_{(\Psi)} = \mathbb{Z}_F \circ \mathbb{F}   ~,
\end{eqnarray}
meanwhile the physical quantities, $\mathbb{A}$ , $\mathbb{F}$ and $\mathbb{W}$ , in the classical mechanics must meet the requirement of the Table 2. Consequently the choices of the auxiliary quantities, $\mathbb{Z}_A$ , $\mathbb{Z}_F$ , and $\mathbb{Z}_W$ , are not arbitrary, and they have to meet the physical conditions in every physical problem.

\section{\label{sec:level1}Dark matter field}

In each complex-quaternion space, it is able to determine the `charge' relevant to some fields, from the definition of the fundamental/adjoint field source. Especially, in the complex-quaternion space $\mathbb{H}_g$ , each of three adjoint field sources is able to exert partially the gravitational influence to other objects. The `charges' in the three adjoint field sources are the competitive candidates for the particles of dark matter field. Further the three adjoint field sources are capable of combining with other fundamental/adjoint field sources to constitute various and complicated particles.

\subsection{\label{sec:level1}One-source particle}

In the complex-sedenion space, it is able to possess simultaneously the field sources of four fundamental fields, and of twelve adjoint fields (see Table 1). Each one-source particle is seized of a single field source, producing one of four interactions merely (Table 5).

In the complex-quaternion space $\mathbb{H}_g$ , the `charge' of one-source particle for the gravitational fundamental field is $m_g^g$ , inducing the gravitational interaction. In the complex 2-quaternion space $\mathbb{H}_e$ , the `charge' of one-source particle for the electromagnetic fundamental field is $m_e^g$ , producing the electromagnetic interaction. In the complex 3-quaternion space $\mathbb{H}_w$ , the `charge' of one-source particle for the W-nuclear  fundamental field is $m_w^g$ , producing the W-nuclear interaction. In the complex 4-quaternion space $\mathbb{H}_s$ , the `charge' of one-source particle for the strong-nuclear  fundamental field is $m_s^g$ , producing the strong-nuclear interaction.

In the complex-quaternion space $\mathbb{H}_g$ , the `charges' of one-source particles for the adjoint fields are, $m_e^e$ , $m_w^w$ , and $m_s^s$ , respectively. Each of them can yield partially the gravitational interaction, so they may be considered as the particles of dark matter. In the complex 2-quaternion space $\mathbb{H}_e$ , the `charges' of one-source particles for the adjoint fields are, $m_g^e$ , $m_w^s$ , and $m_s^w$ , respectively. Each of them may produce partially the electromagnetic interaction. In the complex 3-quaternion space $\mathbb{H}_w$ , the `charges' of one-source particles for the adjoint fields are, $m_g^w$ , $m_e^s$ , and $m_s^e$, respectively. Each of them will generate partially the W-nuclear interaction. In the complex 4-quaternion space $\mathbb{H}_s$ , the `charges' of one-source particles for the adjoint fields are, $m_g^s$ , $m_e^w$ , and $m_w^e$ , respectively. Each of them can create partially the strong-nuclear interaction.

The above states that each of four interactions may consist of the contributions of four different fundamental/adjoint field sources, in the complex-sedenion space $\mathbb{K}$ . As a result, the measured value of each interaction, in the laboratory, may be the superposition of certain influences, coming from various one-source particles. By means of the measurement of fundamental/adjoint field strengths, occasionally it may be possible to distinguish the contributions coming from different one-source particles. However, it will be quite difficult and even impossible to do that, under most situations.

\subsection{\label{sec:level1}Two-source particle}

The two-source particle is possessed of two fundamental/adjoint field sources, producing two sorts of interactions simultaneously, including the charged particle and so forth. These two field sources of a two-source particle may be situated in either the same complex-quaternion space, or two different complex-quaternion spaces.

At present, only a part of two-source particles may be measured, due to the lack of effective measurement methods, such as, the charged particle ( $m_g^g$ and $m_e^g$ ) . In terms of some two-source particles, merely one of two field sources can be measured. For example, only the field source of fundamental field can be measured, for some two-source particles ( $m_g^g$ and $m_e^e$ ; or $m_e^g$ and $m_e^s$ ; and so on ), which consist of one fundamental field source and an adjoint field source. However, some two-source particles are impossible to be measured temporarily, for instance, the two-source particles, which comprise two sorts of adjoint field sources ( $m_g^e$ and $m_e^w$ ; and so on).

In the above, the classical mechanics, described with the complex-sedenions, predicts the existences and gravitational influences of three `charges' ( $m_e^e$ , $m_w^w$ , or $m_s^s$ ) of adjoint fields, in the complex-quaternion space $\mathbb{H}_g$ , in terms of the two-source particles. Meanwhile, the astronomical scientists found the existence of dark matters to be able to exert the gravitational influences, although the scholars have not yet found a few appropriate measurement methods. It means that it is necessary to create a new research domain of the measurement methods in the physics.

\subsection{\label{sec:level1}Three-source particle}

The three-source particle is in possession of three fundamental/adjoint field sources, generating three sorts of interactions simultaneously, including the lepton and so forth. These three field sources of a three-source particle may be located in the same complex-quaternion space, or two or three different complex-quaternion spaces.

The `charges' of a three-source particle may contain the mass ( $m_g^g$ ), electric charge ( $m_e^g$ ), and W charge ( $m_w^g$ ), such as, the charged lepton and so forth. And the `charges' of a three-source particle can comprise the mass ( $m_g^g$ ), electric charge ( $m_e^g$ ), and strong charge ( $m_s^g$ ) ; as well as the electric charge ( $m_e^g$ ), W charge ( $m_w^g$ ), and strong charge ( $m_s^g$ ). In contrast to the variety of two-source particles, the varieties of three-source particles will be much more. At the present time, only the fundamental field sources may be measured in the three-source particles, due to the lack of measurement methods. Undoubtedly, the varieties of three-source particles, which are able to be measured, must be very little temporarily, especially these three-source particles connected with the adjoint field sources.

Because the W-nuclear field and strong-nuclear field both are short-range, a part of particles have to be measured within a short distance, increasing inevitably the difficulty of measurement in the experiments. And it implies that there may be multitudinous undetected particles, surrounding the human being right now. The situation of these undetected particles is just similar to that of the `atmosphere', which was undiscovered yet before several centuries.

\subsection{\label{sec:level1}Four-source particle}

The four-source particle is seized of four fundamental/adjoint field sources, yielding four sorts of interactions simultaneously, including the quark and so forth. These four field sources of a four-source particle may be located in the same complex-quaternion space, or two or three and even four different complex-quaternion spaces.

The `charges' of a four-source particle may consist of the mass ( $m_g^g$ ), electric charge ( $m_e^g$ ), W charge ( $m_w^g$ ), and strong charge ( $m_s^g$ ), such as, the quark and so forth. Also the `charges' of a four-source particle may contain other types of combinations of field sources, especially various combinations among the fundamental field sources and adjoint field sources. Moreover, the types of four-source particles will be much more than that of the three-source particles. In the absence of appropriate measurement methods, only the fundamental field sources may be measured in the four-source particles nowadays. Apparently, the types of four-source particles, which are able to be measured, should be very little temporarily, especially these four-source particles relevant to the adjoint field sources.

Further, there may be certain multi-source particles, with more than four fundamental/adjoint field sources. Take into accounting the short-range property regarding the W-nuclear field and strong-nuclear field, it is possible that there may be numerous undetectable one-source, two-source, and three-source (and four-source, and even multi-source) particles, lingering around the existing microscopic particles all the time. In other words, the dark matter particles and other undiscovered particles have been surrounding not only the celestial bodies but also the human being since a long time, constituting one new type of `atmosphere'.

In terms of the electromagnetic interaction in the preceding particles, it is able to possess the electromagnetic fundamental field and three electromagnetic adjoint fields simultaneously, in the complex-sedenion space $\mathbb{K}$ . Even if there are only the complex-sedenion electromagnetic (fundamental/adjoint) quantum-field potentials, the existences of the three electromagnetic adjoint field potentials enable still them to constitute the Yang-Mills equations for the non-Abelian gauge field, described with the complex-sedenions.

\begin{table}[h]
\caption{There are four sorts of fundamental interactions in the complex sedenion spaces, including their field potentials of fundamental fields and adjoint fields.}
%\begin{ruledtabular}
\centering
\begin{tabular}{@{}lllc@{}}
\hline\hline
interaction         &  macroscopic scale                                          &  microscopic scale                                      &  space                \\
\hline
gravitational       &  gravitational fundamental                                  &  gravitational fundamental                                                      \\
~~~~interaction     &  ~~field potential $\mathbb{A}_g^g$                         &  ~~quantum-field potential $\mathbb{A}_{(\Psi)g}^g$~~~~ &  $\mathbb{H}_g$       \\
                    &  gravitational adjoint                                      &  gravitational adjoint                                                          \\
                    &  ~~field potential $\mathbb{A}_g^e$                         &  ~~quantum-field potential $\mathbb{A}_{(\Psi)g}^e$     &  $\mathbb{H}_e$       \\
                    &  gravitational adjoint                                      &  gravitational adjoint                                                          \\
                    &  ~~field potential $\mathbb{A}_g^w$                         &  ~~quantum-field potential $\mathbb{A}_{(\Psi)g}^w$     &  $\mathbb{H}_w$       \\
                    &  gravitational adjoint                                      &  gravitational adjoint                                                          \\
                    &  ~~field potential $\mathbb{A}_g^s$                         &  ~~quantum-field potential $\mathbb{A}_{(\Psi)g}^s$     &  $\mathbb{H}_s$       \\
\hline
electromagnetic~~~~ &  electromagnetic fundamental~~~~                            &  electromagnetic fundamental                                                    \\
~~~~interaction     &  ~~field potential $\mathbb{A}_e^g$                         &  ~~quantum-field potential $\mathbb{A}_{(\Psi)g}^g$     &  $\mathbb{H}_e$       \\
                    &  electromagnetic adjoint                                    &  electromagnetic adjoint                                                        \\
                    &  ~~field potential $\mathbb{A}_e^e$                         &  ~~quantum-field potential $\mathbb{A}_{(\Psi)g}^e$     &  $\mathbb{H}_g$       \\
                    &  electromagnetic adjoint                                    &  electromagnetic adjoint                                                        \\
                    &  ~~field potential $\mathbb{A}_e^w$                         &  ~~quantum-field potential $\mathbb{A}_{(\Psi)g}^w$     &  $\mathbb{H}_s$       \\
                    &  electromagnetic adjoint                                    &  electromagnetic adjoint                                                        \\
                    &  ~~field potential $\mathbb{A}_e^s$                         &  ~~quantum-field potential $\mathbb{A}_{(\Psi)g}^s$     &  $\mathbb{H}_w$       \\
\hline
W-nuclear           &  W-nuclear fundamental                                      &  W-nuclear fundamental                                                          \\
~~~~interaction     &  ~~field potential $\mathbb{A}_w^g$                         &  ~~quantum-field potential $\mathbb{A}_{(\Psi)w}^g$     &  $\mathbb{H}_w$       \\
                    &  W-nuclear adjoint                                          &  W-nuclear adjoint                                                              \\
                    &  ~~field potential $\mathbb{A}_w^e$                         &  ~~quantum-field potential $\mathbb{A}_{(\Psi)w}^e$     &  $\mathbb{H}_s$       \\
                    &  W-nuclear adjoint                                          &  W-nuclear adjoint                                                              \\
                    &  ~~field potential $\mathbb{A}_w^w$                         &  ~~quantum-field potential $\mathbb{A}_{(\Psi)w}^w$     &  $\mathbb{H}_g$       \\
                    &  W-nuclear adjoint                                          &  W-nuclear adjoint                                                              \\
                    &  ~~field potential $\mathbb{A}_w^s$                         &  ~~quantum-field potential $\mathbb{A}_{(\Psi)w}^s$     &  $\mathbb{H}_e$       \\
\hline
strong-nuclear      &  strong-nuclear fundamental                                 &  strong-nuclear fundamental                                                     \\
~~~~interaction     &  ~~field potential $\mathbb{A}_s^g$                         &  ~~quantum-field potential $\mathbb{A}_{(\Psi)s}^g$     &  $\mathbb{H}_s$       \\
                    &  strong-nuclear adjoint                                     &  strong-nuclear adjoint                                                         \\
                    &  ~~field potential $\mathbb{A}_s^e$                         &  ~~quantum-field potential $\mathbb{A}_{(\Psi)s}^e$     &  $\mathbb{H}_w$       \\
                    &  strong-nuclear adjoint                                     &  strong-nuclear adjoint                                                         \\
                    &  ~~field potential $\mathbb{A}_s^w$                         &  ~~quantum-field potential $\mathbb{A}_{(\Psi)s}^w$     &  $\mathbb{H}_e$       \\
                    &  strong-nuclear adjoint                                     &  strong-nuclear adjoint                                                         \\
                    &  ~~field potential $\mathbb{A}_s^s$                         &  ~~quantum-field potential $\mathbb{A}_{(\Psi)s}^s$     &  $\mathbb{H}_g$       \\
\hline\hline
\end{tabular}
%\end{ruledtabular}
\end{table}

\section{\label{sec:level1}Yang-Mills equations}

In the complex-sedenion space, when the term, $\mathbb{Z}_W \circ \mathbb{W}^\star / ( \hbar v_0 )$ , in the operator $\mathbb{D}_Z$ can be degraded into the term, $g_A \mathbb{Z}_W^\prime \circ \mathbb{A}_{(\Psi)}$ , the complex-sedenion quantum-field potential, quantum-field strength, and quantum-field source will be degenerated into that of the non-Abelian gauge field. Further, making use of the transformations of complex-quaternion spaces, the field equations of non-Abelian gauge field are able to be degraded into the existing Yang-Mills equations for the non-Abelian gauge field.

\subsection{\label{sec:level1}Non-Abelian gauge field}

In the complex-sedenion space, the electromagnetic quantum-field potential $\mathbb{A}_{(\Psi)e}$ is in possession of the components in the four complex-quaternion spaces, $\mathbb{H}_g$ , $\mathbb{H}_e$ , $\mathbb{H}_w$ , and $\mathbb{H}_s$ , including $\mathbb{A}_{(\Psi)e}^g$ , $\mathbb{A}_{(\Psi)e}^e$ , $\mathbb{A}_{(\Psi)e}^w$ , and $\mathbb{A}_{(\Psi)e}^s$ . Apparently, when there is only the electromagnetic quantum-field potential, its components are still able to constitute the non-Abelian gauge field.

Under certain circumstances, the term, $k_{eg} \mathbb{Z}_{We} \circ \mathbb{W}_e^\star / ( \hbar v_0 )$ , of the operator $\mathbb{D}_Z$ may be degenerated into the term, $k_{eg} g_{Ae} \mathbb{Z}_{We}^\prime \circ \mathbb{A}_{(\Psi)e}$ , in the complex-sedenion space. Therefore the definition of complex-sedenion quantum-field strength, for the electromagnetic quantum-field, will be approximately written as,
\begin{eqnarray}
\mathbb{F}_{(\Psi)e} = ( i k_{eg} g_{Ae} \mathbb{Z}_{We}^\prime \circ \mathbb{A}_{(\Psi)e} + \lozenge ) \circ \mathbb{A}_{(\Psi)e}  ~ ,
\end{eqnarray}
where $g_{Ae}$ is a coefficient. $\mathbb{Z}_{We}$ and $\mathbb{Z}_{We}^\prime$ are two auxiliary quantities.

In contrast to the existing electroweak field, the components, $\mathbb{A}_{(\Psi)e}^e$ , $\mathbb{A}_{(\Psi)e}^w$, and $\mathbb{A}_{(\Psi)e}^s$ , of the electromagnetic quantum-field potential, in the three complex-quaternion spaces, $\mathbb{H}_g$ , $\mathbb{H}_s$ , and $\mathbb{H}_w$ , can be considered as three weak-nuclear quantum-field potentials. By means of the transformations of complex-quaternion spaces, the electromagnetic quantum-field potentials and field strengths, described with the complex-sedenions, are able to be degenerated into the field equations for the existing electroweak field.

On the basis of the preceding analysis, one can conclude that the electromagnetic quantum-field, described with the complex-sedenions, is compatible with the existing electroweak field, and even both of them give mutual support to each other, to a certain extent. On the one hand, the components of the electromagnetic quantum-field potential, in the three complex-quaternion spaces, $\mathbb{H}_g$ , $\mathbb{H}_s$ , and $\mathbb{H}_w$, can be considered as three weak-nuclear quantum-field potentials. Therefore the electroweak theory seems to imply that it is reasonable the existence of the electromagnetic adjoint quantum-field, in the complex-sedenion space. On the other hand, the electroweak theory revealed that the existing weak-nuclear field is not a fundamental field, and is just an ingredient of the electroweak field. It could even be said that the weak-nuclear field is just a component of the electromagnetic quantum-field, described with the complex-sedenions. Obviously, the latter description is comparatively brief and convenient.

Similarly, in the complex-sedenion space, the strong-nuclear (or gravitational) quantum-field potential is seized of the components in the four complex-quaternion spaces, constituting the non-Abelian gauge field. Especially, the strong-nuclear quantum-field potential $\mathbb{A}_{(\Psi)s}$ possesses the components in the four complex-quaternion spaces, $\mathbb{H}_g$ , $\mathbb{H}_e$ , $\mathbb{H}_w$ , and $\mathbb{H}_s$ , including $\mathbb{A}_{(\Psi)s}^g$ , $\mathbb{A}_{(\Psi)s}^e$ , $\mathbb{A}_{(\Psi)s}^w$ , and $\mathbb{A}_{(\Psi)s}^s$ . In other words, when there are only the strong-nuclear quantum-field potentials, their components are capable of constituting the non-Abelian gauge field, meeting the requirement of the equations in the Table 4.

By means of the multiplicative closure (Appendix D) of sedenions, the complex-sedenion space is able to explore simultaneously four sorts of fundamental interactions. And that the four interactions have to exist in the complex-sedenion space simultaneously. They are similar to the four surfaces of one tetrahedral, and none of four surfaces can be dispensable. If the three adjoint quantum-fields of the complex-sedenion electromagnetic quantum-field can be considered as the existing weak-nuclear field, there must be one new fundamental interaction, that is, `W-nuclear interaction'. The latter replaces the weak-nuclear interaction, in the complex-sedenion space.

\subsection{\label{sec:level1}Space transformation}

If the electromagnetic fundamental quantum-field can be transformed from the complex 2-quaternion space, $\mathbb{H}_e$, into the complex-quaternion space, $\mathbb{H}_g$ , it will facilitate us to find that there are certain similarities between the definition of field strength in the electroweak field and that of quantum-field strength in the electromagnetic quantum-field, described with the complex-sedenions.

In the complex-sedenion space, when the term, $\mathbb{Z}_W \circ \mathbb{W}^\star / ( \hbar v_0 )$ , in the operator $\mathbb{D}_Z$ can be reduced into the term, $g_A \mathbb{Z}_W^\prime \circ \mathbb{A}_{(\Psi)}$ , the definition of the complex-sedenion quantum-field strength,
\begin{eqnarray}
\mathbb{F}_{(\Psi)} = \{ i \mathbb{Z}_W \circ \mathbb{W}^\star / ( \hbar v_0 ) + \lozenge \} \circ \mathbb{A}_{(\Psi)}  ~,
\end{eqnarray}
will be written approximatively as,
\begin{eqnarray}
\mathbb{F}_{(\Psi)} = ( i g_A \mathbb{Z}_W^\prime \circ \mathbb{A}_{(\Psi)} + \lozenge ) \circ \mathbb{A}_{(\Psi)}  ~ .
\end{eqnarray}

Making use of a transformation, $\mathbb{A}_{(\Psi)}^\prime = \textbf{\emph{I}}_{e0} \circ \mathbb{A}_{(\Psi)}$ , it is able to achieve a new complex-sedenion quantum-field potential, $\mathbb{A}_{(\Psi)}^\prime$ , which is written as,
\begin{eqnarray}
\mathbb{A}_{(\Psi)}^\prime = k_{eg} \mathbb{A}_{(\Psi)g}^\prime + \mathbb{A}_{(\Psi)e}^\prime + k_{sg} \mathbb{A}_{(\Psi)w}^\prime + k_{wg} \mathbb{A}_{(\Psi)s}^\prime ~ ,
\end{eqnarray}
where $\mathbb{A}_{(\Psi)e}^\prime = \textbf{\emph{I}}_{e0} \circ \mathbb{A}_{(\Psi)g}$ . $\mathbb{A}_{(\Psi)g}^\prime = \textbf{\emph{I}}_{e0} \circ \mathbb{A}_{(\Psi)e}$ . $\mathbb{A}_{(\Psi)s}^\prime = \textbf{\emph{I}}_{e0} \circ \mathbb{A}_{(\Psi)w}$ . $\mathbb{A}_{(\Psi)w}^\prime = \textbf{\emph{I}}_{e0} \circ \mathbb{A}_{(\Psi)s}$ . And that the ingredients of the quantum-field potential, $\mathbb{A}_{(\Psi)}^\prime$ , in the four complex-quaternion spaces are as follows,
\begin{eqnarray}
&& \mathbb{A}_{(\Psi)e}^\prime = i \textbf{\emph{I}}_{e0} \mathbb{A}_{(\Psi)g0} - \Sigma \textbf{\emph{I}}_{ek} \mathbb{A}_{(\Psi)gk}     ~ ,
\\
&& \mathbb{A}_{(\Psi)g}^\prime = - i \textbf{\emph{I}}_{g0} \mathbb{A}_{(\Psi)e0} + \Sigma \textbf{\emph{I}}_{gk} \mathbb{A}_{(\Psi)ek}      ~ ,
\\
&& \mathbb{A}_{(\Psi)s}^\prime = i \textbf{\emph{I}}_{s0} \mathbb{A}_{(\Psi)w0} + \Sigma \textbf{\emph{I}}_{sk} \mathbb{A}_{(\Psi)wk}    ~  ,
\\
&& \mathbb{A}_{(\Psi)w}^\prime = - i \textbf{\emph{I}}_{w0} \mathbb{A}_{(\Psi)s0} - \Sigma \textbf{\emph{I}}_{wk} \mathbb{A}_{(\Psi)sk}   ~ .
\end{eqnarray}

By all appearances, the electromagnetic quantum-field potential, $\mathbb{A}_{(\Psi)e}$ , in the complex 2-quaternion space, $\mathbb{H}_e$ , has been transformed into the electromagnetic quantum-field potential, $\mathbb{A}_{(\Psi)g}^\prime$ , in the complex-quaternion space, $\mathbb{H}_g$ . Meanwhile, other three ingredients, $\mathbb{A}_{(\Psi)g}$ , $\mathbb{A}_{(\Psi)w}$, and $\mathbb{A}_{(\Psi)s}$ , of the quantum-field potential, $\mathbb{A}_{(\Psi)}$ , are transformed respectively into three new ingredients, $\mathbb{A}_{(\Psi)e}^\prime$ , $\mathbb{A}_{(\Psi)s}^\prime$ , and $\mathbb{A}_{(\Psi)w}^\prime$ , of the quantum-field potential, $\mathbb{A}_{(\Psi)}^\prime$ .

Subsequently, in the complex-sedenion space, the definition of complex-sedenion quantum-field strength, $\mathbb{F}_{(\Psi)}^\prime$ , can be written approximately as,
\begin{eqnarray}
\mathbb{F}_{(\Psi)}^\prime = ( i g_A^\prime \mathbb{Z}_W^{\prime\prime} \circ \mathbb{A}_{(\Psi)}^\prime + \lozenge ) \circ \mathbb{A}_{(\Psi)}^\prime  ~ .
\end{eqnarray}
where, $g_A^\prime$ is a coefficient, and $\mathbb{Z}_W^{\prime\prime}$ is one auxiliary quantity.

Certainly, one may observe the quantum-field potential, $\mathbb{A}_{(\Psi)}^\prime = \textbf{\emph{I}}_{e0} \circ \mathbb{A}_{(\Psi)}$, from others' perspective. For instance, in the complex-quaternion space, $\mathbb{H}_g$ , there are four physical quantities, $( i \textbf{\emph{I}}_{g0} \mathbb{A}_{(\Psi)g0} + \Sigma \textbf{\emph{I}}_{gk} \mathbb{A}_{(\Psi)gk} )$ , $( i \textbf{\emph{I}}_{g0} \mathbb{A}_{(\Psi)e0} + \Sigma \textbf{\emph{I}}_{gk} \mathbb{A}_{(\Psi)ek} )$ , $( i \textbf{\emph{I}}_{g0} \mathbb{A}_{(\Psi)w0} + \Sigma \textbf{\emph{I}}_{gk} \mathbb{A}_{(\Psi)wk} )$ , $( i \textbf{\emph{I}}_{g0} \mathbb{A}_{(\Psi)s0} + \Sigma \textbf{\emph{I}}_{gk} \mathbb{A}_{(\Psi)sk} )$ , with three unit matrices, $\textbf{\emph{I}}_{e0} $ , $\textbf{\emph{I}}_{w0}$ , and $\textbf{\emph{I}}_{s0}$ . This point of view is consistent with that in the electroweak theory.

According to the multiplication rule of sedenions, three unit vectors, $\textbf{\emph{I}}_{g1}$, $\textbf{\emph{I}}_{g2}$ , and $\textbf{\emph{I}}_{g3}$ , of the quaternions meet the demand of the rule for matrix multiplication of the Pauli matrices. Also three unit vectors, $\textbf{\emph{I}}_{e0}$ , $\textbf{\emph{I}}_{w0}$ ,
and $\textbf{\emph{I}}_{s0}$, of the sedenions satisfy the requirement of the rule for matrix multiplication of the Pauli matrices. Apparently, the three unit vectors of the sedenions are equivalent to the generators of Lie algebra in the electroweak theory. Moreover, the coefficient $g_A^\prime $ in Eq.(54) for the microscopic scale is distinct from the coefficient $\mu$ in Eq.(3) for the macroscopic scale, and that the two coefficients exist in different equations. Similarly, the hypercharge and electric charge are two different coefficients, appearing in different equations, so there is conflict free between the two coefficients.

Under certain conditions, the definition of quantum-field strength, Eq.(54), described with the complex-sedenions, will accord with that of field strength for the non-Abelian gauge field in the electroweak theory naturally. a) In the complex-sedenion operator, $\lozenge$ , it is able to consider merely the contribution of term, $\lozenge_g$ . That is, the contributions, coming from the rest of operator, $\lozenge$, can be neglected. b) Sometimes, the contributions coming from three terms, $g_{Ae}^\prime \mathbb{A}_{(\Psi)e}^\prime$ , $k_{sg} g_{Aw}^\prime \mathbb{A}_{(\Psi)w}^\prime$, and $k_{wg} g_{As}^\prime \mathbb{A}_{(\Psi)s}^\prime$ ,
are considerable approximately. Even the contribution, coming from each of the three terms, may be equivalent to that of the term, $k_{eg} g_{Ag}^\prime \mathbb{A}_{(\Psi)g}^\prime$ also. Herein $g_{Ag}^\prime$ , $g_{Ae}^\prime$ , $g_{Aw}^\prime$, and $g_{As}^\prime$ are coefficients. c) Marking the letters, $g$, $e$, $w$, and $s$, of the superscript and subscript in the paper as the numbers, $1$, $2$, $3$, and $4$, will enable the marking method of the complex-sedenion quantum-field to close to that of electroweak field, facilitating to contrast the similarities and differences between the two theories.

Especially, when there is merely the electromagnetic quantum-field potential, Eq.(54) in the above is accordant with the definition of field strength in the electroweak theory. After the transforming, from the point of view of the complex-sedenion space, the electromagnetic field within the electroweak field will situate in the complex-quaternion space, $\mathbb{H}_g$ , while three weak-nuclear fields (and generators of Lie algebra) within the electroweak field situate in three other complex-quaternion spaces, $\mathbb{H}_e$ , $\mathbb{H}_w$ , and $\mathbb{H}_s$ , respectively. And one can observe distinctly the equivalency between the electroweak field and the electromagnetic fundamental/adjoint quantum-fields, in the complex-sedenion space. From the perspective of interactions, the electromagnetic interaction includes the electromagnetic fundamental field and adjoint quantum-field. As the function of the electromagnetic fundamental/adjoint quantum-fields, the electromagnetic quantum-field must belong to the electromagnetic interaction. In other words, in the complex-sedenion space, the electroweak field belongs to the electromagnetic interaction also, as the equivalent of the electromagnetic quantum-field. All of a sudden, it is found that the study of electroweak theory is so closely interrelated with the electromagnetic quantum-field theory, described with the complex-sedenions, by means of the transformations of coordinate systems and spaces.

Further, in the complex-octonion space $\mathbb{O}$ , for the classical electromagnetic field on the macroscopic scale, it is capable of deducing the definitions of the electromagnetic strength and electromagnetic source (see Ref.[18]), from the field equations, $\lozenge_g \circ \mathbb{A}_e = \mathbb{F}_e$ , and $\lozenge_g^\ast \circ \mathbb{F}_e = - \mu \mathbb{S}_e$ , respectively. Similarly, in the complex-octonion space $\mathbb{O}$ , for the electromagnetic quantum-field on the microscopic scale, there may be one comparatively simple situation that the contribution of term, $\{ i \mathbb{Z}_W \circ \mathbb{W}^\star / ( \hbar v_0 ) \}$ can be neglected approximately. Therefore one can derive the definitions of the electromagnetic quantum strength and quantum source, from the field equations, $\lozenge_g \circ \mathbb{A}_{(\Psi)e} = \mathbb{F}_{(\Psi)e}$ , and $\lozenge_g^\ast \circ \mathbb{F}_{(\Psi)e} = - \mu \mathbb{S}_{(\Psi)e}$ , respectively, for the simple cases.

The preceding discussion illuminates that it is able to define the complex-sedenion quantum-field strength, from the complex-sedenion quantum-field potential. This method can be extended into the definition of complex-sedenion quantum-field source. And the definitions of the complex-sedenion quantum-field potential and field source can be degenerated into the Yang-Mills equations of non-Abelian gauge field in the electroweak theory (Table 6).

In terms of the non-Abelian gauge field, described with the complex-sedenions, it is able to apply the complex-quaternion wavefunction to explore the physical quantities, relevant to the fundamental field and adjoint field, in each complex-quaternion space. In contrast to the wavefunction, described with the complex-numbers, the complex-quaternion wavefunction will be seized of three new degrees of freedom extraordinarily, which can be regarded as the color degrees of freedom.

\begin{table}[h]
\caption{Some field equations of the non-Abelian gauge field in the complex-sedenion quantum mechanics, when the term, $\mathbb{Z}_W \circ \mathbb{W}^\star / ( \hbar v_0 )$, can be reduced into the term, $g_A \mathbb{Z}_W^\prime \circ \mathbb{A}_{(\Psi)}$. In terms of the electromagnetic quantum-field, these field equations can be degenerated into the Yang-Mills equations for the electroweak field.}
%\begin{ruledtabular}
\centering
\begin{tabular}{@{}ll@{}}
\hline\hline
quantum~physics~quantity~~~~~~&  definition                                                                                                                         \\
\hline
quantum~field~potential       &  $\mathbb{A}_{(\Psi)} = i ( i g_A \mathbb{Z}_W^\prime \circ \mathbb{A}_{(\Psi)} + \lozenge )^\star \circ \mathbb{X}_{(\Psi)}  $     \\

quantum~field~strength        &  $\mathbb{F}_{(\Psi)} = ( i g_A \mathbb{Z}_W^\prime \circ \mathbb{A}_{(\Psi)} + \lozenge )  \circ \mathbb{A}_{(\Psi)}  $            \\
quantum~field~source          &  $\mu \mathbb{S}_{(\Psi)} = - ( i g_A \mathbb{Z}_W^\prime \circ \mathbb{A}_{(\Psi)} + \lozenge ) ^\ast \circ \mathbb{F}_{(\Psi)} $  \\
%%%
quantum~torque                &  $\mathbb{W}_{(\Psi)} = - v_0 ( i g_A \mathbb{Z}_W^\prime \circ \mathbb{A}_{(\Psi)} + \lozenge )  \circ \mathbb{L}_{(\Psi)}  $      \\
quantum~force                 &  $\mathbb{N}_{(\Psi)} = - ( i g_A \mathbb{Z}_W^\prime \circ \mathbb{A}_{(\Psi)} + \lozenge )  \circ \mathbb{W}_{(\Psi)} $           \\
\hline\hline
\end{tabular}
%\end{ruledtabular}
\end{table}

\section{\label{sec:level1}Color degrees of freedom}

In certain circumstances for the complex-quaternion space $\mathbb{H}_g$ , a physical problem may allow a three-dimensional unit vector $\textbf{\emph{i}}_q$ , of the complex-quaternion wavefunction $\mathbb{A}_{(\Psi)g}$ , to be degraded into one three-dimensional unit vector $\textbf{\emph{i}}_q^\prime$, which is independent of the unit vector $\textbf{\emph{i}}_q$ . In contrast to the imaginary unit, the unit vector $\textbf{\emph{i}}_q^\prime$ possesses three new degrees of freedom. Consequently, the complex-quaternion wavefunction $\mathbb{A}_{(\Psi)g}$ is able to be degenerated and separated into three complex-number wavefunctions, with three new degrees of freedom, in the conventional quantum mechanics. These additional degrees of freedom are so-called color degrees of freedom.

\subsection{\label{sec:level1}New degrees of freedom}

In terms of the physical problems, we may stand a chance of achieving a few degrees of freedom, in case the complex-quaternion wavefunctions can be degenerated into the complex-number wavefunctions approximately. In the complex-quaternion space $\mathbb{H}_g$ , there may be two sorts of degenerations connected with the complex-quaternion wavefunctions. a) Imaginary unit. In some simple physical problems, the three-dimensional unit vector $\textbf{\emph{i}}_q$ , of the complex-quaternion wavefunction $\mathbb{A}_{(\Psi)g}$ , can be directly degraded into the imaginary unit, $i$ . Therefore, the complex-quaternion wavefunction is able to be degraded into the conventional wavefunction, described with the complex-numbers (see Appendix A). b) Unit vector. In some other physical problems, there is also another possibility that the three-dimensional unit vector $\textbf{\emph{i}}_q$, of the complex-quaternion wavefunction $\mathbb{A}_{(\Psi)g}$ , can only be degraded into one new three-dimensional unit vector $\textbf{\emph{i}}_q^\prime$ , which is independent of the unit vector $\textbf{\emph{i}}_q$. This new three-dimensional unit vector $\textbf{\emph{i}}_q^\prime$ is also seized of the properties of the imaginary unit, and is in possession of three new degrees of freedom, in contrast to the imaginary unit. These new degrees of freedom can be considered as the color degrees of freedom. Subsequently the complex-quaternion wavefunction can be simplified and separated into three complex-number wavefunctions, with three new degrees of freedom, in the conventional quantum mechanics.

In some simple circumstances, in the complex-quaternion space $\mathbb{H}_g$ , the direction properties of the complex-quaternion wavefunction is incapable of playing a major role, and even can be neglected totally. In this case, the three-dimensional unit vector $\textbf{\emph{i}}_q$ , in the complex-quaternion wavefunction, $\mathbb{A}_{(\Psi)g}$, will be degraded into the imaginary unit $i$ . And the complex-quaternion wavefunction $\mathbb{A}_{(\Psi)g}$ is able to be degenerated the classical wavefunction, described with the complex-numbers, in the conventional quantum mechanics. In case the term, $\mathbb{Z}_W \circ \mathbb{W}^\star$ , can be simplified into the term, $\mathbb{W}^\star$ , it is capable of achieving the Dirac wave equation, from the field equations in the Table 6. When the quantum torque is equal to zero, the equation, $\mathbb{W}_{(\Psi)} = 0$, can be reduced into the conventional Dirac wave equation, described with the complex-numbers, for the colorless quantum states in the physical problems.

In some other complicated circumstances, in the complex-quaternion space $\mathbb{H}_g$, the direction property of complex-quaternion wavefunction is only able to play a partial role, so it can merely be neglected partially, although the direction property cannot play a major role either. In this case, the three-dimensional unit vector $\textbf{\emph{i}}_q$ , of the complex-quaternion wavefunction $\mathbb{A}_{(\Psi)g}$ , can only be degraded into a three-dimensional unit vector, $\textbf{\emph{i}}_q^\prime$ , which is independent of the unit vector $\textbf{\emph{i}}_q$ . Herein $\textbf{\emph{i}}_q^\prime = \Sigma f_k \textbf{\emph{i}}_{q(k)}^\prime$ . $\textbf{\emph{i}}_{q(1)}^\prime$ , $\textbf{\emph{i}}_{q(2)}^\prime$ , and $\textbf{\emph{i}}_{q(3)}^\prime$ are three unit vectors, which are perpendicular to each other. $f_k$ is a coefficient. Subsequently, the complex-quaternion wavefunction is able to be degenerated into the complex-number wavefunction, with three new degrees of freedom, in the conventional quantum mechanics. That is, one complex-quaternion wavefunction can be simplified into three `degenerated' complex-number wavefunctions, which are independent of each other and relevant to the unit vector, $\textbf{\emph{i}}_{q(k)}^\prime$. And these three `degenerated' complex-number wavefunctions are identical to each other, except for the unit vector, $\textbf{\emph{i}}_{q(k)}^\prime$ .

Further, in order to differentiate the three `degenerated' complex-number wavefunctions, it is necessary to introduce a new variable to mark them. Therefore this new variable, relevant to the unit vector $\textbf{\emph{i}}_{q(k)}^\prime$ , can be regarded as the color degree of freedom. It stands for the three basis vectors of the unit vector $\textbf{\emph{i}}_{q(k)}^\prime$ . This new argument, that is, the color degree of freedom, enables the complex-quaternion wavefunction to be written as three conventional complex-number wavefunctions equivalently, while the unit vector, $\textbf{\emph{i}}_{q(k)}^\prime$, may be reduced into the imaginary unit, $i$, further. It means that the existing `three colors' can be merged into the complex-quaternion wavefunction directly. Just like the hypothesis of `phlogiston', the assumption of color charge is redundant too. The historical mission of color charge seems to ascertain merely the validity of complex-quaternion wavefunction, in the complex-quaternion space $\mathbb{H}_g$ .

Apparently, the above approximate method, in the complex-quaternion space $\mathbb{H}_g$ , is possible to be extended into some comparatively complicated approximate methods in three complex-quaternion spaces, $\mathbb{H}_e$ , $\mathbb{H}_w$ , and $\mathbb{H}_s$ .

\subsection{\label{sec:level1}Color confinement}

In the complex-quaternion space $\mathbb{H}_g$ , in terms of the gravitational quantum-field potential $\mathbb{A}_{(\Psi)g}$ , in case the three-dimensional unit vector $\textbf{\emph{i}}_q$ , of the complex-quaternion wavefunction $\mathbb{A}_{(\Psi)g}$ , can be degraded into an independent three-dimensional unit vector, $\textbf{\emph{i}}_q^\prime$ , the complex-quaternion wavefunction $\mathbb{A}_{(\Psi)g}$ will be degenerated into three classical complex-number wavefunctions, in the conventional quantum mechanics. This is equivalent to increasing three degrees of freedom, for the classical complex-number wavefunctions, in the conventional quantum mechanics, and it is able to apply the `three colors', $R_g$ , $G_g$ , and $B_g$ , to mark them. Apparently, this method can be extended into some complex-quaternion wavefunctions, in the three complex-quaternion spaces, $\mathbb{H}_e$ , $\mathbb{H}_w$ , and $\mathbb{H}_s$ .

In the complex 2-quaternion space $\mathbb{H}_e$ , in terms of the electromagnetic quantum-field potential $\mathbb{A}_{(\Psi)e}$ , if the three-dimensional unit vector $\textbf{\emph{i}}_{q(e)}$ , of the complex 2-quaternion wavefunction $\mathbb{A}_{(\Psi)e}$ , can be reduced into an independent three-dimensional unit vector, $\textbf{\emph{i}}_{q(e)}^\prime$ , also the complex 2-quaternion wavefunction $\mathbb{A}_{(\Psi)e}$ may be degenerated into three classical complex-number wavefunctions, in the conventional quantum mechanics. This is equivalent to increasing three degrees of freedom, for the classical complex-number wavefunctions of the conventional quantum mechanics, in the complex 2-quaternion space $\mathbb{H}_e$ , and it is able to apply the `three colors', $R_e$ , $G_e$ , and $B_e$ , to mark them.

However, the `three colors', $R_e$ , $G_e$ , and $B_e$ , in the complex 2-quaternion space $\mathbb{H}_e$ , are different from the `three colors', $R_g$ , $G_g$ , and $B_g$ , in the complex-quaternion space $\mathbb{H}_g$ . Therefore, when there are simultaneously two complex-quaternion wavefunctions, $\mathbb{A}_{(\Psi)g}$ and $\mathbb{A}_{(\Psi)e}$ , the classical complex-number wavefunctions of the conventional quantum mechanics will increase six degrees of freedom, or `six colors', accordingly, in the complex-octonion space $\mathbb{O}$.

By analogy with the above, each of two complex-quaternion spaces, $\mathbb{H}_w$ and $\mathbb{H}_s$, will result in three new degrees of freedom. Apparently the
`three colors', $R_w$ , $G_w$ , and $B_w$ , in the complex 3-quaternion space $\mathbb{H}_w$ , are also different from the `three colors', $R_s$ , $G_s$ , and $B_s$ , in the complex 4-quaternion space $\mathbb{H}_s$ . Consequently, in the complex-sedenion space, when the quantum-field potential $\mathbb{A}_{(\Psi)}$ is degenerated into the classical complex-number wavefunctions of the conventional quantum mechanics, it is necessary to increase totally twelve degrees of freedom, or `twelve colors' (Table 7).

The preceding research states that it is possible to increase twelves degrees of freedom, for the physical problems, when the wavefunction, described with the complex-sedenions, is degenerated into the classical complex-number wavefunctions, in the conventional quantum mechanics. In other words, the color degrees of freedom are only the spatial dimensions of the unit vector in the complex-sedenion wavefunction, rather than any property of physical substance. The color degrees of freedom are invisible and a sort of equivalent description, while it is able to measure the contribution of the color degrees of freedom. Consequently, the color confinement will be effective for all time. And even one may deem that the assumption of color charge is unnecessary.

Making a comparison and analysis of preceding studies, it is found that the QCD is so closely interrelated with the quantum mechanics described with the complex-sedenions to a certain extent, in terms of the color degrees of freedom and color confinement. a) Color degrees of freedom. The quantum mechanics, described with the complex-sedenions, contemplates that it is necessary to introduce the color degrees of freedom, increasing the number of complex-number wavefunctions for certain physical quantities, exploring accurately the physical phenomena of the quarks. This is a standpoint accordant with the QCD. b) Essence of `three colors'. The QCD assumed that there must be the color charge, which is similar to the electric charge, elucidating the existence of the color degrees of freedom, although it is unable to validate the authenticity of the color charge until now. Nevertheless, the quantum mechanics, described with the complex-sedenions, deems that the color degrees of freedom are just three spatial dimensions of the unit vector in the complex-quaternion wavefunction. As a result, the color degrees of freedom must possess one precise number. Obviously, the paper is very noticeably different from the QCD, in terms of the explanation for the essence of `three colors'. The QCD believes the `three colors' are relevant to a physical substance in essence, while the paper contemplates that the `three colors' are only the spatial dimensions. c) Color confinement. In the QCD, the color confinement is a rule summarized from the analysis of some physical phenomena, elucidating the invisibility of quarks, exploring the physical phenomena of quarks and leptons. However, in the quantum mechanics described with the complex-sedenions, the color degrees of freedom are just the spatial dimensions, so the color confinement must be effective always. These two theories both accept the color confinement, although they provide two different explanations.

\begin{table}[h]
\caption{There are twelve color degrees of freedom in the quantum mechanics described with the complex-sedenions, when the complex-sedenion quantum-field potential is degenerated into the classical complex-number wavefunctions of the conventional quantum mechanics.}
%\begin{ruledtabular}
\centering
\begin{tabular}{@{}ll@{}}
\hline\hline
complex-quaternion~space~~~~~~&  color~degree~of~freedom        \\
\hline
$\mathbb{H}_g$                &  $R_g$ , $G_g$ , $B_g$          \\
$\mathbb{H}_e$                &  $R_e$ , $G_e$ , $B_e$          \\
$\mathbb{H}_w$                &  $R_w$ , $G_w$ , $B_w$          \\
$\mathbb{H}_s$                &  $R_s$ , $G_s$ , $B_s$          \\
\hline\hline
\end{tabular}
%\end{ruledtabular}
\end{table}

\section{\label{sec:level1}Conclusions and discussions}

Each of four fundamental fields is in possession of its individual complex-quaternion space. As a result, these four fundamental fields are seized of four complex-quaternion spaces, which are independent of each other. Further the four complex-quaternion spaces, which are perpendicular to each other, are able to combine together to become one complex-sedenion space. In other words, the complex-sedenion space can be applied to describe simultaneously the physical properties of gravitational field, electromagnetic field, W-nuclear field, and strong-nuclear field. In the complex-sedenion space, the complex-sedenion operator and field strength will constitute a composite operator, exploring the field equations of the classical mechanics on the macroscopic scale. And the field equations may be degenerated into the electromagnetic field equations and gravitational field equations and so forth.

The complex-quaternion physical quantities of each fundamental field can be rewritten as the exponential form. By means of the concepts of the auxiliary quantity and function, the product of the complex-quaternion physical quantity and auxiliary quantity is able to be rewritten as the exponential wavefunction, which may be degenerated into the conventional wavefunction in the existing quantum mechanics. Similarly, making use of the auxiliary quantity, it is able to deduce the complex-sedenion field equations of the quantum mechanics on the microscopic scale, from the above field equations of the classical mechanics. Under certain approximate conditions, two of quantum-field equations will be degenerated respectively into the complex-number Dirac wave equation and Yang-Mills equation in the non-Abelian gauge field.

In the complex-quaternion space $\mathbb{H}_g$ , when the three-dimensional unit vector, $\textbf{\emph{i}}_q$, can be degraded into the imaginary unit $i$ , the complex-quaternion wavefunction will be degenerated into complex-number wavefunction, in the conventional quantum mechanics. What is more important is that there may be another approximate method, for certain physical problems. That is, the three-dimensional unit vector, $\textbf{\emph{i}}_q$, can be degraded into the three-dimensional unit vector, $\textbf{\emph{i}}_q^\prime$ , which is independent of the unit vector $\textbf{\emph{i}}_q$ . It means that the degrees of freedom of the three-dimensional unit vector, $\textbf{\emph{i}}_q^\prime$ , are three more than that of the imaginary unit $i$ . In other words, the wavefunction, relevant to the three-dimensional unit vector $\textbf{\emph{i}}_q^\prime$ , has three degrees of freedom more than that connected with the imaginary unit $i$ . In case the three-dimensional unit vector,
$\textbf{\emph{i}}_q^\prime = \Sigma f_k \textbf{\emph{i}}_{q(k)}^\prime$ , was mistaken for the imaginary unit $i$, it is necessary to introduce `three colors' to mark the three new degrees of freedom of the wavefunction, relevant to the three-dimensional unit vector $\textbf{\emph{i}}_q^\prime$ . For the same reason, there may be similar situations, in the complex-quaternion spaces, $\mathbb{H}_e$ , $\mathbb{H}_w$, and $\mathbb{H}_s$ . Therefore there will be `twelve colors' totally, in the complex-sedenion space. And the color degrees of freedom are merely the spatial dimensions of the unit vector in the wavefunction, rather than any property of physical substance.

In the complex-sedenion space, there are four fundamental field sources, and twelve adjoint field sources. In each of four complex-quaternion spaces, there are one fundamental field source, and three adjoint field sources. Especially, in the complex-quaternion space $\mathbb{H}_g$ , one fundamental field source is $m_g^g$, and three adjoint field sources are $m_e^e$ , $m_w^w$, and $m_s^s$ , respectively. These three adjoint field sources are able to generate a part of gravitational effects, so they can be considered as the candidates for the particles of dark matter. Further these three adjoint field sources can be combined with some other fundamental/adjoint field sources to become the two-source, three-source, or four-source (and even multi-source) particles. Consequently, there may be many sorts of particles to possess the properties of dark matter. These undetectable particles of dark matter may be ubiquitous, surrounding not only the celestial bodies but also the particles of ordinary matter, encompassing the human being.

It should be noted the paper discussed only some simple cases about the wavefunctions and quantum-field equations, described with the complex-sedenions. However it clearly states that the quantum-field equations can be degenerated into the Dirac wavefunction and Yang-Mills equations in the non-Abelian gauge field. Later, the complex-sedenion wavefunction can be applied to elucidate the color degrees of freedom and color confinement and so forth in the non-Abelian gauge field. And the color degrees of freedom are merely the spatial dimensions of the unit vector in the complex-sedenion wavefunction, rather than any property of physical substance, unpuzzling the color confinement essentially. In the following study, it is going to apply the complex-sedenions to explore the varieties and contributions of the color degrees of freedom, discussing the influence of the color confinement on some physical phenomena.

\appendix

\section{Wave function}

In the subspace $\mathbb{H}_g$ of the complex-octonion space $\mathbb{O}$ , the quaternion angular momentum is written as,
\begin{eqnarray}
\mathbb{L}_g = L_{g0} + i \textbf{L}_g^i + \textbf{L}_g ~ ,
\end{eqnarray}
where $L_{g0} = ( u_0 P_{g0} + \textbf{u} \cdot \textbf{P}_g ) + k_{eg}^{~~2} ( \textbf{U}_0 \circ \textbf{P}_{e0} + \textbf{U} \cdot \textbf{P}_e )$, $\textbf{L}_g = ( \textbf{u} \times \textbf{P}_g + k_{eg}^{~~2} \textbf{U} \times \textbf{P}_e )$, $\textbf{L}_g^i = ( P_{g0} \textbf{u} - u_0 \textbf{P}_g ) + k_{eg}^{~~2} ( \textbf{U} \circ \textbf{P}_{e0} - \textbf{U}_0 \circ \textbf{P}_e )$. The part, $L_{g0}$ , covers the term, $\textbf{u} \cdot \textbf{P}_g$ . The part, $\textbf{L}_g^i$ , contains the term, $P_{g0} \textbf{u}$ , and is similar to the electric dipole moment. The orbital angular momentum, $\textbf{L}_g$ , includes the term, $\textbf{u} \times \textbf{P}_g$ , and is similar to the magnetic dipole moment. $\mathbb{U}_g = i u_0 + \textbf{u} $ . $\mathbb{U}_e = i \textbf{U}_0 + \textbf{U} $ . $\mathbb{P}_g = i P_{g0} + \textbf{P}_g $ . $\mathbb{P}_e = i \textbf{P}_{e0} + \textbf{P}_e $ . $\textbf{L}_g = \Sigma L_{gk} \textbf{\emph{i}}_k$ . $\textbf{L}_g^i = \Sigma L^i_{gk} \textbf{\emph{i}}_k$ . $L_{gj}$ and $L^i_{gk}$ are all real.

The quaternion wavefunction, relevant to the quaternion angular momentum, is defined as, $\Psi_{Lg} = \mathbb{L}_g / \hbar$ , and
\begin{eqnarray}
\Psi_{Lg} = i L_9 exp \{ \textbf{\emph{i}}_9 ( \pi / 2 ) \} + L_q exp ( \textbf{\emph{i}}_q \alpha_q )   ~,
\end{eqnarray}
with
\begin{eqnarray}
\alpha_q = arc cos \{ [ ( P_{g0} \textbf{\emph{i}}_0 ) \cdot ( r_0 \textbf{\emph{i}}_0 )
+ ( \Sigma P_{gk} \textbf{\emph{i}}_k ) \cdot ( \Sigma r_k \textbf{\emph{i}}_k ) ] / ( \hbar L_q ) \}  ~,
\nonumber
\end{eqnarray}
where $\textbf{L}_g^i / \hbar = L_9 exp \{ \textbf{\emph{i}}_9 ( \pi / 2 ) \}$ , $( L_{g0} + \textbf{L}_g ) / \hbar = L_q exp ( \textbf{\emph{i}}_q \alpha_q )$ . $\textbf{\emph{i}}_9$ and $\textbf{\emph{i}}_q$ both are unit vectors in the complex-quaternion space $\mathbb{H}_g$ . $\textbf{\emph{i}}_9^{~2} = - 1$. $\textbf{\emph{i}}_q^{~2} = - 1$. $\alpha_q$ is real.

In case there are, $\mathbb{X}_g = 0$, $ \Sigma ( L_{gj}^i )^2 \ll \Sigma ( L_{gj} )^2$ , and $\mid L_{g0} \mid / \{ \Sigma ( L_{gj} )^2 \}^{1/2} \ll 1$, the angle $\alpha_q$ will be reduced to the angle $\alpha_q^\prime$ . Meanwhile the wavefunction $\Psi_{Lg}$ is degenerated approximately to the wavefunction,
\begin{eqnarray}
\Psi_{Lg}^\prime = L_q \textbf{\emph{i}}_q \circ exp ( - \textbf{\emph{i}}_q \alpha_q^\prime )  ~,
\end{eqnarray}
with
\begin{eqnarray}
\alpha_q^\prime = [ ( P_{g0} \textbf{\emph{i}}_0 ) \cdot ( r_0 \textbf{\emph{i}}_0 ) + ( \Sigma P_{gk} \textbf{\emph{i}}_k ) \cdot ( \Sigma r_k \textbf{\emph{i}}_k ) ] / ( \hbar L_q ) ~.
\nonumber
\end{eqnarray}

By means of the transformation, $ \Psi = - \textbf{\emph{i}}_q \circ \Psi_{Lg}^\prime$ , the wavefunction $\Psi$ can be written as,
\begin{eqnarray}
\Psi = L_q exp ( - \textbf{\emph{i}}_q \alpha_q^\prime )  ~.
\end{eqnarray}
Further, from the other transformation, $\Psi^{\prime\prime} = - \textbf{z} \circ \Psi_{Lg}^\prime$ , it is able to achieve another wavefunction,
\begin{eqnarray}
\Psi^{\prime\prime} = (- \textbf{z} ) \circ \{ \textbf{\emph{i}}_q \circ exp ( - \textbf{\emph{i}}_q \alpha_q^\prime ) \} ~,
\end{eqnarray}
where $\textbf{z} = \Sigma z_k \textbf{\emph{i}}_k$ , $L_q = 1$ , $z_k$ is a dimensionless real number.

The above reveals what plays an important role in the quantum theory is the wavefunction, $\Psi^{\prime\prime} = - \textbf{z} \circ \Psi_{Lg}^\prime$ , rather than $\Psi_{Lg}^\prime$ . Generally, from the wavefunction $\Psi_L$ , it is able to define an octonion wavefunction,
\begin{eqnarray}
\Psi_{ZL} = - \mathbb{Z}_L \circ \Psi_L ~,
\end{eqnarray}
where $\mathbb{Z}_L$ is an auxiliary quantity, and also an octonion dimensionless quantity, including $\textbf{\emph{i}}_g$ and $\textbf{z}$ . $\Psi_{ZL}$ covers $\Psi$ and $\Psi^{\prime\prime}$ and so on.

Similarly, in the quantum mechanics described with the complex-quaternions, other physical quantities and various wavefunctions, in the complex-quaternion space $\mathbb{H}_g$ , can be transformed into that in the conventional quantum mechanics. And that the method can be extended into other three complex-quaternion spaces.

\section{Dirac wave equation}

In certain circumstances, the wavefunction relevant to the octonion torque can be written as,
\begin{eqnarray}
\Psi_{WL} = - v_0 \{ i \mathbb{W}^\star / ( \hbar v_0 ) + \lozenge \} \circ ( \mathbb{Z}_L \circ \mathbb{L} )  ~ ,
\end{eqnarray}
or
\begin{eqnarray}
\Psi_{WL} = - ( i \mathbb{W}^\star + \hbar v_0 \lozenge ) \circ \Psi_{ZL}  ~ ,
\end{eqnarray}
where $ \Psi_{ZL} = ( \mathbb{Z}_L \circ \mathbb{L} ) / \hbar $ .

The octonion torque, $\mathbb{W}$ , comprises the contributions coming from the electromagnetic and gravitational fields. And the contribution cases of the octonion angular momentum, $\mathbb{L}$, are similar to that of $\mathbb{W}$ . In case $\Psi_{WL} = 0$, one wave equation relevant to the octonion torque will be derived from the above. Further this wave equation can be degenerated into the Dirac wave equation. If there is no applied electromagnetic and gravitational fields, $\Psi_{WL} = 0$ will be simplified to the wave equation, for the particle with the nonzero rest mass in a free motion state.

In case there are not only the electromagnetic potential but also gravitational potential, the octonion torque will be reduced to,
\begin{eqnarray}
\mathbb{W} \approx i W_{g0}^i + \textbf{W}_g  ~ ,
\end{eqnarray}
with
\begin{eqnarray}
&& W_{g0}^i \approx  k_p P_{g0} v_0 + k_{eg}^{~~2} ( - \textbf{A}_{e0} \circ \textbf{P}_{e0} ) - A_{g0} P_{g0}  ~ ,
\\
&& \textbf{W}_g \approx k_{eg}^{~~2} ( \textbf{A}_e \circ \textbf{P}_{e0} ) + P_{g0} \textbf{A}_g   ~ ,
\end{eqnarray}
where $\mathbb{A}_g = i A_{g0} + \textbf{A}_g $ . $\mathbb{A}_e = i \textbf{A}_{e0} + \textbf{A}_e $ . $P_{g0} = m^\prime v_0$ . $\textbf{P}_{e0} = ( \mu_e / \mu_g ) q V_0 \textbf{\emph{I}}_0 $ . $m^\prime$ is the density of gravitational mass, while $q$ is the density of electric charge. The speed of light, $v_0$ , is the scalar part of quaternion velocity, $\mathbb{V}_g = v_0 \partial_0 \mathbb{R}_g$. $\textbf{V}_0$ is the scalar-like part of 2-quaternion velocity, $\mathbb{V}_e = v_0 \partial_0 \mathbb{R}_e $ , with $\textbf{V}_0 = V_0 \textbf{\emph{I}}_0$ . In general, $V_0 / v_0 \approx 1 $ .

As a result, the octonion torque can be written as,
\begin{eqnarray}
\mathbb{W}  \approx  i k_p m^\prime v_0^{~2} - i ( k_{eg}^{~~2} \textbf{A}_{e0} \circ \textbf{P}_{e0} +  A_{g0} P_{g0} ) + k_{eg}^{~~2} \textbf{A}_e \circ \textbf{P}_{e0} +  P_{g0} \textbf{A}_g    ~ ,
\end{eqnarray}
where either of two terms, $\mathbb{A}_g$ and $P_{g0}$ , comprises the contribution of gravitational fields. Meanwhile $\mathbb{A}_e$ and $\textbf{P}_{e0}$ both involve the contribution of electromagnetic fields.

When $\Psi_{WL} = 0$, substituting the above in the wavefunction, Eq.(B2), regarding the octonion torque, will deduce the complex-octonion Dirac wave equation as follows,
\begin{eqnarray}
&& \{  i \hbar v_0 \partial_0 - ( k_{eg}^{~~2} \textbf{A}_{e0} \circ \textbf{P}_{e0} +  A_{g0} P_{g0} ) +  \hbar v_0 \nabla
\nonumber
\\
&& ~~~~~
 +  i ( k_{eg}^{~~2} \textbf{A}_e \circ \textbf{P}_{e0} +  P_{g0} \textbf{A}_g  )  + k_p m^\prime v_0^{~2}  \} \circ \Psi_{ZL} = 0  ~ ,
\end{eqnarray}
for the charged particle in the electromagnetic and gravitational fields.

It means that either of gravitational potential and electromagnetic potential makes a contribution to the wave equation regarding the octonion torque, when there are the gravitational and electromagnetic fields simultaneously.

\section{Schr\"{o}dinger wave equation}

Making use of one appropriate operator, $\lozenge_{DS}$ , it is able to derive the Schr\"{o}dinger wave equation from the Dirac wave equation. When this operator acts on the wavefunction, Eq.(B2), from the left side, one new wavefunction can be written as,
\begin{eqnarray}
\Psi_{WL}^\prime = \lozenge_{DS} \circ \Psi_{WL}  ~ ,
\end{eqnarray}
or
\begin{eqnarray}
\Psi_{WL}^\prime =  - v_0 \{  i \mathbb{W}_{DS} / ( \hbar v_0 ) + \lozenge \} \circ \{ [ i \mathbb{W}^\star / ( \hbar v_0 ) + \lozenge ] \circ ( \mathbb{Z}_L \circ \mathbb{L} ) \} ~ ,
\end{eqnarray}
where $ \lozenge_{DS} = i \mathbb{W}_{DS} / ( \hbar v_0 )+ \lozenge $ , with $ \mathbb{W}_{DS} = \mathbb{W}^\star + i ( 2 k_p m^\prime v_0^{~2} ) $ .

It is well known that the quaternions possess the associative and noncommutative properties. Meanwhile the octonions have the nonassociative and noncommutative properties, although they still satisfy the weak associative property. That is, two octonions, $\mathbb{Z}$ and $\mathbb{Y}$ , obey the following weak associative property.
\begin{eqnarray}
\mathbb{Z} \circ ( \mathbb{Z} \circ \mathbb{Y} ) = ( \mathbb{Z} \circ \mathbb{Z} ) \circ \mathbb{Y}  ~ .
\end{eqnarray}

Certainly the weak associative property of octonions and the associative property of quaternions both will facilitate us to simplify some wavefunctions. Consequently the above complex-octonion wavefunction is able to be written as,
\begin{eqnarray}
&& \Psi_{WL}^\prime = - v_0 \{ [ i \mathbb{W}^\star / ( \hbar v_0 ) + \lozenge ] \circ [ i \mathbb{W}^\star / ( \hbar v_0 ) + \lozenge ] \} \circ ( \mathbb{Z}_L \circ \mathbb{L} )
\nonumber
\\
&&~~~~~~~~~~
- v_0 [ - 2 k_p m^\prime v_0^{~2} / ( \hbar v_0 ) ] \circ [ i \mathbb{W}^\star / ( \hbar v_0 ) + \lozenge ] \circ ( \mathbb{Z}_L \circ \mathbb{L} )  ~ ,
\end{eqnarray}
and then the octonion wavefunction, $ \Psi_{WL}^\prime $ , can be simplified into,
\begin{eqnarray}
\Psi_{WL}^\prime \approx && - v_0 \{  [ E_w / ( \hbar v_0 ) + i \partial_0 ]^2 + [ i \textbf{W}_g / ( \hbar v_0 ) + \nabla ] \cdot [ i \textbf{W}_g / ( \hbar v_0 ) + \nabla ]
\nonumber
\\
&& - [ k_p m^\prime v_0^{~2} / ( \hbar v_0 ) ]^2 + i k ( k_{eg}^{~~2} \textbf{P}_{e0} \circ \textbf{B} + P_{g0} \textbf{b} ) / ( \hbar v_0 ) \}  \circ (\mathbb{Z}_L \circ \mathbb{L})   ~ ,
\end{eqnarray}
where the fourth term connects with the dimension of radius vector, $k$ , apparently. The energy part can be rewritten as, $W_{g0}^i = k_p m^\prime v_0^{~2} + E_w$ . The term $E_w$ includes $( \textbf{A}_0 \circ \textbf{P}_{e0} )$ and so on, while the term $\textbf{W}_g$ contains $( \textbf{A}_e \circ \textbf{P}_{e0} )$ and so forth.

In the above, the electromagnetic strength is, $\mathbb{F}_e = \textbf{F}_{e0} + \textbf{F}_e $ , while the gravitational strength is, $\mathbb{F}_g = F_{g0} + \textbf{F}_g $ . For the sake of convenience the paper adopts the gauge conditions, $\textbf{F}_{e0} = 0$ , and $F_{g0} = 0$ . Therefore, the electromagnetic strength is reduced to, $ \textbf{F}_e = i \textbf{E} / v_0 + \textbf{B} $ . And the gravitational strength is reduced to, $ \textbf{F}_g = i \textbf{g} / v_0 + \textbf{b} $ . Herein, $\textbf{E}$ is the electric field intensity. $\textbf{B}$ is the magnetic flux density. $\textbf{g}$ is the gravitational acceleration. The term, $\textbf{b}$ , is one component of gravitational strength, and is similar to the magnetic flux density.

When $\Psi_{WL}^\prime = 0$, we can deduce the Schr\"{o}dinger wave equation, described with the complex-octonion, for the charged particle. And the Schr\"{o}dinger wave equation can be written as follows,
\begin{eqnarray}
&& \{ [ E_w / ( \hbar v_0 ) + i \partial_0 ]^2 + [ i \textbf{W}_g / ( \hbar v_0 ) + \nabla ] \cdot [ i \textbf{W}_g / ( \hbar v_0 ) + \nabla ]
\nonumber
\\
&& ~~ - [ k_p m^\prime v_0^{~2} / ( \hbar v_0 ) ]^2 + i k ( k_{eg}^{~~2} \textbf{P}_{e0} \circ \textbf{B} + P_{g0} \textbf{b} ) / ( \hbar v_0 )  \} \circ \Psi_{ZL} = 0 ~ .
\end{eqnarray}

Now we can discuss the wavefunction, $\Psi^{\prime\prime}$ , which is one component of the wavefunction $ \Psi_{ZL} $ , in the space $\mathbb{H}_g$ . If the direction of unit vector, $\textbf{\emph{i}}_q$ , is not able to play a major role in the wavefunction $\Psi^{\prime\prime}$ , this unit vector will be replaced by the imaginary unit, $i$ . So the wavefunction $ \Psi^{\prime\prime} $ may be chosen approximately as,
\begin{eqnarray}
\Psi^{\prime\prime} = \Psi (\textbf{r}) exp ( - i E_s t / \hbar ) ~ ,
\end{eqnarray}
where $E_s$ is the energy, and $\Psi (\textbf{r})$ is one complex vector function.

Substituting the above into Eq.(C6) infers the wave equation,
\begin{eqnarray}
&& \{ - [ \textbf{W}_g - i ( \hbar v_0 ) \nabla ] \cdot [ \textbf{W}_g - i ( \hbar v_0 ) \nabla ] - ( k_p m^\prime v_0^{~2} )^2
\nonumber
\\
&& ~~~~~
+ ( E_w + k_p m^\prime v_0^{~2} + E^\prime )^2 + i k ( k_{eg}^{~~2} \textbf{P}_{e0} \circ \textbf{B}  + P_{g0} \textbf{b} ) \} \circ \Psi (\textbf{r}) = 0 ~ ,
\end{eqnarray}
where $ E_s = k_p m^\prime v_0^{~2} + E^\prime $, $ k_p m^\prime v_0^{~2} \gg E_w $, and  $ k_p m^\prime v_0^{~2} \gg E^\prime $.

Consequently the above will be simplified into,
\begin{eqnarray}
&& \{ - [ \textbf{W}_g - i ( \hbar v_0 ) \nabla ] \cdot [ \textbf{W}_g - i ( \hbar v_0 ) \nabla ]
\nonumber
\\
&&~~~~~
+ ( 2 k_p m^\prime v_0^{~2} ) ( E_w + E^\prime ) + i k ( k_{eg}^{~~2} \textbf{P}_{e0} \circ \textbf{B} + P_{g0} \textbf{b} ) \} \circ \Psi (\textbf{r}) = 0 ~ ,
\end{eqnarray}
further we obtain,
\begin{eqnarray}
&& E^\prime \Psi (\textbf{r}) = \{ [ \textbf{W}_g - i ( \hbar v_0 ) \nabla ] \cdot [ \textbf{W}_g - i ( \hbar v_0 ) \nabla ] /  ( 2 k_p m^\prime v_0^{~2} )
\nonumber
\\
&& ~~~~~~~~~~~~~~
- E_w + i k q V_0 \hbar \textbf{\emph{I}}_0 \circ \textbf{B} / ( 2 k_p m^\prime v_0 ) + i k \hbar \textbf{b} / ( 2 k_p )  \} \circ \Psi (\textbf{r}) ~ ,
\end{eqnarray}
where last two terms both are the torque rather than the energy, in the complex-octonion space. The term, $ \textbf{L}_{e0(q)} \circ \textbf{B} $, is the torque between the magnetic flux density, $\textbf{B}$ , with the quantized term, $\textbf{L}_{e0(q)} = k q V_0 \hbar \textbf{\emph{I}}_0 / ( 2 k_p m^\prime v_0 ) $ . Meanwhile the term, $ L_{g0(q)} \circ \textbf{b} $ , is the torque between the gravitational strength component $\textbf{b}$ with the quantized term, $ L_{g0(q)} = k \hbar / ( 2 k_p ) $ .

\section{Multiplicative closure}

The algebra of sedenions is noncommutative and nonassociative, but it is not an alternative algebra. The sedenions are in possession of the properties of multiplicative closure, containing the octonions, quaternions, complex-numbers and reals (Table 8). The sedenion multiplication table is in Table 8.
\begin{table}[h]
\caption{\label{tab:table1}Sedenion multiplication table ($\emph{\textbf{I}}_{g0}$ = 1).}
\centering
\begin{tabular}{c|cccc|cccc||cccc|cccc}
\hline\hline
$ $ & $1$ & $\emph{\textbf{I}}_{g1}$  & $\emph{\textbf{I}}_{g2}$ & $\emph{\textbf{I}}_{g3}$  & $\emph{\textbf{I}}_{e0}$ & $\emph{\textbf{I}}_{e1}$ & $\emph{\textbf{I}}_{e2}$  & $\emph{\textbf{I}}_{e3}$
& $\emph{\textbf{I}}_{w0}$ & $\emph{\textbf{I}}_{w1}$ & $\emph{\textbf{I}}_{w2}$  & $\emph{\textbf{I}}_{w3}$ & $\emph{\textbf{I}}_{s0}$  & $\emph{\textbf{I}}_{s1}$ & $\emph{\textbf{I}}_{s2}$ & $\emph{\textbf{I}}_{s3}$
\\
\hline
$1$ & $1$ & $\emph{\textbf{I}}_{g1}$  & $\emph{\textbf{I}}_{g2}$ & $\emph{\textbf{I}}_{g3}$  & $\emph{\textbf{I}}_{e0}$  & $\emph{\textbf{I}}_{e1}$ & $\emph{\textbf{I}}_{e2}$  & $\emph{\textbf{I}}_{e3}$
& $\emph{\textbf{I}}_{w0}$ & $\emph{\textbf{I}}_{w1}$ & $\emph{\textbf{I}}_{w2}$  & $\emph{\textbf{I}}_{w3}$ & $\emph{\textbf{I}}_{s0}$  & $\emph{\textbf{I}}_{s1}$ & $\emph{\textbf{I}}_{s2}$ & $\emph{\textbf{I}}_{s3}$
\\
$\emph{\textbf{I}}_{g1}$ & $\emph{\textbf{I}}_{g1}$ & $-1$ & $\emph{\textbf{I}}_{g3}$  & $-\emph{\textbf{I}}_{g2}$ & $\emph{\textbf{I}}_{e1}$ & $-\emph{\textbf{I}}_{e0}$ &  $-\emph{\textbf{I}}_{e3}$ & $\emph{\textbf{I}}_{e2}$
& $\emph{\textbf{I}}_{w1}$ & $-\emph{\textbf{I}}_{w0}$ & $-\emph{\textbf{I}}_{w3}$  & $\emph{\textbf{I}}_{w2}$ & $-\emph{\textbf{I}}_{s1}$  & $\emph{\textbf{I}}_{s0}$ & $\emph{\textbf{I}}_{s3}$ & $-\emph{\textbf{I}}_{s2}$
\\
$\emph{\textbf{I}}_{g2}$ & $\emph{\textbf{I}}_{g2}$ & $-\emph{\textbf{I}}_{g3}$ & $-1$ & $\emph{\textbf{I}}_{g1}$  & $\emph{\textbf{I}}_{e2}$  & $\emph{\textbf{I}}_{e3}$ & $-\emph{\textbf{I}}_{e0}$ & $-\emph{\textbf{I}}_{e1}$
& $\emph{\textbf{I}}_{w2}$ & $\emph{\textbf{I}}_{w3}$ & $-\emph{\textbf{I}}_{w0}$  & $-\emph{\textbf{I}}_{w1}$ & $-\emph{\textbf{I}}_{s2}$  & $-\emph{\textbf{I}}_{s3}$ & $\emph{\textbf{I}}_{s0}$ & $\emph{\textbf{I}}_{s1}$
\\
$\emph{\textbf{I}}_{g3}$ & $\emph{\textbf{I}}_{g3}$ & $\emph{\textbf{I}}_{g2}$ & $-\emph{\textbf{I}}_{g1}$ & $-1$ & $\emph{\textbf{I}}_{e3}$  & $-\emph{\textbf{I}}_{e2}$ & $\emph{\textbf{I}}_{e1}$  & $-\emph{\textbf{I}}_{e0}$
& $\emph{\textbf{I}}_{w3}$ & $-\emph{\textbf{I}}_{w2}$ & $\emph{\textbf{I}}_{w1}$  & $-\emph{\textbf{I}}_{w0}$ & $-\emph{\textbf{I}}_{s3}$  & $\emph{\textbf{I}}_{s2}$ & $-\emph{\textbf{I}}_{s1}$ & $\emph{\textbf{I}}_{s0}$
\\
\hline
$\emph{\textbf{I}}_{e0}$ & $\emph{\textbf{I}}_{e0}$ & $-\emph{\textbf{I}}_{e1}$ & $-\emph{\textbf{I}}_{e2}$ & $-\emph{\textbf{I}}_{e3}$ & $-1$ & $\emph{\textbf{I}}_{g1}$ & $\emph{\textbf{I}}_{g2}$  & $\emph{\textbf{I}}_{g3}$
& $\emph{\textbf{I}}_{s0}$ & $\emph{\textbf{I}}_{s1}$ & $\emph{\textbf{I}}_{s2}$  & $\emph{\textbf{I}}_{s3}$ & $-\emph{\textbf{I}}_{w0}$  & $-\emph{\textbf{I}}_{w1}$ & $-\emph{\textbf{I}}_{w2}$ & $-\emph{\textbf{I}}_{w3}$
\\
$\emph{\textbf{I}}_{e1}$ & $\emph{\textbf{I}}_{e1}$ & $\emph{\textbf{I}}_{e0}$ & $-\emph{\textbf{I}}_{e3}$ & $\emph{\textbf{I}}_{e2}$  & $-\emph{\textbf{I}}_{g1}$ & $-1$ & $-\emph{\textbf{I}}_{g3}$ & $\emph{\textbf{I}}_{g2}$
& $\emph{\textbf{I}}_{s1}$ & $-\emph{\textbf{I}}_{s0}$ & $\emph{\textbf{I}}_{s3}$  & $-\emph{\textbf{I}}_{s2}$ & $\emph{\textbf{I}}_{w1}$  & $-\emph{\textbf{I}}_{w0}$ & $\emph{\textbf{I}}_{w3}$ & $-\emph{\textbf{I}}_{w2}$
\\
$\emph{\textbf{I}}_{e2}$ & $\emph{\textbf{I}}_{e2}$ & $\emph{\textbf{I}}_{e3}$ & $\emph{\textbf{I}}_{e0}$  & $-\emph{\textbf{I}}_{e1}$ & $-\emph{\textbf{I}}_{g2}$ & $\emph{\textbf{I}}_{g3}$  & $-1$ & $-\emph{\textbf{I}}_{g1}$
& $\emph{\textbf{I}}_{s2}$ & $-\emph{\textbf{I}}_{s3}$ & $-\emph{\textbf{I}}_{s0}$  & $\emph{\textbf{I}}_{s1}$ & $\emph{\textbf{I}}_{w2}$  & $-\emph{\textbf{I}}_{w3}$ & $-\emph{\textbf{I}}_{w0}$ & $\emph{\textbf{I}}_{w1}$
\\
$\emph{\textbf{I}}_{e3}$ & $\emph{\textbf{I}}_{e3}$ & $-\emph{\textbf{I}}_{e2}$ & $\emph{\textbf{I}}_{e1}$  & $\emph{\textbf{I}}_{e0}$  & $-\emph{\textbf{I}}_{g3}$ & $-\emph{\textbf{I}}_{g2}$ & $\emph{\textbf{I}}_{g1}$  & $-1$
& $\emph{\textbf{I}}_{s3}$ & $\emph{\textbf{I}}_{s2}$ & $-\emph{\textbf{I}}_{s1}$  & $-\emph{\textbf{I}}_{s0}$ & $\emph{\textbf{I}}_{w3}$  & $\emph{\textbf{I}}_{w2}$ & $-\emph{\textbf{I}}_{w1}$ & $-\emph{\textbf{I}}_{w0}$
%---%
\\
\hline\hline
$\emph{\textbf{I}}_{w0}$ & $\emph{\textbf{I}}_{w0}$ & $-\emph{\textbf{I}}_{w1}$ & $-\emph{\textbf{I}}_{w2}$  & $-\emph{\textbf{I}}_{w3}$  & $-\emph{\textbf{I}}_{s0}$  & $-\emph{\textbf{I}}_{s1}$ & $-\emph{\textbf{I}}_{s2}$ & $-\emph{\textbf{I}}_{s3}$
& $-1$ & $\emph{\textbf{I}}_{g1}$  & $\emph{\textbf{I}}_{g2}$ & $\emph{\textbf{I}}_{g3}$ & $\emph{\textbf{I}}_{e0}$  & $\emph{\textbf{I}}_{e1}$ & $\emph{\textbf{I}}_{e2}$  & $\emph{\textbf{I}}_{e3}$
\\
$\emph{\textbf{I}}_{w1}$ & $\emph{\textbf{I}}_{w1}$ & $\emph{\textbf{I}}_{w0}$ & $-\emph{\textbf{I}}_{w3}$  & $\emph{\textbf{I}}_{w2}$  & $-\emph{\textbf{I}}_{s1}$  & $\emph{\textbf{I}}_{s0}$ & $\emph{\textbf{I}}_{s3}$ & $-\emph{\textbf{I}}_{s2}$
& $-\emph{\textbf{I}}_{g1}$ & $-1$ & $-\emph{\textbf{I}}_{g3}$ & $\emph{\textbf{I}}_{g2}$ & $-\emph{\textbf{I}}_{e1}$  & $\emph{\textbf{I}}_{e0}$ & $\emph{\textbf{I}}_{e3}$  & $-\emph{\textbf{I}}_{e2}$
\\
$\emph{\textbf{I}}_{w2}$ & $\emph{\textbf{I}}_{w2}$ & $\emph{\textbf{I}}_{w3}$ & $\emph{\textbf{I}}_{w0}$  & $-\emph{\textbf{I}}_{w1}$  & $-\emph{\textbf{I}}_{s2}$  & $-\emph{\textbf{I}}_{s3}$ & $\emph{\textbf{I}}_{s0}$ & $\emph{\textbf{I}}_{s1}$
& $-\emph{\textbf{I}}_{g2}$  & $\emph{\textbf{I}}_{g3}$ & $-1$ & $-\emph{\textbf{I}}_{g1}$ & $-\emph{\textbf{I}}_{e2}$  & $-\emph{\textbf{I}}_{e3}$ & $\emph{\textbf{I}}_{e0}$  & $\emph{\textbf{I}}_{e1}$
\\
$\emph{\textbf{I}}_{w3}$ & $\emph{\textbf{I}}_{w3}$ & $-\emph{\textbf{I}}_{w2}$ & $\emph{\textbf{I}}_{w1}$  & $\emph{\textbf{I}}_{w0}$  & $-\emph{\textbf{I}}_{s3}$  & $\emph{\textbf{I}}_{s2}$ & $-\emph{\textbf{I}}_{s1}$ & $\emph{\textbf{I}}_{s0}$
& $-\emph{\textbf{I}}_{g3}$  & $-\emph{\textbf{I}}_{g2}$ & $\emph{\textbf{I}}_{g1}$ & $-1$ & $-\emph{\textbf{I}}_{e3}$  & $\emph{\textbf{I}}_{e2}$ & $-\emph{\textbf{I}}_{e1}$  & $\emph{\textbf{I}}_{e0}$
\\ \hline
$\emph{\textbf{I}}_{s0}$ & $\emph{\textbf{I}}_{s0}$  & $\emph{\textbf{I}}_{s1}$ & $\emph{\textbf{I}}_{s2}$ & $\emph{\textbf{I}}_{s3}$ & $\emph{\textbf{I}}_{w0}$ & $-\emph{\textbf{I}}_{w1}$ & $-\emph{\textbf{I}}_{w2}$  & $-\emph{\textbf{I}}_{w3}$
& $-\emph{\textbf{I}}_{e0}$ & $\emph{\textbf{I}}_{e1}$ & $\emph{\textbf{I}}_{e2}$ & $\emph{\textbf{I}}_{e3}$ & $-1$ & $-\emph{\textbf{I}}_{g1}$ & $-\emph{\textbf{I}}_{g2}$  & $-\emph{\textbf{I}}_{g3}$
\\
$\emph{\textbf{I}}_{s1}$ & $\emph{\textbf{I}}_{s1}$  & $-\emph{\textbf{I}}_{s0}$ & $\emph{\textbf{I}}_{s3}$ & $-\emph{\textbf{I}}_{s2}$ & $\emph{\textbf{I}}_{w1}$ & $\emph{\textbf{I}}_{w0}$ & $\emph{\textbf{I}}_{w3}$  & $-\emph{\textbf{I}}_{w2}$
& $-\emph{\textbf{I}}_{e1}$ & $-\emph{\textbf{I}}_{e0}$ & $\emph{\textbf{I}}_{e3}$ & $-\emph{\textbf{I}}_{e2}$ & $\emph{\textbf{I}}_{g1}$ &$-1$ &  $\emph{\textbf{I}}_{g3}$  & $-\emph{\textbf{I}}_{g2}$
\\
$\emph{\textbf{I}}_{s2}$ & $\emph{\textbf{I}}_{s2}$  & $-\emph{\textbf{I}}_{s3}$ & $-\emph{\textbf{I}}_{s0}$ & $\emph{\textbf{I}}_{s1}$ & $\emph{\textbf{I}}_{w2}$ & $-\emph{\textbf{I}}_{w3}$ & $\emph{\textbf{I}}_{w0}$  & $\emph{\textbf{I}}_{w1}$
& $-\emph{\textbf{I}}_{e2}$ & $-\emph{\textbf{I}}_{e3}$ & $-\emph{\textbf{I}}_{e0}$ & $\emph{\textbf{I}}_{e1}$ & $\emph{\textbf{I}}_{g2}$ & $-\emph{\textbf{I}}_{g3}$  & $-1$ & $\emph{\textbf{I}}_{g1}$
\\
$\emph{\textbf{I}}_{s3}$ & $\emph{\textbf{I}}_{s3}$  & $\emph{\textbf{I}}_{s2}$ & $-\emph{\textbf{I}}_{s1}$ & $-\emph{\textbf{I}}_{s0}$ & $\emph{\textbf{I}}_{w3}$ & $\emph{\textbf{I}}_{w2}$ & $-\emph{\textbf{I}}_{w1}$  & $\emph{\textbf{I}}_{w0}$
& $-\emph{\textbf{I}}_{e3}$ & $\emph{\textbf{I}}_{e2}$ & $-\emph{\textbf{I}}_{e1}$ & $-\emph{\textbf{I}}_{e0}$ & $\emph{\textbf{I}}_{g3}$ & $\emph{\textbf{I}}_{g2}$  & $-\emph{\textbf{I}}_{g1}$ & $-1$
\\
\hline\hline
\end{tabular}
\end{table}

\section*{Competing Interests}

The author declares that there is no conflict of interest regarding the publication of this paper.

\begin{acknowledgments}
The author is indebted to the anonymous referees for their valuable comments on the previous manuscript. And I thank Dr. Wei Ye, one of my Ph.D. classmates, for the interest in this research and financial support. This project was supported partially by the National Natural Science Foundation of China under grant number 60677039.
\end{acknowledgments}


\begin{references}



\bibitem{color1}
      H. C. Chandola, H. C. Pandey, H. Nandan,
      ``Topology of QCD vacuum and color confinement",
      {\it Canadian Journal of Physics\/},
      vol. 80, no. 7, pp.745--754, 2002.

\bibitem{color2}
      A. Nakamura, T. Saito,
      ``Color confinement in Coulomb gauge QCD",
      {\it Progress of Theoretical Physics\/},
      vol. 115, no. 1, pp.189--200, 2006.

\bibitem{color3}
      M. Chaichian, K. Nishijima,
      ``Does color confinement imply massive gluons?"
      {\it European Physical Journal C\/},
      vol. 47, no. 3, pp.737--743, 2006.

\bibitem{color4}
      M. Eto, L. Ferretti, K. Konishi, G. Marmorini, M. Nitta, K. Ohashi, W. Vinci, N. Yokoi,
      ``Non-Abelian duality from vortex moduli: A dual model of color-confinement",
      {\it Nuclear Physics B\/},
      vol. 780, no. 3, pp.161--187, 2007.

\bibitem{color5}
      F. Wang, J.-L. Ping, H.-R. Pang, L.-Z. Chen,
      ``Color confinement multi quark resonance",
      {\it Nuclear Physics A\/},
      vol. 790, no. 1-4, pp.493--497, 2007.

\bibitem{color6}
      T. Suzuki, K. Ishiguro, Y. Koma, T. Sekido,
      ``Gauge-independent Abelian mechanism of color confinement in gluodynamics",
      {\it Physical Review D\/},
      vol. 77, no. 3, Article ID 034502, 2008.

\bibitem{color7}
      T. Suzuki, M. Hasegawa, K. Ishiguro, Y. Koma, T. Sekido,
      ``Gauge invariance of color confinement due to the dual Meissner effect caused by Abelian monopoles",
      {\it Physical Review D\/},
      vol. 80, no. 5, Article ID 054504, 2009.

\bibitem{color8}
      A. Yamamoto, H. Suganuma,
      ``Relevant energy scale of color confinement from lattice QCD",
      {\it Physical Review D\/},
      vol. 79, no. 5, Article ID 054504, 2009.

\bibitem{color9}
      M.-L. Yu, M.-M. Xu, Z.-Y. Liu, L.-S. Liu,
      ``Liquid property of sQGP obtained from a bond percolation model with color confinement",
      {\it Journal of Physics G\/},
      vol. 36, no. 12, Article ID 125001, 2009.

\bibitem{color10}
      J. Braun, H. Gies, J. M. Pawlowski,
      ``Quark confinement from color confinement",
      {\it Physics Letters B\/},
      vol. 684, no. 4-5, pp.262--267, 2010.

\bibitem{color11}
      S. M. Troshin, N. E. Tyurin,
      ``Unitarity and the color confinement",
      {\it Modern Physics Letters A\/},
      vol. 25, no. 40, pp.3363--3370, 2010.

\bibitem{color12}
      R. Kitano,
      ``Hidden local symmetry and color confinement",
      {\it Journal of High Energy Physics\/},
      vol. 2011, no. 11, Article ID 124, 2011.

\bibitem{color13}
      H. C. Pandey, H. C. Chandola, H. Dehnen,
      ``Color confinement and finite temperature QCD phase transition",
      {\it International Journal of Modern Physics A\/},
      vol. 19, no. 02, pp.271--285, 2004.

\bibitem{color14}
      S. J. Gates, K. Stiffler,
      ``Adinkra `color' confinement in exemplary off-shell constructions of 4D, N = 2 supersymmetry representations",
      {\it Journal of High Energy Physics\/},
      vol. 2014, no. 7, Article ID 51, 2014.

\bibitem{color15}
      S. J. Brodsky, G. F. de Teramond, H. G. Dosch,
      ``Light-front holographic QCD and color confinement",
      {\it International Journal of Modern Physics A\/},
      vol. 29, no. 21, Article ID 1444013, 2014.

\bibitem{color16}
      S. J. Brodsky,
      ``Light-front holography, color confinement, and supersymmetric features of QCD",
      {\it Few-Body Systems\/},
      vol. 57, no. 8, pp.703--715, 2016.

\bibitem{color17}
      D. E. Kharzeev, E. M. Levin,
      ``Color confinement and screening in the $\theta$-vacuum",
      {\it Physical Review Letters\/},
      vol. 114, no. 24, Article ID 242001, 2015.

%

\bibitem{weng1}
      Z.-H. Weng,
      ``Some properties of dark matter field in the complex octonion space",
      {\it International Journal of Modern Physics A\/},
      vol. 30, no. 35, Article ID 1550212, 2015.

%

\bibitem{quaternion1}
      J. D. Edmonds,
      ``Quaternion wave equations in curved space-time",
      {\it International Journal of Theoretical Physics\/},
      vol. 10, no. 2, pp.115--122, 1974.

\bibitem{quaternion2}
      F. A. Doria,
      ``A Weyl-like equation for the gravitational field",
      {\it Lettere al Nuovo Cimento\/},
      vol. 14, no. 13, pp.480--482, 1975.

\bibitem{quaternion3}
      J. G. Winans,
      ``Quaternion physical quantities",
      {\it Foundations of Physics\/},
      vol. 7, no. 5-6, pp.341--349, 1977.

\bibitem{quaternion4}
      W. M. Honig,
      ``Quaternionic electromagnetic wave equation and a dual charge-filled space",
      {\it Lettere al Nuovo Cimento\/},
      vol. 19, no. 4, pp.137--140, 1977.

\bibitem{quaternion5}
      A. Singh,
      ``Quaternionic form of the electromagnetic-current equations with magnetic monopoles",
      {\it Lettere al Nuovo Cimento\/},
      vol. 31, no. 5, pp.145--148, 1981.

\bibitem{quaternion6}
      S. P. Brumby and G. C. Joshi,
      ``Global effects in quaternionic quantum field theory",
      {\it Foundations of Physics\/},
      vol. 26, no. 12, pp.1591--1599, 1996.

\bibitem{quaternion7}
      S. P. Brumby, B. E. Hanlon and G. C. Joshi,
      ``Implications of quaternionic dark matter",
      {\it Physics Letters B\/},
      vol. 401, no. 3-4, pp.247--253, 1997.

\bibitem{quaternion8}
      V. Majernik,
      ``Quaternionic formulation of the classical fields",
      {\it Advances in Applied Clifford Algebras\/},
      vol. 9, no. 1, pp.119--130, 1999.

\bibitem{quaternion9}
      V. Majernik,
      ``The energy density of the quaternionic field as dark energy in the universe",
      {\it General Relativity and Gravitation\/},
      vol. 36, no. 9, pp.2139--2149, 2004.

\bibitem{quaternion10}
      H. T. Anastassiu, P. E. Atlamazoglou and D. I. Kaklamani,
      ``Application of bicomplex (quaternion) algebra to fundamental electromagnetics: a lower order alternative to the Helmholtz equation",
      {\it IEEE Transactions on Antennas \& Propagation\/},
      vol. 51, no. 8, pp.2130--2136, 2003.

\bibitem{quaternion11}
      S. M. Grusky, K. V. Khmelnytskaya and V. V. Kravchenko,
      ``On a quaternionic Maxwell equation for the time-dependent electromagnetic field in a chiral medium",
      {\it Journal of Physics A\/},
      vol. 37, no. 16, pp.4641--4647, 2004.

\bibitem{quaternion12}
      K. Morita,
      ``Quaternions, Lorentz group and the Dirac theory",
      {\it Progress of Theoretical Physics\/},
      vol. 117, no. 3, pp.501--532, 2007.

\bibitem{quaternion13}
      A. S. Rawat and O. P. S. Negi,
      ``Quaternion gravi-electromagnetism",
      {\it International Journal of Theoretical Physics\/},
      vol. 51, no. 3, pp.738--745, 2012.

\bibitem{quaternion14}
      A. J. Davies,
      ``Quaternionic Dirac equation",
      {\it Physical Review D\/},
      vol. 41, no. 8, pp.2628--2630, 1990.

%

\bibitem{weng2}
      Z.-H. Weng,
      ``Dynamic of astrophysical jets in the complex octonion space",
      {\it International Journal of Modern Physics D\/},
      vol. 24, no. 10, Article ID 1550072, 2015.

%

\bibitem{octonion1}
      R. Penney,
      ``Octonions and Dirac equation",
      {\it American Journal of Physics\/},
      vol. 36, no. 10, pp.871--873, 1968.

\bibitem{octonion2}
      S. Marques-Bonham,
      ``The Dirac equation in a non-Riemannian manifold III: An analysis using the algebra of quaternions and octonions",
      {\it Journal of Mathematical Physics\/},
      vol. 32, no. 5, pp.1383--1394, 1991.

\bibitem{octonion3}
      S. De Leo and K. Abdel-Khalek,
      ``Octonionic Dirac equation",
      {\it Progress of Theoretical Physics\/},
      vol. 96, no. 4, pp.833--845, 1996.

\bibitem{octonion4}
      J. Koplinger,
      ``Dirac equation on hyperbolic octonions",
      {\it Applied Mathematics and Computation\/},
      vol. 182, no. 1, pp.443--446, 2006.

\bibitem{octonion5}
      M. Gogberashvili,
      ``Octonionic electrodynamics",
      {\it Journal of Physics A\/},
      vol. 39, no. 22, pp.7099--7104, 2006.

\bibitem{octonion6}
      V. L. Mironov and S. V. Mironov,
      ``Octonic first-order equations of relativistic quantum mechanics",
      {\it International Journal of Modern Physics A\/},
      vol. 24, no. 22, pp.4157--4167, 2009.

\bibitem{octonion7}
      V. L. Mironov and S. V. Mironov,
      ``Octonic second-order equations of relativistic quantum mechanics",
      {\it Journal of Mathematical Physics\/},
      vol. 50, no. 1, Article ID 012302, 2009.

\bibitem{octonion8}
      V. L. Mironov and S. V. Mironov,
      ``Octonic representation of electromagnetic field equations",
      {\it Journal of Mathematical Physics\/},
      vol. 50, no. 1, Article ID 012901, 2009.

\bibitem{octonion9}
      A. R. Dundarer,
      ``Multi-instanton solutions in eight-dimensional curved space",
      {\it Modern Physics Letters A\/},
      vol. 6, no. 5, pp.409--415, 2011.

\bibitem{octonion10}
      S. Demir and M. Tanisli,
      ``A compact biquaternionic formulation of massive field equations in gravi-electromagnetism",
      {\it European Physical Journal - Plus\/},
      vol. 126, no. 11, pp.1--12, 2011.

\bibitem{octonion11}
      S. Demir,
      ``Hyperbolic octonion formulation of gravitational field equations",
      {\it International Journal of Theoretical Physics\/},
      vol. 52, no. 1, pp.105--116, 2013.

\bibitem{octonion12}
      S. Demir, M. Tanisli and T. Tolan,
      ``Octonic gravitational field equations",
      {\it International Journal of Modern Physics A\/},
      vol. 28, no. 21, Article ID 1350112, 2013.

\bibitem{octonion13}
      C. Castro,
      ``On octonionic gravity, exceptional Jordan strings and nonassociative ternary Gauge field theories",
      {\it International Journal of Geometric Methods in Modern Physics\/},
      vol. 9, no. 2, Article ID 1250021, 2012.

\bibitem{octonion14}
      S. Furui,
      ``The flavor symmetry in the standard model and the triality symmetry",
      {\it International Journal of Modern Physics A\/},
      vol. 27, no. 27, Article ID 1250158, 2012.

\bibitem{octonion15}
      S. Furui,
      ``Axial anomaly and the triality symmetry of octonion",
      {\it Few-Body Systems\/},
      vol. 54, no. 11, pp.2097--2111, 2013.

\bibitem{octonion16}
      S. Furui,
      ``Axial anomaly and the triality symmetry of leptons and hadrons",
      {\it Few-Body Systems\/},
      vol. 55, no. 11, pp.1083--1097, 2014.

\bibitem{octonion17}
      P. Kalauni and J. C. A. Barata,
      ``Reconstruction of symmetric Dirac-Maxwell equations using nonassociative algebra",
      {\it International Journal of Theoretical Physics\/},
      vol. 12, no. 3, Article ID 1550029, 2015.

\bibitem{octonion18}
      B. C. Chanyal, P. S. Bisht, and O. P. S. Negi,
      ``Generalized split-octonion electrodynamics",
      {\it International Journal of Theoretical Physics\/},
      vol. 50, no. 6, pp.1919--1926, 2011.

\bibitem{octonion19}
      B. C. Chanyal, P. S. Bisht and O. P. S. Negi,
      ``Octonionic non-abelian gauge theory",
      {\it International Journal of Theoretical Physics\/},
      vol. 52, no. 10, pp.3522--3533, 2013.

\bibitem{octonion20}
      B. C. Chanyal, V. K. Sharma and O. P. S. Negi,
      ``Octonionic gravi-electromagnetism and dark matter",
      {\it International Journal of Theoretical Physics\/},
      vol. 54, no. 10, pp.3516--3532, 2015.

%

\bibitem{sedenion1}
      J. Koplinger,
      ``Gravity and electromagnetism on conic sedenions",
      {\it Applied Mathematics and Computation\/},
      vol. 188, no. 1, pp.948--953, 2007.

\bibitem{sedenion2}
      J. Koplinger,
      ``Hypernumbers and relativity",
      {\it Applied Mathematics and Computation\/},
      vol. 188, no. 1, pp.954--969, 2007.

\bibitem{sedenion3}
      J. Koplinger,
      ``Signature of gravity in conic sedenions",
      {\it Applied Mathematics and Computation\/},
      vol. 188, no. 1, pp.942--947, 2007.

\bibitem{sedenion4}
      V. L. Mironov and S. V. Mironov,
      ``Sedeonic generalization of relativistic quantum mechanics",
      {\it International Journal of Modern Physics A\/},
      vol. 24, no. 32, pp.6237--6254, 2009.

\bibitem{sedenion5}
      S. V. Mironov and V. L. Mironov,
      ``Sedeonic equations of massive fields",
      {\it International Journal of Theoretical Physics\/},
      vol. 54, no. 1, pp.153--168, 2015.

\bibitem{sedenion6}
      V. L. Mironov and S. V. Mironov,
      ``Gauge invariance of sedeonic equations for massive and massless fields",
      {\it International Journal of Theoretical Physics\/},
      vol. 55, no. 7, pp.3105--3119, 2016.

\bibitem{sedenion7}
      S. Demir and M. Tanisli,
      ``Sedenionic formulation for generalized fields of dyons",
      {\it International Journal of Theoretical Physics\/},
      vol. 51, no. 4, pp.1239--1252, 2012.

\bibitem{sedenion8}
      B. C. Chanyal,
      ``Sedenion unified theory of gravi-electromagnetism",
      {\it Indian Journal of Physics\/},
      vol. 88, no. 11, pp.1197--1205, 2014.

\bibitem{sedenion9}
      B. C. Chanyala,
      ``Dual octonion electrodynamics with the massive field of dyons",
      {\it Journal of Mathematical Physics\/},
      vol. 57, no. 3, Article ID 033503, 2016.

%

\bibitem{weng3}
      Z.-H. Weng,
      ``Forces in the complex octonion curved space",
      {\it International Journal of Geometric Methods in Modern Physics\/},
      vol. 13, no. 6, Article ID 1650076, 2016.




\end{references}
\end{document}